%% file: paper.tex
\documentclass[letterpaper,twocolumn,10pt]{article}
\usepackage{usenix-2020-09}
\usepackage[total={7in,9in},top=1in,left=0.75in,right=0.75in,bottom=1in,headsep=0.5in]{geometry}

\usepackage{physics}
\usepackage{amsmath}
\usepackage{tikz}
\usepackage{mathdots}
\usepackage{comment}
\usepackage{yhmath}
\usepackage{cancel}
\usepackage{color}
\usepackage{siunitx}
\usepackage{array}
\usepackage{multirow}
\usepackage{amssymb}
\usepackage{gensymb}
\usepackage{tabularx}
\usepackage{extarrows}
\usepackage{booktabs}
\usetikzlibrary{fadings}
\usetikzlibrary{patterns}
\usetikzlibrary{shadows.blur}
\usetikzlibrary{shapes}

\usepackage[leftline=false,
            rightline=false,
            linewidth=.6pt,
            innerrightmargin=0pt,
            innerleftmargin=0pt,
            innertopmargin=0pt,
            innerbottommargin=0pt
]{mdframed}

\usepackage{enumitem}
\usepackage{amsmath}
\usepackage{caption}
\usepackage{subcaption}
\usepackage{graphicx}
\usepackage[utf8]{inputenc}
\usepackage{pgfplots}
\DeclareUnicodeCharacter{2212}{−}
\usepgfplotslibrary{groupplots,dateplot}
\usetikzlibrary{patterns,shapes.arrows}
\pgfplotsset{compat=newest}

\usepackage{url}
\usepackage{booktabs}

\usepackage{listings}
\usepackage{lstautogobble}
\usepackage{color}
\usepackage{xcolor}
\usepackage{zi4}

\usepackage{hyperref}
\usepackage{cleveref}

\definecolor{bluekeywords}{rgb}{0.13, 0.13, 1}
\definecolor{greencomments}{rgb}{0, 0.6, 0}
\definecolor{redstrings}{rgb}{0.9, 0, 0}
\definecolor{graynumbers}{rgb}{0.8353, 0.8353, 0.8353}
\definecolor{codegreen}{rgb}{0,0.6,0}
\definecolor{codegray}{rgb}{0.5,0.5,0.5}
\definecolor{codepurple}{rgb}{0.58,0,0.82}
\definecolor{backcolour}{rgb}{0.95,0.95,0.92}
\definecolor{backcolour}{rgb}{0.9568, 0.9568, 0.9568}
\definecolor{magenta}{rgb}{0.9294, 0.007, 0.549}
\definecolor{attackRed}{RGB}{228,72,56}

\lstdefinestyle{mystyle}{
    backgroundcolor=\color{backcolour},
    commentstyle=\color{greencomments},
    keywordstyle=\color{magenta},
    numberstyle=\tiny\color{codegray},
    stringstyle=\color{codepurple},
    basicstyle=\ttfamily\footnotesize,
    breakatwhitespace=false,
    breaklines=true,
    captionpos=b,
    keepspaces=true,
    numbers=left,
    numbersep=5pt,
    showspaces=false,
    showstringspaces=false,
    showtabs=false,
	  frameround=tttt,
    tabsize=2,
    columns=[c]fullflexible,
    escapeinside={(*@}{@*)},
    xleftmargin=8pt,
		autogobble=true
}

\renewcommand{\textbf}[1]{{\sffamily\selectfont\bfseries #1}}

\makeatletter
\renewcommand{\paragraph}{%
  \@startsection{paragraph}{4}%
  {\z@}{1ex \@plus .2ex \@minus .2ex}{-1em}%
  {\normalfont\normalsize\bfseries}%
}
\makeatother

\DeclareRobustCommand{\circled}[1]{\tikz[baseline=(char.base)]{
            \node[shape=circle,draw,inner sep=1pt] (char) {#1};}}

\makeatletter
\newcommand{\crefnames}[3]{%
  \@for\next:=#1\do{%
    \expandafter\crefname\expandafter{\next}{#2}{#3}%
  }%
}
\makeatother
\crefnames{part,chapter,section}{\S}{\S\S}
\crefnames{para,paragraph,subparagraph}{\P}{\P\P}
\crefnames{listing}{Listing}{Listings}
\crefnames{sublisting}{Listing}{Listings}
\crefnames{figure}{Figure}{Figures}
\crefnames{table}{Table}{Tables}
\crefnames{appendix}{Appendix}{Appendices}

\mdfdefinestyle{mdstyle}{
    backgroundcolor=gray!10,
    innerleftmargin=0.1cm,
    innerrightmargin=0.1cm,
    innertopmargin=0.1cm,
    innerbottommargin=0.1cm,
    leftline=true,
    rightline=true,
}

\newcommand\highlightbox[1]{
  \begin{mdframed}[style=mdstyle]
    #1
  \end{mdframed}
}

\begin{document}

\title{Reverse Engineering the Apple M1 Conditional Branch Predictor\\for Out-of-Place Spectre Mistraining}

\author{
{\rm Adam Tuby}\\
Tel Aviv University\\
{\rm \texttt{\small{adamtuby@gmail.com}}}
\and
{\rm Adam Morrison}\\
Tel Aviv University\\
{\rm \texttt{\small{mad@cs.tau.ac.il}}}
}

\date{}

\maketitle

\begin{abstract}

Spectre v1 information disclosure attacks, which exploit CPU conditional branch misprediction, remain unsolved in deployed software.
Certain Spectre v1 gadgets can be exploited only by \emph{out-of-place mistraining}, in which the attacker
controls a victim branch's prediction, possibly from another address space, by training a branch that aliases with the victim
in the branch predictor unit (BPU) structure.
However, constructing a BPU-alias for a victim branch is hard.
Consequently, practical out-of-place mistraining attacks use brute-force searches to randomly achieve aliasing.
To date, such attacks have been demonstrated only on Intel x86 CPUs.

This paper explores the vulnerability of Apple M-Series CPUs to practical out-of-place Spectre v1 mistraining.
We show that brute-force out-of-place mistraining fails on the M1.
We analytically explain the failure is due to the search space size, assuming (based on Apple patents) that the M1 CPU uses a variant of the TAGE conditional branch predictor.
Based on our analysis, we design a new BPU-alias search technique with reduced search space.
Our technique requires knowledge of certain M1 BPU parameters and mechanisms, which we reverse engineer.
We also use our newfound ability to perform out-of-place Spectre v1 mistraining to test if the M1 CPU implements hardware mitigations against cross-address space out-of-place mistraining---and find evidence for partial mitigations.

\end{abstract}

\section{Introduction}

A Spectre v1 attack exploits CPU conditional branch prediction to
leak data from a victim program's memory~\cite{Kocher2018spectre}.
The attack trains the CPU's branch prediction unit (BPU) to mispredict a branch in an instruction sequence in the victim,
referred to as a \emph{Spectre gadget}.
The gadget's mis-speculated execution is such that it reads data from an attacker-controlled location in the victim's address
space and leaks it over some microarchitectural covert channel~\cite{flush+,last_level_cache_practical,smother,MaliciousMMU} before being
squashed by the CPU.
The attacker obtains the leaked data by listening on the covert channel, possibly on another process or core.

Spectre v1 stands out in the class of transient execution attacks~\cite{canella2019systematic}, which
leak architecturally-inaccessible data to the attacker via the microarchitectural traces left by ``transient'' doomed-to-squash
instructions.
Spectre v1 is unique among these attacks in that, on one hand, CPU designers do not intend to mitigate it in hardware~\cite{intel-spectre-v1-fix,intel-spectre-v1-ceo,amd-spectre-v1-fix},
while on the other hand, it is not comprehensively mitigated in deployed software,%
\footnote{To our knowledge, comprehensive software mitigations such as speculative load hardening (SLH)~\cite{SLH} are not deployed in operating systems or browsers, likely due to their performance penalty~\cite{OfekConfusion}.}
resulting in recurring discoveries of vulnerable gadgets in Spectre-hardened browsers and operating system~(OS) kernels~\cite{johannesmeyer_kasper_2022,OfekConfusion,spookjs,iLeakage}.%
\footnote{This is in contrast to Meltdown-type attacks~\cite{Fallout,Lipp2018meltdown,RAGE,CrossTalk,ridl,CacheOut,Schwarz2019ZombieLoad,foreshadow,lvi}, which exploit CPU bugs that get fixed,
and other speculative execution attacks~\cite{Kocher2018spectre,spectre_returns,ret2spec,spec_variant_four,wikner_retbleed_2022,oleksenko2023hide,trujillo_inception_2023,wikner_phantom_2023},
for which mitigations are deployed in software~(e.g., \cite{retpoline,LinuxMitigations}) and/or hardware~(e.g., \cite{ibrs,csv2,ArmSpectreBHB,IntelSpectreBHB}).}

\begin{figure}
\begin{lstlisting}[language = C,
				   frame=single,
				   firstnumber = 1,
				   escapeinside={(*@}{@*)},
				   autogobble=true,
				   emph={memcpy, strlen, call, idx},
				   emphstyle={\color{attackRed}},
				   caption={Bounds check bypass, susceptible to in-place mistraining. ({\color{attackRed}Red} values are untrusted.)},
				   label=lst:spectre_v1_gadget_1]
if (idx < array1_len) { // bounds check (mispredicts true)
	x = array2[array1[idx] * 4096]; // leak array1[idx]
}
\end{lstlisting}
\vspace{-10pt}
\end{figure}

Most Spectre v1 exploits~\cite{Kocher2018spectre,iLeakage,spookjs,leakyPage} use \emph{in-place mistraining}~\cite{canella2019systematic}, in which the attacker influences a victim branch's prediction by causing the victim to execute that branch.
For example, in the classic Spectre v1 ``array bounds check bypass'' gadget~(\cref{lst:spectre_v1_gadget_1}), the attacker invokes the victim
(e.g., by calling a system call, for an OS victim) with a valid input such that the victim's execution trains the BPU to predict the bounds check passing.
The attacker then mounts the attack by invoking the victim with a malicious input, for which the victim performs and leaks the result of an out-of-bounds
access whose check mispredicts as in-bound.

Certain Spectre gadgets, however, cannot be exploited using in-place mistraining.
Such gadgets can only be exploited by \emph{out-of-place mistraining}~\cite{canella2019systematic}, in which the attacker
controls a victim branch's prediction by training an \emph{aliasing} branch whose BPU entry is the same as the victim's branch---possibly from another address space (e.g., attacker process vs. OS).
\Cref{lst:spectre_v1_gadget_2} shows an example of such a gadget, which is a simplified speculative
type confusion gadget~\cite{OfekConfusion}.
This gadget contains two mutually-exclusive \texttt{if} blocks: the first block places an untrusted value into a variable \texttt{x} and the second leaks the contents of \texttt{array1[x]}.
If mis-speculation causes both blocks to execute, the contents of an attacker-determined location is leaked.
However, such an execution cannot be achieved with in-place mistraining, as it requires mistraining to put the BPU into a state which is not consistent with any valid victim execution (making the same prediction for two branches whose conditions are mutually exclusive).

\begin{figure}
\begin{lstlisting}[language = C,
				   frame=single,
				   firstnumber = 1,
				   escapeinside={(*@}{@*)},
				   autogobble=true,
				   emph={val},
				   emphstyle={\color{attackRed}},
				   caption={Type confusion gadget, requires out-of-place mistraining. ({\color{attackRed}Red} values are untrusted.)},
				   label=lst:spectre_v1_gadget_2]
x = 0;
if (cond) { // mispredicts true
    x = val; // attacker-controlled value (e.g., input)
}
if (!cond) { // predicts true, x now attacker-controlled
	y = array2[array1[x] * 4096]; // leak array1[x]
}
\end{lstlisting}
\vspace{-10pt}
\end{figure}

While out-of-place mistraining expands the Spectre attack surface, it is challenging to perform.
Creating BPU-aliasing with a victim branch requires finding a collision both in the BPU's branch-to-entry mapping and in the victim BPU entry's tag, both of which are proprietary in-silicon functions of the victim branch's virtual address and the global branching history leading to the branch~\cite{HalfAndHalf,TAGE,OGEHL}.
Out-of-place mistraining attacks thus require brute-force searches to randomly find an alias~\cite{OfekConfusion,BHI} and/or reverse engineering of the BPU's structure and in-silicon hash functions~\cite{Indirector}.%
\footnote{References~\cite{BHI,Indirector} perform out-of-place mistraining of the indirect branch predictor for Spectre v2 attacks, but the underlying challenge is equivalent to that of mistraining the conditional branch predictor for a Spectre v1 attack.}

To date, attacks exploiting out-of-place mistraining have been demonstrated only on x86 CPUs.
However, Apple computing devices---used by millions of consumers---have switched from x86 CPUs to Apple's M-Series CPUs in recent years.
Apple's proprietary M-Series microarchitectures have received relatively little research scrutiny~\cite{pp1js0,BranchDifferent,leakyPage,Pacman,M1Explore,DougallM1,Augury,gofetch}, none of which explores out-of-place mistraining.

M-Series CPUs are ARM64-based.
Out-of-place mistraining has been shown possible on certain ARM64 CPUs~\cite{canella2019systematic}, but not on Apple's M-Series.
Moreover, this demonstration was a proof-of-concept done by mirroring the victim's code space in the attacker process.
But this synthetic approach is not possible in a real attack---such as a process attacking the OS kernel---which requires brute-force searching and/or reverse engineering of the BPU, as explained above.
This is not a theoretical gap: while Barberis et al.~\cite{BHI} successfully performed brute-force out-of-place mistraining on Intel CPUs, they report brute-search failing on ARM64 CPUs.
Another intriguing aspect of M-Series CPUs is that Apple has patented hardware mitigations against out-of-place mistraining of indirect branches for Spectre v2 attacks~\cite{AppleSecurityPatent}---but not against
conditional branch predictor mistraining for Spectre v1 attacks.

This paper's goal is to study the vulnerability of M-Series CPUs to out-of-place mistraining of the conditional branch predictor.
We tackle the following research questions:

\highlightbox{Are Apple's M-Series CPUs vulnerable to brute-force out-of-place Spectre v1 mistraining?
Do M-Series CPUs block out-of-place Spectre v1 mistraining in hardware?}

Importantly, our goal is not to mount a full Spectre attack against a real victim, which also requires finding a vulnerable gadget.
Rather, we seek to discover if out-of-place mistraining remains part of the Spectre v1 attack toolbox on M-Series CPUs.
This can broaden the scope of potential Spectre gadgets, motivate work on finding them, and inform mitigation decisions (e.g., extending hardware mitigations to the conditional branch predictor).

We study the Apple M1 CPU, which features an ARM64-based  big.LITTLE design~\cite{BigLittle} of high-performance ``Firestorm'' cores and energy-efficient ``Icestorm'' cores (both of which we study).
Spectre v1 attacks have been demonstrated on the M1, both as proofs-of-concept~\cite{BranchDifferent,leakyPage} and in practice~\cite{leakyPage,spookjs}, but only by using in-place mistraining.

We make the following four contributions:

\paragraph{\circled{1} Brute-force mistraining on the M1 is costly~(\cref{sec:motivation})}
We empirically find that brute-force out-of-place mistraining on the M1 can take 1\,B attempts to succeed.
Hypothesizing (based on Apple patents~\cite{AppleBPUPatent,AppleSecurityPatent}) that the M1 CPU uses a variant of the TAGE conditional branch predictor~\cite{ISL-TAGE},
we explain this finding by analyzing the complexity of brute-force out-of-place mistraining against TAGE.
TAGE consists of a hierarchy of predictors, called \emph{components}, that make predictions based on the outcome history of all branches and on the predicted branch's history.
We show that this multi-component structure typically induces a search space that is larger than the tag collision search space targeted by prior brute-force searches~\cite{OfekConfusion,BHI}.

\paragraph{\circled{2} Forcing mistraining of the last component~(\cref{sec:LPC Gadget})}
Our analysis shows that when a branch's predictions are produced by the last TAGE component (predicting based on the longest history), brute-force search complexity
reduces to searching for a tag collision in that component.
Although we cannot control which TAGE component predicts the victim branch, we can still exploit this insight.
We devise a \emph{last provider component} (LPC) primitive to perform brute-force search in a way that deterministically forces each random mistraining attempt to update a random entry of the \emph{last} TAGE component.
This turns the brute-force search into a search for an alias of the victim branch in the last component.
Finding such an aliasing branch means that mistraining succeeds, as TAGE favors predictions based on longer histories and will thus use the aliasing branch's entry when predicting the victim branch.
Our evaluation shows that using the LPC primitive reduces brute-force mistraining complexity by $16\times$--$32\times$.
However, implementing the LPC primitive requires knowledge of certain BPU parameters and mechanisms.

\paragraph{\circled{3} Reverse engineering the M1 conditional BPU~(\cref{sec:RE})}
We reverse engineer the organization of the M1's BPU branch history register (BHR): the attributes of branch instructions that affect the BHR,  its history length and bit width, and parts of its update policy.
Our findings enable a simple technique for constructing an LPC primitive on M1 cores.

\paragraph{\circled{4} Conditional branch predictor isolation~(\cref{sec:BPU Isolation})}
We use the ability to perform out-of-place Spectre v1 mistraining, enabled by our reverse engineering, to test if M1 cores implement BPU isolation mechanisms that mitigate cross-address space out-of-place mistraining.
Curiously, we find that userspace/kernel BPU isolation appears to exist on the Icestorm microarchitecture \emph{but not on Firestorm}. We also find that no cross-process BPU isolation exists on either microarchitecture.

\paragraph{Responsible disclosure notice}
We responsibly disclosed our findings to Apple. Apple acknowledged the disclosure but did not share information about a planned response.

\section{Background}

\subsection{Apple M1}
\label{sec:apple_m_series_dual_core}

Apple M1 is a family of ARM64-based system-on-a-chip (SoCs) for desktop computers.
Apple M1 processors use a big.LITTLE design~\cite{BigLittle}, with two types of core microarchitectures: high-performance ``Firestorm'' cores and energy-efficient ``Icestorm'' cores.
We study the original M1 processor, which has four Firestorm cores and four Icestorm cores.

The ARM64 architecture used by M1 cores has four execution privilege levels, called \emph{exception levels}. Usermode programs execute in the least privileged execution mode, EL0. The kernel executes in the supervisor privilege level, EL1.

\subsection{Branch Prediction}

Modern superscalar out-of-order (OoO) cores~\cite{tomasulo1967efficient}, such as Apple's M1 cores~\cite{DougallM1}, decouple instruction fetch from execution.
The processor \emph{fetches} instructions from memory in program order, \emph{dispatching} them to an execution backend.
The backend is responsible for executing a fixed-size sliding \emph{instruction window} of the latest fetched instructions.
Instructions \emph{retire} in program order, exiting the instruction window once they have executed and committing their results to architectural state.
The backend computes dependencies between the instructions and schedules instruction execution regardless of their position in the instruction window, based only on availability of hardware execution resources and dependency constraints.

Processors rely on branch prediction~\cite{hennessy2011computer} to avoid stalling instruction fetching when encountering a branch instruction.
Because the correct instruction to fetch next is the branch's target address, which is unknown until the branch executes, the processor instead predicts the branch's target address.
Once the branch executes, the prediction's correctness is validated. If the prediction was correct, the branch instruction can eventually retire, with the relevant predictor structures updated to tune future predictions.
Otherwise, on a misprediction, the backend's state is rolled back to its last known valid state (which was checkpointed upon making the prediction), thereby squashing the mispredicted branch and all subsequent (mis-speculated) instructions in the instruction window.

The processor's branch prediction unit (BPU) is responsible for branch prediction, namely, predicting~(1)~whether the fetched instruction is a branch (before decoding), (2)~whether the branch is taken or not, and~(3)~the branch's target address.
BPUs typically employ different structures for each of these tasks. We focus on the conditional branch predictor, which predicts branch direction; the branch target buffer (BTB) is responsible for identifying branches and predicting their targets.

\subsection{TAGE Branch Predictor}
\label{sec:tage_background}

Based on Apple patent filings~\cite{AppleBPUPatent,AppleSecurityPatent}, we hypothesize that M1 processors employ a
TAgged GEometric history length (TAGE)~\cite{ISL-TAGE} conditional branch predictor. TAGE consists of a hierarchy of predictors,
called \emph{components}, that make predictions based on ``global'' histories (of all past branch outcomes) of
different lengths in addition to the ``local'' history of the predicted branch's past outcomes.
This multi-component structure enables TAGE to accommodate branches with varying correlation degrees to different history
lengths.

\begin{figure}
    \centering
    \captionsetup{justification=centering}
    \resizebox{\columnwidth}{!}{\input{./diagrams/tage.tex}}
    \caption{Abstract representation of the TAGE predictor.}
	\label{fig:tage_abstract}
\end{figure}
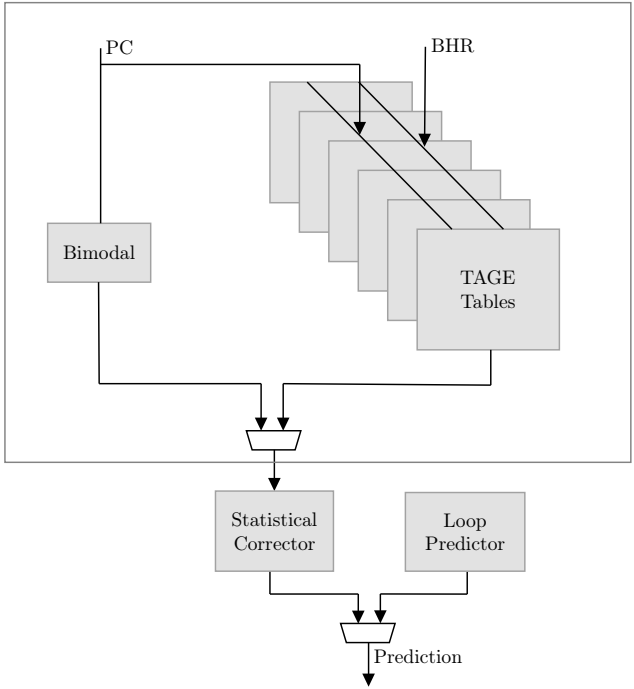

\Cref{fig:tage_abstract} depicts TAGE's structure. The branch history register (BHR) is a large (hundreds-bit range) shift
register that tracks the history of recently-executed branches, consisting of their outcomes, source and target addresses, etc.
TAGE's components consists of a default base predictor and a series of tagged tables.
The base predictor is a bimodal predictor~\cite{bimodal}, indexed directly by the predicted branch's address (PC), which
predicts based only on the branch's local history.
The tagged tables are indexed by different hash functions of the branch address and global histories of geometrically-increasing
lengths.
Table entries contain a (typically 3-bit) saturating counter whose most significant bit provides the prediction as well as a tag (derived
from the hash value) and a ``useful'' counter, which guides replacement decisions and is updated based on prediction accuracy.

At prediction time, TAGE simultaneously queries the base predictor and the tagged components.
Each tagged component is a set-associative structure, associated with a unique hash function that maps the branch's PC and a
suffix of the BHR to a set and tag value. The component provides the prediction made by the entry in the set whose tag value
matches the tag; if no such match exists, the component does not provide a prediction.
TAGE provides the prediction of the matching tagged component with the longest history or of the base predictor, if
no tagged component had a match. The component that provides the prediction is called the \emph{provider component}.

On branch resolution, the providing entry's saturating prediction counter is updated based on the branch's outcome
(incremented when the outcome is taken; decremented if not-taken).
If the prediction was incorrect, TAGE assumes that relying on a longer history will improve prediction accuracy.
TAGE thus tries to allocate an entry for the branch in a component using a longer history than the provider component,
if possible.
TAGE considers higher-level components which have entries (that the history and PC hash to) with a zero useful counter,
and probabilistically allocates one of these entries for the branch, biasing the selection toward components with shorter
histories.

Entry useful counters are updated each time the entry provides TAGE's overall prediction.
The entry's prediction is compared to the prediction that would have been made if the entry did not match.
If these predictions differ, the entry's useful counter is incremented or decremented, depending on whether
it predicted correctly or not. Useful counters are also decayed periodically, so that an entry cannot be
considered useful indefinitely.

To further refine its predictions, TAGE can be complemented with a loop predictor and a statistical corrector~\cite{ISL-TAGE}.
The loop predictor aims to anticipate shifts in the outcome of a loop's backwards branch.
The statistical corrector tracks statistically correlated branches that are not correctly predicted by the TAGE predictor.

\section{Threat Model}

We consider an unprivileged (non-root) attacker running on an M1-based Apple computer, iPhone, or iPad.
For example, the attacker can be malicious code embedded in a benign-looking application, even from the official App Store~\cite{MaliciousApps}.

The attacker's goal is to obtain information located in the address space of some victim program, either the macOS kernel or another userspace application.
We assume that the attacker can trigger execution of a Spectre gadget in the victim.
For instance, if the victim is the macOS kernel, the attacker can invoke a system call whose handling executes the gadget.
We consider only Spectre v1 attacks and do not assume the system is vulnerable to any other transient execution attack.

We assume that the attacker knows the address (in the victim address space) of their target data.
If the target data has a fixed address (statically-allocated data) in the victim, our assumption requires the attacker to either (1) break any victim address space layout randomization (ASLR) protections using orthogonal attacks~\cite{KASLR-break}
or (2) exploit the Spectre gadget to scan the victim's address space until identifying the desired data~\cite{Blindside,BHI,SLAM}.
If the target data is dynamically allocated and has no a priori known address, it can still be located by scanning the victim's address space.

\section{Motivation: Out-of-Place Mistraining} \label{sec:motivation}

Our motivation is to explore practical out-of-place branch mistraining~\cite{canella2019systematic} for Spectre v1 attacks on the Apple M1.
Prior work~\cite{BHI,OfekConfusion} on Intel CPUs performs out-of-place branch mistraining by brute-force, without relying on the BPU's structure~(\cref{sec:brute force def}).
However, brute-force on the M1 is costly due to its TAGE predictor's sophisticated structure~(\cref{sec:brute force fails}).

To address this problem, \cref{sec:LPC Gadget} describes how to make out-of-place mistraining possible given knowledge about certain aspects of the M1's BPU structure, which we reverse engineer in~\cref{sec:RE}.
We further leverage our reverse engineering results in~\cref{sec:BPU Isolation} to shed light on whether the BPU includes out-of-place
mistraining mitigation mechanisms~\cite{ExynosBPU} alluded to in an Apple patent~\cite{AppleSecurityPatent}.

\subsection{Brute-Force Out-of-Place Mistraining} \label{sec:brute force def}

The goal of out-of-place mistraining is to create BPU \emph{aliasing}, so that the prediction the BPU learns about an attacker \emph{mistraining} branch $b_A$ will also be the prediction made when the victim executes a branch $b_V$ in a Spectre gadget.

We define \emph{brute-force mistraining} as randomizing the attributes that determine a branch's BPU entry and tag until the mistraining succeeds.
In TAGE as well as other branch predictors, these attributes consist of the mistraining branch's virtual address (PC value) and its branch history (BHR value)~\cite{ITTAGE,ISL-TAGE,BHI,Demystifying,HalfAndHalf}. Brute-force mistraining is performed with the following steps:
\begin{enumerate}[leftmargin=*]
\item Place a sequence of interlinked branches at addresses picked uniformly at random, where the mistraining branch $b_A$ is the branch terminating the sequence.
\item Train the BPU on the mistraining branch by repeatedly executing the branch sequence.
\item Check if BPU aliasing occurred by trying the Spectre attack. It the attack succeeds, stop; otherwise, retry with a new sequence.
\end{enumerate}

Branch history injection (BHI~\cite{BHI}) performs exactly this mistraining, but its goal is to mistrain the branch target buffer (BTB) to mispredict the target of a victim indirect branch. On Intel CPUs, the BTB is a direct-mapped tagged structure~\cite{Demystifying}, where a branch's entry index and tag are computed from the PC and BHR values. Therefore, BTB aliasing succeeds if the attacker's PC and BHR values get mapped to the same index and tag as the victim branch's PC and BHR values.

A speculative type confusion attack against the Linux eBPF subsystem~\cite{OfekConfusion} performs a similar, but weaker, brute-force search to mistrain the conditional branch predictor. In that attack, the attacker controls the victim's branch history value. Therefore, the attack only randomizes the mistraining branch's PC value. However, we are interested in the general setting, where the attacker has no control of the victim branch's PC and BHR values, and thus focus on the brute-force search described above.

\subsection{Brute-Force Mistraining on the M1 is Costly} \label{sec:brute force fails}

Here, we show that brute-force out-of-place mistraining without fully reverse engineering the BPU's structure---%
such as the mistraining in prior work~\cite{BHI,OfekConfusion} targeting Intel CPUs---is costly to perform on the M1~(\cref{sec:brute force experiment}).
We posit that this complexity is due to the M1's use of a sophisticated TAGE predictor~(\cref{sec:tage_background}), alluded to in Apple's patents~\cite{AppleBPUPatent,AppleSecurityPatent}.
To substantiate this claim, \cref{sec:brute force theory} theoretically analyzes the complexity of brute-force out-of-place mistraining against a TAGE predictor and shows that TAGE induces a search space that is larger than the tag collision search space.
This analysis also grounds our technique for performing out-of-place mistraining using reverse engineered knowledge about the BPU structure~(\crefrange{sec:LPC Gadget}{sec:RE}).

\subsubsection{Experiment: Brute-Force Mistraining on the M1} \label{sec:brute force experiment}

We attempt brute force out-of-place mistraining on the Apple M1.
We adapt BHI's released code, which targets indirect branch prediction, to conditional branches.
\Cref{table:brute-force} reports the number of successful mistraining attempts over a 26-hour period of continuous mistraining attempts, as well as the probable size of the search space for randomly finding BPU aliasing implied by these observations.
To estimate the search space size, we rely on the fact that if the search space size is $2^k$, then the number of mistraining successes of $n$ attempts follows the Binomial distribution $Bin(n,\dfrac{1}{2^k})$, as the mistraining attempts are independent Bernouli trials (they have the same success probability $\dfrac{1}{2^k}$ and a boolean success/failure outcome).
We can therefore use the Chernoff bound to obtain an upper bound on the probability of the experiment's observed outcome for each possible search space size, and report the size for which the upper bound becomes a trivial $\approx 1$ (for other sizes, the bound is typically infinitesimally small).
Overall, the search space sizes of $2^{27}$--$2^{30}$ implied by our experiment are orders of magnitudes larger than the search space sizes of $2^{14}$--$2^{17}$ reported by BHI for Intel cores~\cite{BHI}.
Our results can also explain BHI's report of brute force search failing on (non-Apple) ARM processors~\cite{BHI}---for the search space sizes we estimate, brute-force search fails unless one performs hundreds of millions of trials.

\begin{table}[t]
\centering
\small
\begin{tabular}{p{.075\textwidth} c c c}
\toprule
\multirow{2}{*}{\textbf{Core}} & \multirow{2}{*}{\textbf{\# Trials}} &  \multirow{2}{*}{\textbf{\# Success}} & \textbf{Probable search} \\
              &                    &                                                        &  \textbf{space size} \\
\midrule
Firestorm & 1\,B & 1 & $2^{30}$ \\
Icestorm & 650\,M & 4 & $2^{27}$\\
\bottomrule
\end{tabular}
\caption{Success of brute force out-of-place mistraining (carried out over a 26-hour period) and the search space size implied by the observed success rate.}
\label{table:brute-force}
\end{table}

\subsubsection{Analysis: Brute-force Mistraining Against TAGE} \label{sec:brute force theory}

TAGE has no unique ``BPU entry'' for a branch, as a branch can have entries in multiple TAGE components.
This section theoretically analyzes the impact of this property on the complexity of brute-force mistraining.

\paragraph{Preliminaries}
We consider a TAGE predictor with $T$ tagged tables.
For simplicity, we assume that all tagged tables have the same structure: $2^s$ sets $\times$ $2^w$ ways, and $t$-bit tags.

We denote the PC value of the victim branch and the BHR value when the victim branch executes by $b_V$ and $h_V$, respectively.
Similarly, we denote the PC and BHR value of the mistraining branch during a mistraining attempt by $b_A$ and $h_A$, respectively.

We say that $b_A$ and $b_V$ \emph{alias in TAGE component $C_i$} if $C_i$'s hash function maps $\left<h_V, b_V\right>$ and $\left<h_A, b_A\right>$ to the same set and tag values.
Because all TAGE tables have the same structure, the probability for $v_A$ and $b_V$ aliasing in $C_i$ is $p_i=p=\dfrac{1}{2^{s+t}}$.

\paragraph{Analysis}
Let $C_i$ denote the TAGE provider component of $b_V$'s prediction.
We assume that $C_i$ is one of the tagged tables (i.e., not the base bimodal predictor).
For the moment, assume that $C_i$ is not the component associated with the longest branch history (i.e., $C_i$ is not the last component).

Consider a random trial performed by the mistraining process.
Suppose that $b_A$ and $b_V$ alias in $C_i$.
When the mistraining starts, $C_i$ contains $b_V$'s prediction.
Thus, during mistraining, $C_i$ will produce (mis)predictions for $b_A$.
TAGE will therefore allocate a new entry for $\left<h_A,b_A\right>$ in a \emph{higher-level} component $C_j$~(\cref{sec:tage_background}).
For mistraining to succeed, $b_A$ and $b_V$ must \emph{also} alias in $C_j$, so that when the victim executes, TAGE will produce the mistrained prediction from $C_j$.
This combined event only happens with probability $p^2$, because $C_j$ and $C_i$ use different indexing hash functions~\cite{ISL-TAGE,AppleBPUPatent}.

Now, suppose that $b_A$ and $b_V$ do not alias in $C_i$ (which happens with probability $1-p$).
In this case, mistraining can only succeed if it ends with (1) $b_A$ being allocated an entry in a component $C_j$ higher than $C_i$ and (2) $b_A$ and $b_V$ aliasing in $C_j$.
(Otherwise, TAGE will pick $C_i$'s prediction for $b_V$ when the victim executes.)
How can event (1) occur? It cannot be that mistraining updates a preexisting entry in $C_j$ (that $b_A$ aliases), as that would imply that $C_j$ was the provider component of $b_V$'s predictions.
It must thus be that during mistraining, TAGE allocates a $C_j$ entry for $b_A$, overwriting its tag.
Recall that TAGE chooses a component to allocate from probabilistically while favoring lower components.
Let $p'_i$ be the probability that TAGE chooses $C_j$ to be higher than $C_i$.
We cannot provide a closed-form expression for $p'_i$, because it depends on the exact state of the BPU, but below we show that $p'_i \approx \dfrac{1}{2^i}$.
Because the probability of $b_A$ and $b_V$ aliasing in the chosen component (event (2)) is $p$, it follows that mistraining succeeds with probability $\approx (1-p)\cdot \dfrac{1}{2^i} \cdot p$ in this case of $b_A$ and $b_V$ not aliasing in $C_i$.

The bound on $p'_i$ follows from the following TAGE mechanism: if $k < l$ and either of $C_k$ and $C_l$ can be allocated, then the probability to pick $C_k$ is twice the probability to pick $C_l$~\cite{TAGE}.
Our analysis assumes that $b_A$'s entry can be allocated from any component---i.e., that the ``useful'' counters of the entries that $b_A$ maps to in each component are equal.
(This assumption is based on the fact that prior mistraining trials randomized the content of the BPU, making all of its entries equally useful.)
It follows that the probability of TAGE picking component $i$ is $\dfrac{2^{T-i}}{2^T-1}$ for $1 \leq i \leq T$.
Thus, the probability of the chosen component being higher than $C_i$ is
\[
p'_i = \sum_{j=i+1}^T \dfrac{2^{T-j}}{2^T-1} =
\dfrac{2^{T-i}-1}{2^T-1} \approx \dfrac{1}{2^i}.
\]

The analysis above shows that due to TAGE's multiple-component design, the probability of a mistraining trial succeeding is $p_{succ} \approx p^2 + (1-p)\cdot \dfrac{1}{2^i} \cdot p$.
Now, the expected number of mistraining trials until mistraining succeeds is $1/p_{succ}$, which means that:
\highlightbox{Brute-force mistraining complexity against TAGE is significantly higher than that of only finding aliasing within $b_V$'s provider component, which requires only $1/p$ trials on average.}

\paragraph{What if $C_i$ is the last component?}
If no higher component than $C_i$ exists, the above analysis does not apply.
Mistraining will succeed with probability $p$, if $b_A$ aliases with $b_V$ in $C_i$.
Unfortunately, only $\tfrac{1}{T}$ of the BPU's tagged entries reside in the last component, so it is not likely that $b_V$'s provider component will be the last component.

\section{Forcing Mistraining of the Last Component} \label{sec:LPC Gadget}

This section describes an improved brute-force mistraining technique that reduces the complexity of out-of-place mistraining.
We assume knowledge of certain TAGE parameters and mechanisms (detailed in~\cref{sec:LPC gadget details}).
We reverse engineer this knowledge in~\cref{sec:RE}.

Our technique is based on the analysis of~\cref{sec:brute force theory}, which shows that when the victim branch's TAGE provider component is the last component (associated with the entire BHR), mistraining will succeed if the mistraining branch aliases with the victim branch in the
last component.
While we cannot control the TAGE provider component of the victim branch, we flip this observation on its head: we show how to mistrain in a way that \emph{deterministically forces} the mistraining branch's provider component to be the last component.

We describe our technique in the form of a \emph{last provider component (LPC) primitive}~(\cref{sec:LPC gadget details}).
The LPC primitive receives a branch slide---an interlinked sequence of branches, such as that generated by a brute-force mistraining trial---and performs mistraining with that branch slide in a way that guarantees the mistrained prediction will be installed in the last TAGE component.

Applying the LPC primitive to the random branch slides generated by brute-force mistraining~(\cref{sec:brute force def}) results in brute-force mistraining whose only success condition is for the mistraining branch to alias with the victim branch in the last component.
When this event occurs, TAGE's selection logic will pick the mistraining branch's malicious prediction when the victim branch executes, as TAGE picks the prediction made by the highest component providing a prediction.
As a result, applying the LPC primitive reduces the complexity of brute-force mistraining to the complexity of achieving aliasing in the last component.

\subsection{The Last Provider Component Primitive} \label{sec:LPC gadget details}

\subsubsection{Design}

The last provider component (LPC) primitive takes the following inputs:
\begin{enumerate}[leftmargin=*]
\item A \emph{branch slide}, i.e., a pointer to an interlinked sequence of branches ending with the mistraining branch. The goal of the branches preceding the mistraining branch is to set the BHR to some value.
\item The desired prediction for the mistraining branch: taken or not-taken.
\end{enumerate}

Let $h_A$ and $b_A$ denote the BHR value after executing the branch slide and the mistraining branch's PC value, respectively.
The goal of the LPC component is to train TAGE so that the last TAGE component will provide a prediction for $\left<h_A,b_A\right>$ and that its prediction will be the desired prediction.

To deterministically force TAGE to install a prediction for the mistrained branch in the last component, the LPC primitive performs two interleaved BPU training sessions,
in which the training branch has opposing outcomes and the BHR value differs only in the last (most significant) bit.
In the session where the mistraining branch has the desired outcome, it executes following the original branch slide.
In the other session, the LPC primitive modifies the branch slide so that its BHR value differs from $h_A$ only in the last bit.
As a result, TAGE can distinguish the two sessions only by using the full BHR, which forces the prediction to be made by the last component.

\begin{figure}[t]
\begin{lstlisting}[language = C,
				   frame=single,
				   firstnumber = 1,
				   escapeinside={(*@}{@*)},
				   autogobble=true,
				   emph={memcpy, strlen, call, bool},
				   emphstyle={\color{magenta}},
				   caption=Focusing on the Last Provider Component,
				   label=lst:lpc-gadget]
// @param br_slide: Pointer to branch slide, whose last member is the mistraining branch
// @param taken: Should the mistraining branch predict taken or not taken
void LPC_primitive(void (*br_slide)(), bool taken) {
  for (int i = 0; i < n_training_rounds; i++) {
    // Train desired prediction:
    // Execute branch slide + mistraining branch.
    execute_mistrain(br_slide, taken);

    // Lines 11-13 train the negated prediction, with
    // the BHR differing only in its last bit.
    flip_last_BHR_bit(br_slide); // modify the branch slide so that the last bit of its BHR value is flipped
    execute_mistrain(br_slide, !taken);
    flip_last_BHR_bit(br_slide); // revert branch slide changes
  }
}
\end{lstlisting}
\end{figure}

Listing~\ref{lst:lpc-gadget} shows the LPC primitive's pseudo code. It repeats the following steps sufficiently many times to train the BPU:
\begin{enumerate}[leftmargin=*]
\item
Execute the mistraining branch with the desired direction following the original branch slide. This is performed by the \texttt{execute\_mistrain} procedure, which receives a pointer to the branch slide and the desired direction.
This procedure sets up the mistraining branch's operand so that it takes the desired direction, jumps to the branch slide, and returns after the branch slide and mistraining branch execute.
\item
Modify the branch slide so that the BHR value resulting from its execution equals to $h_A$ with the last bit flipped. This is performed by the \texttt{flip\_last\_BHR\_bit} procedure. Exactly how to modify the branch slide to
achieve this effect is microarchitecture-specific and requires knowledge of TAGE parameters and mechanisms. We describe the procedure for the M1 in~\cref{sec:LPC impl}.
\item
Execute the mistraining branch with the opposite direction following the modified branch slide.
\item
Restore the branch slide to its original content.
\end{enumerate}

\subsubsection{Implementation} \label{sec:LPC impl}

Implementing the LPC primitive requires knowing microarchitecture-specific details about the TAGE structure, so that the primitive can modify
a branch slide in a way that flips the last bit of its BHR value.
First, we need to know the \emph{BHR update policy}: (1) which attributes of branches affect the BHR and (2) how. For (1), we need to know which subset of a branch's PC, operands, and outcome affects the BHR, which bits are used to update the BHR, and if there are additional relevant attributes.
For (2), we need to know how a branch's attributes are propagated into the BHR, e.g., by how many bits the BHR is shifted for each branch. We may also need to know if changing an attribute can affect multiple bits in the BHR, and how.

In addition to the BHR's update policy, we also need to know its \emph{width in bits}, which we refer to as the history length $HL$. $HL$ is necessary to determine which branch(s) in the branch slide to modify. For example, suppose the BHR update policy is such that the BHR is shifted by one bit for each executed branch.
Then we must require a branch slide of length $\geq HL+1$ (the last branch is the mistraining branch, so $HL$ extra branches are needed to fix a BHR value).
In fact, there is no point in a branch slide of length $k > HL+1$, because the first $k-(HL+1)$ branches cannot impact the BHR's value when the mistraining branch executes.

\paragraph{M1 implementation}
Assuming a branch slide of $HL+1$ branches, our reverse engineering~(\cref{sec:Effect of PC Address Bits}) shows that flipping the last bit of the BHR value can be achieved
by moving the first branch in the branch slide 4 bytes ahead or behind in memory.
This changes the second bit of the branch's PC value, which changes the last bit of the value held in the BHR when the last (mistraining) branch executes.

\subsubsection{Evaluation}

We demonstrate the efficacy of the LPC primitive by repeating the brute force out-of-place mistraining experiment~(\cref{sec:brute force experiment}) while employing the M1 implementation of the LPC primitive.
\Cref{table:brute-force-lpc} reports the results (search space sizes are calculated as in~\cref{sec:brute force experiment}).
Compared to a standard brute force search, a brute force search employing our LPC primitive succeeds with $32\times$ and $16\times$ fewer attempts on the Firestorm and Icestorm microarchitectures, respectively.

\begin{table}
\centering
\small
\begin{tabular}{p{.075\textwidth} c c c}
\toprule
\multirow{2}{*}{\textbf{Core}} & \multirow{2}{*}{\textbf{\# Trials}} &  \multirow{2}{*}{\textbf{\# Success}} & \textbf{Search space size} \\
              &                    &                                                        &  \textbf{(and improvement)} \\
\midrule
Firestorm & 1.1\,B & 33 & $2^{25}$\ \ \ ($32\times$) \\
Icestorm & 565\,M & 61 & $2^{23}$\ \ \  ($16\times$) \\
\bottomrule
\end{tabular}
\caption{Effect of the LPC primitive on the brute force out-of-place mistraining complexity.}
\label{table:brute-force-lpc}
\end{table}

\section{Reverse Engineering the M1 Conditional BPU} \label{sec:RE}

This section describes our reverse engineering of the M1 processor's conditional branch predictor.
Our analysis uses carefully-crafted microbenchmarks whose branch misprediction behavior exposes different aspects of the branch predictor's structure~(\cref{sec:Branch Aliasing Recognition}),
based on which we reverse engineer the branch history register's (BHR) organization~(\cref{BHR Organization}).

Our analysis assumes that M1 cores employ a TAGE conditional branch predictor~(\cref{sec:tage_background}).
This hypothesis is based on Apple patent filings~\cite{AppleBPUPatent,AppleSecurityPatent} and we confirm it experimentally.

We analyze both Firestorm and Icestorm cores. Unless otherwise indicated, presented findings are relevant to both core microarchitectures.

\subsection{Experimental Setup}

\paragraph{Platform}

We use an M1 Mac Mini desktop computer with 16\,GiB of memory, running macOS 12.4 (XNU kernel version 21.5.0).
We use the development kernel variant from Apple's kernel debug kit (KDK).

To observe branch mispredictions, our experiments use platform-specific performance monitor count (PMC) registers.
To enable our microbenchmarks to configure the performance monitoring unit (PMU), we load a kernel extension (kext) that accesses PMU control registers on behalf of a process.
By default, the PMC registers we use are only accessible in the supervisor execution privilege level (El1).
While our kext can be used to make the PMC registers userspace-accessible, we discovered that the macOS kernel frequently
resets the PMU configuration, reverting PMC register access back to privileged-only.
This behavior would limit the duration of our experiments: if a PMU configuration reset occurs during an experiment run, the process would die when it next tries to read a PMC register.
We thus need to permanently enable userspace PMC access, as some of our experiments execute for several hours.
We achieve this by manually patching the kernel, modifying every instruction that updates PMU control registers to avoid clearing our kext's PMU configuration.
The patching uses a search-and-replace of bytes in the kernel's machine code.

\paragraph{Execution Environment}
To minimize noise, we (1) run experiments one at a time; (2) bind the experiment to a specific core, so that it does not get context-switched to a different core
during its execution\footnote{We use the macOS \texttt{kern.sched\_thread\_bind\_cpu} sysctl call, which is available on macOS development kernels booted with the \texttt{enable\_skstb=1} argument.};
and (3) count only PMC events that occur during userspace execution, so that we do not incorrectly count kernel events that occur if the
core receives an external interrupt.

\subsection{Detecting Aliased Branches} \label{sec:Branch Aliasing Recognition}

Our experimental methodology is based on the ability to detect if two $\left<\text{history},\text{branch}\right>$ pairs alias in the BPU (i.e., in every TAGE component),
where by ``history'' we refer to the BHR value after executing some branch slide before the branch.
Aliasing occurs when prediction for both $\left<\text{history},\text{branch}\right>$ pairs is provided by the same TAGE component $C$ and they map to the same set and tag in $C$.
Thus, the prediction learned about one pair becomes the prediction made about the other pair.

Given a TAGE BPU whose entries maintain $c$-bit saturated counters for predictions, we determine if two pairs $\left<h,b\right>$ and $\left<h',b'\right>$
alias in the BPU with the following algorithm:

\begin{enumerate}[leftmargin=*,itemsep=0em]
	\item Fix $b$ and $b'$ with different outcomes (taken or not-taken).
	\item Repeatedly:
    \begin{enumerate}[leftmargin=*]
        \item Execute $2^{c+1}$ iterations of $\left<h,b\right>$.
        \item Execute $2^{c+1}$ iterations of $\left<h',b'\right>$.
    \end{enumerate}
    \item Aliasing occurs if the branch misprediction rate is $\approx 25$\%.
\end{enumerate}

Suppose the same TAGE component $C$ provides predictions for $b$ and $b'$ and the pairs alias in $C$.
Then executing $2^{c+1}$ iterations of a pair saturates the counter in their shared entry in $C$ to the maximum value fitting the outcome of the branch.
Switching to executing the other pair will then saturate the counter to the opposite side.
During the first $2^{c-1}$ iterations---before the counter's most significant bit flips---it will mispredict.
If $C$ is the last component, it will continue mispredicting, and so the misprediction rate will be near $\dfrac{2^{c-1}}{2^{c+1}}=25$\%.

However, if $C$ is not the last component, its mispredictions will drive TAGE to use higher-level components to predict the branches' outcomes.
Still, if the pairs alias in the BPU (i.e., in every component), the above scenario will repeat in any component, and eventually the last
component will become both branches' provider component.
Consequently, over enough iterations for this process to converge, the branches' misprediction rate will be $\approx 25$\%.

On the other hand, if the branches do not alias in the BPU, then TAGE will with high probability succeed in distinguishing them.
Thus, over enough iterations for convergence, the branches' misprediction rate will be 0\%.

\Cref{sec:Branch Aliasing Recognition Details} fills in the missing pieces needed to implement this algorithm: counting branch mispredictions~(\cref{sec:PMC Event Encodings})
and finding the width of the M1 BPU's saturating prediction counters, which we find is $c=3$~(\cref{sec:counter}).

\subsection{BHR Organization}
\label{BHR Organization}

Here, we reverse engineer the BHR's organization: the attributes of branch instructions that affect the BHR~(\crefrange{sec:Effect of PC Address Bits}{sec:Offset-Tuning}), its history length~(\cref{BHR History Length}) and bit width~(\cref{BHR Width}), and parts of its update policy~(\cref{sec:BHR Update Policy}).

\subsubsection{Methodology}
\label{sec:BHR-Exp-Scheme}

\begin{figure}
    \centering
    \resizebox{0.7\columnwidth}{!}{\input{./diagrams/bhr-baseline-experiment.tex}}
    \caption{Experiment template.}
	\label{BHR-Exp-Scheme}
\end{figure}
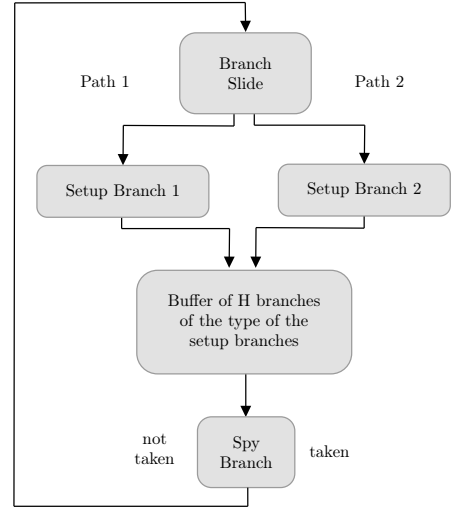

To test the effect of different branch attributes (e.g., instruction type, PC bits) on the BHR, we use the experimental template depicted in~\cref{BHR-Exp-Scheme}.
The experiment runs two execution paths that are identical save for a single branch instruction that differs in the attribute whose effect we wish to test.
Each execution path consists of the following parts:
\begin{enumerate}[leftmargin=*,itemsep=0em]
\item A shared branch slide large enough to bring any reasonably-sized BHR to a fixed (although unknown to us) state. The branch slide consists of both direct and indirect branches, as the BHR could be influenced by both types of branches.
\item A path-specific \emph{setup branch}. This is the branch that differs in the attribute tested. The branch slide falls through to the setup branch, so that even if the setup branches differ in some PC address bits, the BHR slide remains unchanged.
\item A shared sequence of $H$ branches of the same type as the setup branch. We vary $H$ to determine, if the attribute affects the BHR, how fast does the effect disappear.
\item A shared conditional \textit{spy branch}, whose outcome differs between the two paths.
\end{enumerate}

We execute each path 16 times in an alternating manner.
Except for (possibly) the setup branches, the address, type, and operands of every instruction in both paths are identical.
As a result, if the difference between the setup branches does not affect the BHR, the spy branch will have the same BHR value in both paths and thus get mapped to the same TAGE entry.
Because the spy branch has opposing outcomes in the two paths, that entry's 3-bit saturating prediction counter will mispredict for the first four executions of each path, resulting in a $4/16=25$\% misprediction rate.
If the difference between the setup branches does affect the BHR, the spy branch will have a different history in each path and should thus enjoy a near zero misprediction rate~(see~\cref{sec:Branch Aliasing Recognition}).

\paragraph{Implementation}
We implement the two logical paths by creating one physical path in memory and editing it at run-time when the experiment switches between paths.
This editing is trivial if the tested attribute is not a PC address bit, as we just need to edit the setup branch's instruction encoding.
In experiments testing effect of a PC address bit~(\cref{sec:Effect of PC Address Bits}), the physical path contains both setup branches separated by nop instructions.
The editing consists of removing the non-executing path's setup branch by overwriting it with nops and undoing the previous overwrite of the executing path's setup branch. This costly design choice is deliberate: avoiding the execution of potentially large amounts of nop instructions would require introducing a branching mechanism dependent on the logical path. Such a mechanism may alter the BHR, thereby compromising the experiment. Indeed, we demonstrate in this section that the BHR is influenced by virtually all attributes of branch instructions. As a result, implementing a branching mechanism to avoid nop instructions seems infeasible.

\paragraph{Comparison point}
As an added validation, we also report results from a baseline ``shadow branch'' whose outcome is identical to the spy branch's outcome.
The shadow branch is also preceded by a BHR-resetting branch slide.
We execute it after each execution of a tested path.
Consequently, the shadow branch is expected to experience 4 mispredictions for its first four out of sixteen executions after we switch between paths, similarly to the expected result for the spy branch if the tested
attribute does not affect the BHR.

\subsubsection{Effect of PC Address Bits} \label{sec:Effect of PC Address Bits}

Here, we test which bits of a branch's PC address affect the BHR.
We use the template of~\cref{BHR-Exp-Scheme}, where the address of the setup branches differs in some bit $i \geq 2$.%
\footnote{We count bit indexes from zero (the least significant bit).}
The underlying physical path contains a $2^i$ byte region of nops (to hold either setup branch), so each iteration in the experiment executes $2^i$-bytes worth of nops, resulting in execution times that grow exponentially with $i$, potentially extending to many hours.
As a result, (1)~we can run only a few experiment repetitions for bits $i \geq 24$, resulting in less pronounced spy branch misprediction rates; and~(2)~we cannot test values of $i$ beyond 30, but we later show through other means that bits $i \geq 32$ do not affect the BHR~(\cref{sec:Offset-Tuning}).
Overall, therefore, we leave only the exact effect of bit $31$ undetermined.

\begin{figure}
	\centering

		\resizebox{\columnwidth}{!}{\input{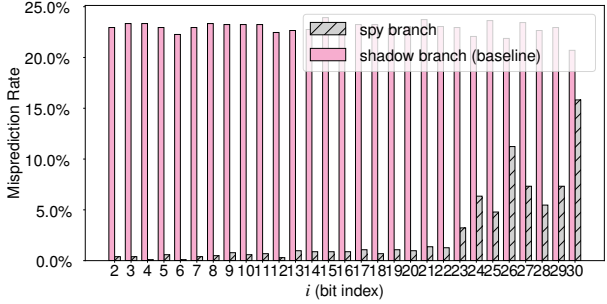}}

	\caption{Effect of conditional branch PC address bits on the BHR on the Firestorm cores (results for $H=0$). For bits that have effect, the spy branch's misprediction rate is $\approx 0$\%.}
	\label{CondBranchBitsBHR}
\end{figure}

\begin{figure}
	\centering

		\resizebox{\columnwidth}{!}{\input{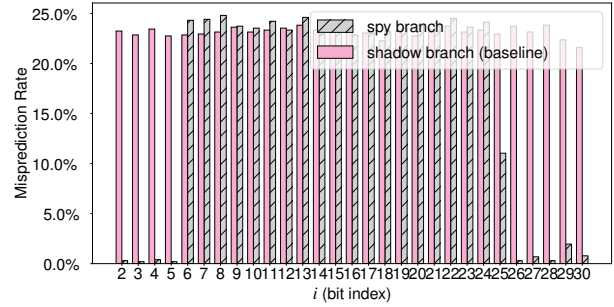}}

	\caption{Effect of indirect branch PC address bits on the BHR on the Firestorm microarchitecture (results for $H=0$). For bits that have effect, the spy branch's misprediction rate is $\approx 0$\%.}
	\label{IndirBranchBitsBHR}
\end{figure}

We first test the BHR effect of conditional branch PC address bits.
\Cref{CondBranchBitsBHR} shows the Firestorm results for $H=0$.
In this case, the setup branch immediately precedes the spy branch, so the results are not susceptible to interference from other branches in the history.
The results show that only bits 2--24 in a conditional branch's PC address affect the BHR.
We similarly test the BHR effect of indirect branch PC address bits.
The results~(\cref{IndirBranchBitsBHR}) indicate that only bits 2--5 and 25--30 affect the BHR.

These findings, combined with upcoming~\cref{sec:BHR Update Policy}, imply a simple technique for constructing an LPC primitive~(see~\cref{sec:LPC impl}).

\subsubsection{Effect of Branch Target Bits} \label{sec:Effect of Indirect Branch Target Bits}

We test which bits of a branch's target bits affect the BHR.
For conditional branches, the target is encoded in the instruction in the form of an immediate jump offset.
For indirect branches, the target is in a register operand.
We use the template of~\cref{BHR-Exp-Scheme}, where the target of the setup branches differs in some bit $i$.
Crucially, however, the address of the first branch $B$ that executes after the setup branch should be identical in both paths.
The underlying physical path thus contains a $2^i$ byte region of nops (to hold the target of either setup branch) that falls through to $B$.
Similarly to~\cref{sec:Effect of PC Address Bits}, this forces us to focus on bits $i \leq 30$, but in~\cref{sec:Offset-Tuning} we show through other means that target bits $i \geq 32$ do not affect the BHR.
Overall, therefore, we leave only the exact affect of target bit $31$ undetermined.

\begin{figure}
	\centering

		\resizebox{\columnwidth}{!}{\input{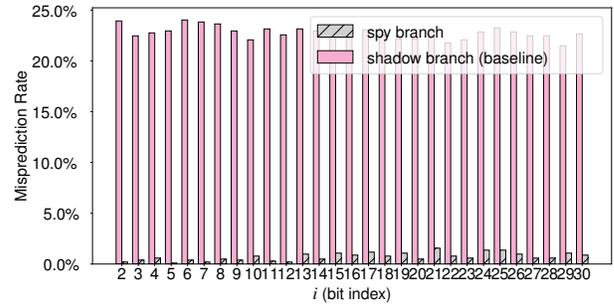}}

	\caption{Effect of indirect branch target address bits on the BHR on the Firestorm cores (results for $H=0$). For bits that have effect, the spy branch's misprediction rate is $\approx 0$\%.}
	\label{IndirTargetBitsBHR}
\end{figure}

We first test the BHR effect of indirect branch target address bits.
\Cref{IndirTargetBitsBHR} shows the Firestorm results for $H=0$, which show that every bit 2--30 affects the BHR.

\begin{figure}
	\centering

		\resizebox{\columnwidth}{!}{\input{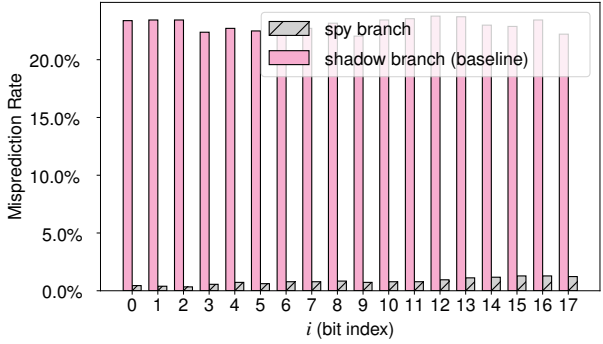}}

	\caption{Effect of conditional branch jump immediate offset bits on the BHR on the Firestorm microarchitecture (results for $H=0$). For bits that have effect, the spy branch's misprediction rate is $\approx 0$\%.}
	\label{CondImmediateBitsBHR}
\end{figure}

We similarly test the BHR effect of conditional branch jump immediate bits.
The immediate is a signed 19-bit field, so we only test the relevant bits.
The results~(\cref{CondImmediateBitsBHR}) indicate that all these bits affect the BHR.

\subsubsection{Non-Influence of High Address Bits} \label{sec:Offset-Tuning}

\begin{figure}[t!]
\begin{lstlisting}[language = {[x86masm]Assembler},
				   frame=single,
				   firstnumber = 1,
				   escapeinside={(*@}{@*)},
				   autogobble=true,
				   emph={adr, br, cmp, b.eq},
				   emphstyle={\color{magenta}},
				   caption=Branch slide updating direct and indirect branch PC and target addresses.,
				   label=lst:offset-tuning-code]
adr x9, #8
br x9           // jump to next instruction
cmp xzr, xzr
b.eq 1f         // jump to next instruction
...             // (sequence repeats)
\end{lstlisting}
\end{figure}

\begin{figure}
	\centering

		\centering
		\resizebox{\columnwidth}{!}{\input{./src/TAGE-SCL/BHR/offset-tuning/results/firestorm.pgf}}

	\caption{Effect of conditional/indirect branch PC and target address high bits on the BHR on the Firestorm cores. For bits that have effect, the spy branch's misprediction rate is $\approx 0$\%.}

	\label{fig:OffsetTuning}
\end{figure}
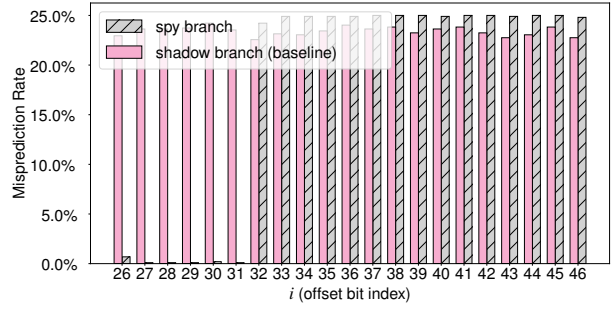

\begin{figure*}[h!]
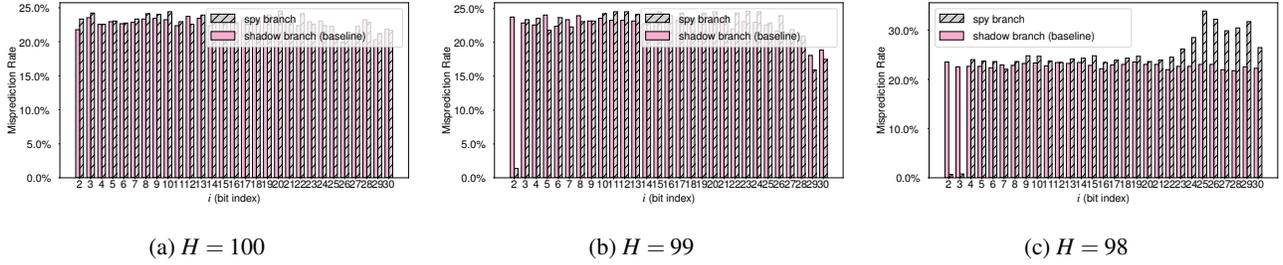

	\centering

	\begin{subfigure}[t]{0.32\textwidth}
		\centering
	\resizebox{\columnwidth}{!}{\input{./src/TAGE-SCL/BHR/conditional-branch-ip-bits/results/firestorm/H_100.pgf}}
		\caption{$H=100$}
	\end{subfigure}
	\begin{subfigure}[t]{0.32\textwidth}
		\centering
		\resizebox{\columnwidth}{!}{\input{./src/TAGE-SCL/BHR/conditional-branch-ip-bits/results/firestorm/H_99.pgf}}
		\caption{$H=99$}
	\end{subfigure}
	\begin{subfigure}[t]{0.32\textwidth}
		\centering
		\resizebox{\columnwidth}{!}{\input{./src/TAGE-SCL/BHR/conditional-branch-ip-bits/results/firestorm/H_98.pgf}}
		\caption{$H=98$}
	\end{subfigure}

	\caption{Effect of conditional branch PC address bits on the BHR on the Firestorm microarchitecture at distance $H$ from the spy branch varies. Icestorm results are qualitatively similar except for the values of $H$ in which the effects disappear.}
	\label{CondBranchBitsBHRLen}
\end{figure*}

We find that the upper 32 bits of conditional/indirect branch PC and target address do not affect the BHR.
To observe this, we change the experimental template of~\cref{BHR-Exp-Scheme} as follows.
To test the effect of bit $i$, we use two physical paths located in memory such that their start addresses differ only in bit $i$.
Each path contains an identical sequence of instructions, consisting of a branch slide followed by a spy branch (i.e., no setup branches and $H=0$).
The branch slide consists of a sequence of indirect and conditional branches to relative targets~(\cref{lst:offset-tuning-code}).
As it executes, updates to the BHR based on branch PC addresses and targets are made, but all these addresses differ only in bit $i$.
As a result, if bit $i$ in any of these attributes affects the BHR, the spy branch will have a different BHR value in each history and
we should see a 0\% misprediction rate for it.
Otherwise, we should see a 25\% misprediction rate.

We test the effects of bits 26--46, as the ARMv8 architecture uses 48-bit virtual addresses by default.
\Cref{fig:OffsetTuning} shows the results for the Firestorm microarchitecture, which show that bits 32--46 do not affect the BHR.
This finding simplifies certain Spectre v1 attacks against the kernel~(see~\cref{EL1_EL0_Isolation}).

\subsubsection{BHR History Length}
\label{BHR History Length}

The $H$ parameter in the experiments of~\crefrange{sec:Effect of PC Address Bits}{sec:Effect of Indirect Branch Target Bits} controls the distance in history (counted in branches) between the attribute
being tested and the spy branch.
Increasing $H$ thus eventually shows the distance at which the attribute stops affecting the BHR, i.e., that attribute's history length.

\Cref{CondBranchBitsBHRLen} demonstrates this effect for the conditional branch PC address bits on the Firestorm cores.
It shows that at $H=100$, no PC address bit affects the BHR; at $H=99$, only bit 2 still has an effect; and only bits 2--3 have an effect at $H=98$.
This trend continues as $H$ decreases until $H=77$, whose results are identical to those of $H=0$ and all intermediate values of $H$ (not shown).
The results are similar for the other attributes tested.
We conclude that the Firestorm's branch history length is 100.
Icestorm results (not shown) indicate that its history length is 60.

\subsubsection{BHR Update Policy}
\label{sec:BHR Update Policy}

The fact that branch attribute bits gradually lose their impact on the BHR as their distance in history increases hints at some ``shift-and-update'' BHR update policy~\cite{VladimirThesis}.
In such a policy, for each executed branch $B$, the BHR is shifted by $s$ bits (losing old history) and then $B$'s attributes are combined
into the BHR by some update operator $\odot$:
\[
\text{shift-and-}\odot_{s}(BHR, B) = (BHR \ll s) \odot attrs(B),
\]
where $attrs(\cdot)$ is a function of the relevant $B$ attribute bits.

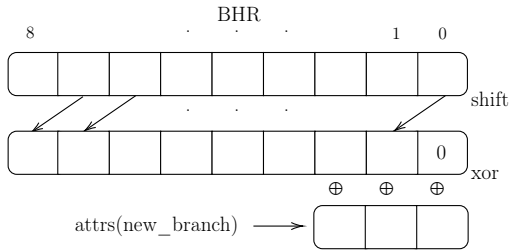
\begin{figure}[t]
	\centering
	\resizebox{0.8\columnwidth}{!}{\input{./diagrams/shift-and-xor.tex}}
	\caption{Example of a shift-and-XOR BHR update policy for a 9-bit BHR, shift value of 1, and a 3-bit attribute.}
	\label{ShiftAndXor}
\end{figure}

For example, \cref{ShiftAndXor} depicts a shift-and-XOR policy on a 9-bit BHR with a shift of 1 bit and a 3-bit attribute.

\begin{figure*}
	\centering

	\begin{subfigure}[!]{0.9\columnwidth}
		\centering
		\resizebox{\columnwidth}{!}{\input{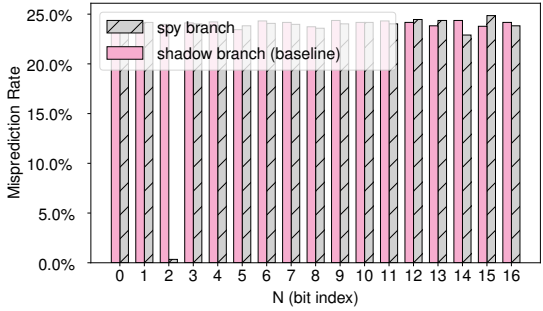}}
		\caption{Shift value $s=1$}
	\end{subfigure}
	\hfill
	\begin{subfigure}[!]{0.9\columnwidth}
		\centering
		\resizebox{\columnwidth}{!}{\input{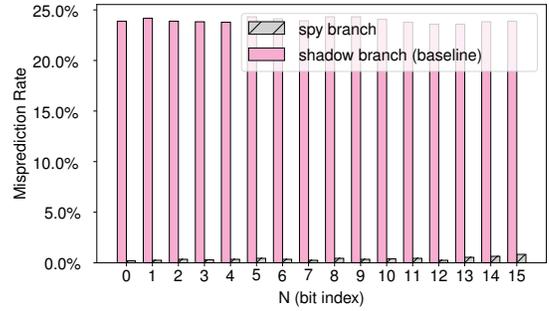}}
		\caption{Shift value $s=2$}
	\end{subfigure}

    \vspace{-8pt}
	\caption{BHR update policy of conditional branch immediate offset bits is shift-and-XOR with a shift value of 1 (Firestorm). High spy branch misprediction rate for bit $i$ and shift value $s$ indicates that when two conditional branches update the BHR consecutively, bit $i+s$ of the first branch's immediate gets XORed with bit $i$ of the second branch's immediate.}
	\label{BHRUpdatePolicy}
    \vspace{-10pt}
\end{figure*}

\begin{figure*}
	\centering

	\begin{subfigure}[t]{0.32\textwidth}
		\centering
		\resizebox{\columnwidth}{!}{\input{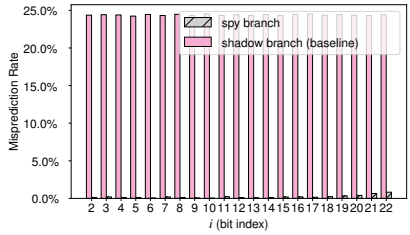}}
		\caption{Ground truth ($i-1$ branches omitted).}
	\end{subfigure}
	\begin{subfigure}[t]{0.32\textwidth}
		\centering
		\resizebox{\columnwidth}{!}{\input{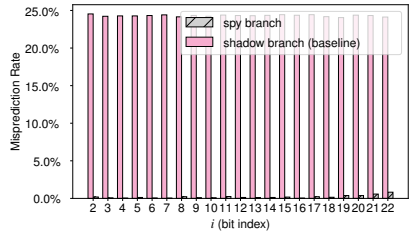}}
		\caption{Non-taken conditional branches.}
	\end{subfigure}
	\begin{subfigure}[t]{0.32\textwidth}
		\centering
		\resizebox{\columnwidth}{!}{\input{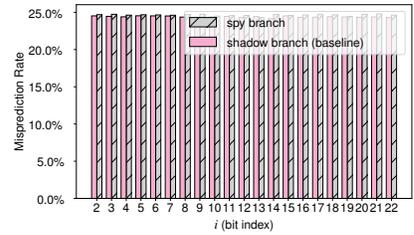}}
		\caption{Unconditional direct branches.}
	\end{subfigure}
    \vspace{-9pt}
	\caption{Effect of conditional branch PC address bit $i$ on the BHR when the BHR is updated by $100-(i-1)$ conditional taken branches followed by $i-1$ branches of a different type before the spy branch (Firestorm). For bits that have effect, the spy branch’s misprediction rate is $\approx 0$\%.}
	\label{fig:BranchTypesUsed}
    \vspace{-10pt}
\end{figure*}

\paragraph{Conditional branch jump offset}
We find that bits 0--1 and 3--16 of a conditional branch's immediate jump offset use a $\text{shift-and-XOR}_1$ update policy.
To prove this policy, we test for which bits $i$ and shift values $s$ the property $P(i,s)$ holds, where $P(i,s)$ says that when two conditional branches update the BHR consecutively, then bit $i+s$ of the first branch's immediate gets XORed with bit $i$ of the second branch's immediate.

We perform the tests by tweaking the experimental template of~\cref{BHR-Exp-Scheme} so that each execution path has \emph{two} setup branches, $B_1$ and $B_2$, that fall through to the spy branch (i.e., $H=0$).
The setup branches are identical except for their immediate fields. One path always has $imm(B_1)=imm(B_2)=1$.
The other path has $imm(B_1)=1+2^i$ and $imm(B_2)=1+2^{i+s}$.
If $P(i,s)$ holds, then bit $i$ in $imm(B_1)$ and bit $i+s$ in $imm(B_2)$ will cancel out when $B_2$ updates the BHR.
Consequently, the spy branch will have the same BHR value in both executions and will have a 25\% misprediction rate.
Otherwise, it will have a near 0\% misprediction rate.

\Cref{BHRUpdatePolicy} shows the test results on the Firestorm core, for all immediate bits and $s=1,2$ (results for $s=3,4$ are similar to $s=2$ and omitted).
The results confirm that all the immediate offset bits except 2 are XORed to the shifted (by one bit) BHR.
We conclude that the immediate bits are not themselves shifted before being XORed by observing that immediate bit $i$ loses its effect on the BHR in the experiment of~\cref{sec:Effect of Indirect Branch Target Bits}
at distance $H=100-i$ from the spy branch~(\cref{CondImmediateBitsBHR-Update} in~\cref{sec:appendix-figures}); if the bit were shifted first, its effect would be lost earlier.

Using similar experiments, we find that a subset of the bits of the other tested branch attributes also use a $\text{shift-and-XOR}_1$ policy. \Cref{table:update-policy} shows our findings.

\begin{table}
\centering
\small
\begin{tabular}{p{.15\textwidth} l c l c}
\toprule
\textbf{Attribute} & \multicolumn{2}{c}{\textbf{Bits}} & \multicolumn{2}{c}{\textbf{Bits shift}} \\
\midrule
Cond. branch PC	               & 2--4,6--24 & (Fig.\ref{BHRUpdatePolicyCondIP})         & 0 & (Fig.\ref{CondBranchBitsBHR-Update}) \\
Indirect branch target         & 2--20 		& (Fig.\ref{BHRUpdatePolicyIndirectTarget}) & 0 & (Fig.\ref{IndirTargetBitsBHR-Update}) \\
\bottomrule
\end{tabular}
\caption{Branch attribute bits that use a $\text{shift-and-XOR}_1$ BHR update policy. Referenced figures appear in~\cref{sec:appendix-figures}. }
\label{table:update-policy}
\end{table}

\paragraph{Non-taken conditional branch policy}
We show that when a conditional branch is not taken, the BHR is \emph{not} shifted. (But it \emph{is} updated in some undetermined manner~(\cref{sec:Conditional Branch Outcome Effect}).)
To test whether a non-taken conditional branch shifts the BHR, we test if when bit $i$ of a branch attribute loses its effect on the BHR at distance $h$ from the spy branch,
does its effect re-appear if we use a non-taken conditional branch as the last branch in the $h$-conditional/indirect branch buffer~(\cref{BHR-Exp-Scheme}).
If so, then the non-conditional branch does not shift the BHR.

In more detail, we tweak the experiment testing the effect of bit $i$ of a conditional branch's PC on the BHR~(\cref{sec:Effect of PC Address Bits}).
We test setup branches that differ in bit $i$ of their PC, for $i \in \left[2,23\right]$.
We execute a sequence of $HL$ branches between the setup and spy branches, consisting of $HL-(i-1)$ conditional taken branches, followed by $i-1$ branches of a different type.

\Cref{fig:BranchTypesUsed} shows the results. As a baseline, \cref{fig:BranchTypesUsed}(a) shows an experiment in which we omit the last $i-1$ branches.
In this case, given that bit $i$ of a conditional branch's PC address should lose its effect at distance $HL-(i-2)$~(\cref{CondBranchBitsBHR-Update}),
we expect to see every bit $i$ affecting the BHR (spy branch misprediction rate of $\approx 0$\%).
Indeed, \cref{fig:BranchTypesUsed}(a) matches expectations.
\cref{fig:BranchTypesUsed}(b) shows the results with $i-1$ non-taken conditional branches. Every bit still affects the BHR, demonstrating that
non-taken conditional branches do not shift the BHR.
Finally, \cref{fig:BranchTypesUsed}(c) shows the results with $i-1$ direct (unconditional) branches. All bits lose their BHR effect, which shows
that direct (unconditional) branches do shift the BHR.

\subsubsection{Non-Taken Conditional Branches Affect the BHR} \label{sec:Conditional Branch Outcome Effect}

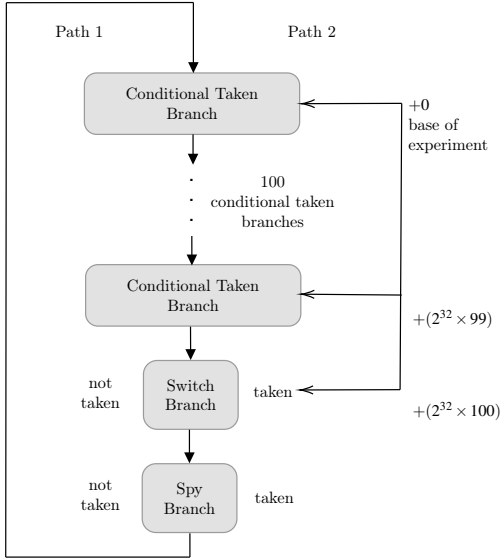
\begin{figure}
	\centering
    \captionsetup{justification=centering}
	\resizebox{0.8\columnwidth}{!}{\input{./diagrams/outcome-effect.tex}}
	\caption{Structure of test for whether a non-taken conditional branch affects the BHR.}
    \vspace{-10pt}
	\label{fig:OutcomeEffectExperiment}
\end{figure}

Although we find that non-taken conditional branches do not shift the BHR, this section finds that they do update it in some undetermined manner.
We use the experiment setup depicted in~\cref{fig:OutcomeEffectExperiment}.
The experiment consists of two logical execution paths, which are represented by a single physical path.
The physical path contains the following branches:
\begin{enumerate}[leftmargin=*]
\item A series of 100 interlinked taken conditional branches, leading to a \emph{switch branch}. The conditional branches are placed sparsely in the virtual address space. They are therefore linked by indirect branches (not shown in \cref{fig:OutcomeEffectExperiment}).
Each conditional branch is followed by an indirect branch that jumps to the next conditional branch.
\item A conditional switch branch, which either jumps to or falls through to a similar indirect branch. This indirect branch jumps to the spy branch.
\item The spy branch, whose outcome differs in each logical path.
\end{enumerate}

The logical paths differ only in the outcome of the spy branch and switch branch, which are identical to each other and toggled between logical executions.
As usual, the experiment executes each logical path 16 times in an alternating manner.

Crucially, all conditional branch PC addresses are 32-bit aligned, separated from each other by $2^o$ bytes, for some offset $o \geq 32$ (i.e., \cref{fig:OutcomeEffectExperiment} shows $o=32$).
All conditional branches have identical immediate jump offsets. Thus, the conditional branches differ from each other only in the high bits of their PC addresses.
Similarly, the interlinking indirect branches differ from each other only in the high bits of their PC and target addresses.
Therefore, as each pair of conditional/indirect branch differs from other pairs only in branch attribute bits that do not affect the BHR~(\cref{sec:Offset-Tuning}), each such pair---including the switch branch's, when its outcome is taken---performs exactly the same BHR update.

In particular, if non-taken conditional branches do not affect the BHR, the BHR value when the spy branch executes is the same in both executions, as the BHR effect of a taken switch branch is indistinguishable from the effect of only the slide's branches.
The spy branch should thus experience a $\approx 25$\% misprediction rate.
If, however, a non-taken conditional branch does affect the BHR in some way, the BPU will be able to distinguish the logical executions.
In this case, the spy branch should enjoy a $\approx 0$\% misprediction rate.

\begin{figure}[t]
	\centering
	\resizebox{\columnwidth}{!}{\input{./src/TAGE-SCL/BHR/conditional-branch-outcome/results/firestorm.pgf}}
	\caption{Result of experiment testing whether non-taken conditional branches affect the BHR, as the offset bit $o$ varies (Firestorm microarchitecture).}
	\label{OutcomeBHR}
\end{figure}
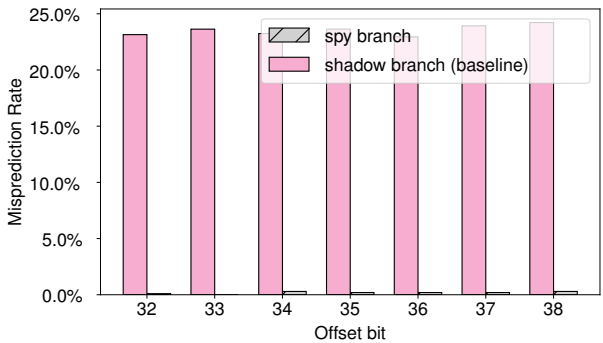

The results of the experiment (\Cref{OutcomeBHR}) confirm that a non-taken conditional branch does update the BHR in some undetermined manner.

\subsubsection{BHR Width}
\label{BHR Width}

The width in bits of a BHR that is shifted by $s$ bits for each history update and maintains a history of length $H$
is $s \cdot H$ bits.
It follows from the results of~\cref{sec:BHR Update Policy} and~\cref{BHR History Length} that the BHR width of Firestorm and Icestorm cores is 100 and 60 bits, respectively.

\section{Conditional Branch Predictor Isolation} \label{sec:BPU Isolation}

Apple has patented a hardware mitigation to block mistraining of indirect branch target addresses in Spectre v2 attacks~\cite{AppleSecurityPatent}.
The mitigation extends BTB entries with a \emph{security tag} derived from an identifier of the security context that created the BTB entry.
BTB predictions are ignored if the predicting entry's security tag does not match the security tag computed for the current security context.
Apple's mechanism thus dynamically \emph{isolates} BTB entries of different security contexts.

The patent describes deriving security tags to block different attack scenarios.
Cross-privilege attacks (e.g., process vs. OS) are blocked by storing the privilege level (EL).
Cross-process attacks are blocked by storing a process identifier.

In this section, we use the ability to perform out-of-place Spectre v1 mistraining to test if the M1 implements similar isolation mechanisms
for conditional branch prediction entries.
We test for BPU isolation between userspace and the kernel~(\cref{EL1_EL0_Isolation}) as well as between processes~(\cref{Process Isolation}).

In summary, we find that~(1)~userspace/kernel BPU isolation appears to exist on the Icestorm microarchitecture%
\footnote{We say ``appears to exist'' because we cannot make a certain determination about attack impossibility based on empirical observations. Theoretically, a different out-of-place mistraining technique might succeed where ours failed.}
\emph{but not on Firestorm} and (2)~no cross-process BPU isolation exists on either microarchitecture.

\subsection{Userspace/Kernel BPU Isolation}
\label{EL1_EL0_Isolation}

We test for userspace/kernel BPU isolation by attempting out-of-place mistraining of a victim kernel branch from userspace.
Because userspace and kernel virtual address spaces are distinguished by their most significant bits, our finding that these bits do not affect the BHR~(\cref{sec:Offset-Tuning})
simplifies userspace/kernel out-of-place mistraining if the attacker knows the kernel execution path leading to the victim path.
In this case, the attacker can mistrain by executing a ``shadow'' of the kernel's execution path in userspace, whose PC and target addresses match the kernel's addresses in the low 32 bits.

For simplicity, we create an artificial setup of such a scenario.
While this setup does not capture the complexities of a Spectre attack in the wild,%
\footnote{For example, although macOS kernel images are publicly available, the kernel employs KASLR, so the attacker has to try mistraining with different shadows to account for different address randomizations.}
it suffices to establish whether a userspace process can mistrain a kernel branch.

For the victim branch, we use a loadable kernel extension (kext).
The kext contains a branch slide, consisting of conditional taken branches and indirect branches followed by the conditional victim branch.
The kext provides a \texttt{sysctl} call which, when invoked, executes the victim branch (preceded by the branch slide) for 16 iterations and returns the observed misprediction rate.
The attacker userspace process performs out-of-place mistraining against a shadow of the victim branch slide and victim branch, and then invokes the victim via the \texttt{sysctl}.

If a core microarchitecture implements userspace/kernel BPU isolation, the attacker's mistraining will fail, and we expect to see the kext report a near-zero
misprediction rate.
Conversely, if mistraining succeeds, we expect to see a 25\% misprediction rate, as the shared BPU entry starts mispredicting based on the attacker's
training and gradually learns the correct victim prediction~(\cref{sec:Branch Aliasing Recognition}).

Figure~\ref{ExecutionModeIsolation} shows the results, indicating that userspace/kernel isolation appears to exist on the Icestorm microarchitecture, but it does not
exist on Firestorm.
To further validate our results, we also successfully use userspace out-of-space mistraining on a Firestorm core to exploit an in-kext Spectre v1 gadget that accesses the cache.

\begin{figure*}
\centering
\begin{minipage}[t]{\columnwidth}
	\centering

	\begin{subfigure}[t]{0.8\columnwidth}
		\centering
		\resizebox{\columnwidth}{!}{\input{./src/TAGE-SCL/Security-Tags/EL0_EL1-Security-Tag/results.pgf}}
	\end{subfigure}

	\caption{Userspace out-of-place mistraining against OS kernel. Misprediction rate of 25\% indicates mistraining success.}
	\label{ExecutionModeIsolation}
    \vspace{-10pt}
\end{minipage}\qquad
\begin{minipage}[t]{\columnwidth}
	\centering

	\begin{subfigure}[t]{0.8\columnwidth}
		\centering
		\resizebox{\columnwidth}{!}{\input{./src/TAGE-SCL/Security-Tags/PID-Security-Tag/results.pgf}}
	\end{subfigure}

	\caption{Cross-process out-of-place mistraining. Misprediction rate of 25\% indicates mistraining success.}
	\label{ProcessIsolation}
    \vspace{-10pt}
\end{minipage}
\end{figure*}

\subsection{Cross-Process BPU Isolation}
\label{Process Isolation}

We test for cross-process BPU isolation by attempting cross-process out-of-place mistraining.
Similarly to~\cref{EL1_EL0_Isolation}, we use an artificial setup, in which the attacker knows the PC values of the victim's branch and branch history.

We run two processes bound to the same core.
Both processes map an identical branch slide into the same address in the address space.
The branch slide terminates with a conditional victim branch (in the victim) or mistraining branch (in the attacker), which have opposing outcomes.
Each process blocks on a file descriptor. Upon receiving an event on the descriptor, it performs 16 iterations of the branch slide.

We alternate unblocking the attacker and victim.
Figure~\ref{ProcessIsolation} shows the results, which indicate that cross-BPU isolation does not exist.
We further validate these results by successfully using cross-process out-of-space mistraining to exploit a Spectre v1 gadget in another process, using the same gadget as in~\cref{EL1_EL0_Isolation}.

\section{Related Work}

\paragraph{Microarchitecture reverse engineering}
The conditional and indirect branch predictors on Intel and AMD x86 CPUs have been extensively reverse engineered for microarchitectural attack research~\cite{HornSpectre,Kocher2018spectre,JumpOverAslr,SetAndForget,ExploreBranchPredictor,Indirector}, software defense against branch-based side-channel attacks~\cite{HalfAndHalf}, and better microarchitectural understanding~\cite{Demystifying,VladimirPaper,VladimirThesis}.
Apple's M-Series CPUs have received significantly less attention.
Researchers have reversed engineered some details of the M1's microarchitectures~\cite{M1Explore,DougallM1} and well as its memory-dependent prefetcher~\cite{Augury,gofetch}, but to our knowledge, the structure of the M1's BPU has not been explored.
Our reverse engineering methodology is similar to that of prior work on Intel BPUs~\cite{HornSpectre,VladimirPaper,HalfAndHalf,Indirector}, but adapted to the unique aspects of the M1, such as unavailability of Hyper-Threading for executing the two experimental paths.

\paragraph{Spectre attacks on Apple's M-Series CPUs}
Hetterich and Schwarz~\cite{BranchDifferent} and Leaky.page~\cite{leakyPage} demonstrated the feasibility of Spectre v1 attacks
on the M-Series CPUs.
PACMAN used a Spectre v1-like attack to leak kernel pointer authentication code (PAC~\cite{PAC}) verification results to userspace~\cite{Pacman}.
Spook.js~\cite{spookjs} and iLeakage~\cite{iLeakage} mounted in-browser Spectre v1 attacks.
These attacks all employ \emph{in-place} mistraining~\cite{canella2019systematic} (i.e., training the victim branch by invoking it), whereas our work enables out-of-place mistraining.

\paragraph{Mitigations}
Out-of-place mistraining of the indirect branch predictor has been mitigated in hardware by introducing BPU flushing mechanisms~\cite{ibrs,csv2} or by isolating entries of different security domains~\cite{AppleSecurityPatent,ExynosBPU}.
To our knowledge, no such mechanisms are available for the conditional branch predictor.
Half\&Half~\cite{HalfAndHalf} is a software-only approach for partitioning the conditional branch predictor into isolated halves (e.g., user-level and OS kernel), but it is applicable only to Intel CPUs.

\section{Conclusion}

We analyze the vulnerability of Apple's M1 CPU to practical out-of-place Spectre v1 mistraining.
We show that brute-force out-of-place mistraining on the M1 is costly due to its TAGE predictor's sophisticated structure.
To enable practical out-of-place mistraining on the M1, we propose the LPC primitive, which reduces the complexity of out-of-place mistraining by $16\times$--$32\times$.
The LPC primitive requires knowledge of certain M1 BPU parameters and mechanisms, which we reverse engineer.
Finally, we use our newfound ability to perform out-of-place Spectre v1 mistraining to test if the M1 CPU implements hardware mitigations for cross-address space out-of-place mistraining.
We find that userspace/kernel BPU isolation appears to exist on Icestorm cores but not on Firestorm cores, and that no cross-process BPU isolation exists on either microarchitecture.

\bibliographystyle{plain}
\bibliography{bibliography}

\appendix
\section{Building Blocks for Detecting Aliased Branches} \label{sec:Branch Aliasing Recognition Details}

\subsection{Identifying Branch Misprediction PMC Events} \label{sec:PMC Event Encodings}

In this section, we determine how to configure the M1 PMU to count conditional and indirect branch mispredictions.
The M1's PMU exposes eight PMC registers, \texttt{PMC2}--\texttt{PMC9}, that can count various \emph{PMC events}.
Which event a PMC counts is determined by a 1-byte event encoding value written to one of two event selection registers, \texttt{PMESR0}
and \texttt{PMESR1}.

The event encoding for counting conditional branch mispredictions are not documented.
We identify them by sweeping the event space of the PMC registers,
looking for encodings that correctly count branch mispredictions for a branch with a known number of branch mispredictions.
We find that event \texttt{0xc5} of \texttt{PMC5} counts the desired event, based on the following experiments.

\begin{figure}[t]
	\centering
    \vspace{-18pt}
	\resizebox{\columnwidth}{!}{\input{./src/MacOS-Environment-Setup/Event-Encoding/cond-branch-mispredictions/PMC5/Random-Condition/results_firestorm/result0xc5.pgf}}
	\caption{Frequency of events counted by \texttt{PMC5} under event encoding \texttt{0xc5} while executing a conditional branch with a random outcome.}
	\label{CondMispredEvent}
\end{figure}
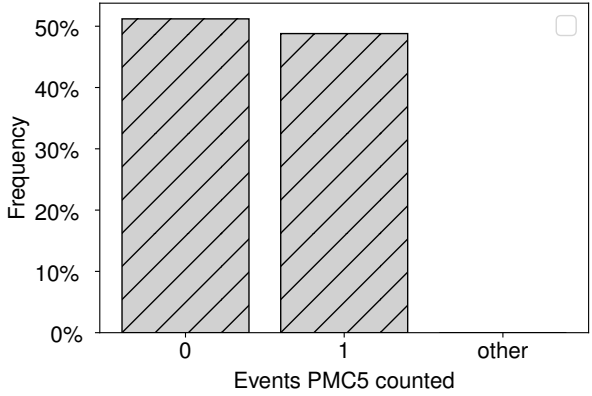

The first experiment continuously executes one conditional branch with a random outcome (taken/not-taken).
This branch is hard to predict, so we expect a 50\% misprediction rate.
\Cref{CondMispredEvent} shows the misprediction rate reported by \texttt{PMC5} event \texttt{0xc5}, which closely matches the expected results.

\begin{figure}[t]
	\centering
    \vspace{-15pt}
	\resizebox{\columnwidth}{!}{\input{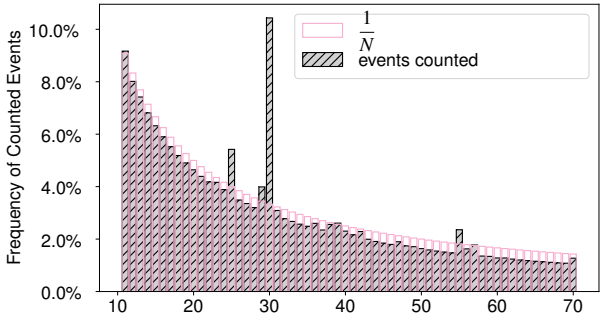}}
	\caption{Frequency of events counted by \texttt{PMC5} under event encoding \texttt{0xc5} for the backwards branch of a loop with $N$ iterations (x-axis).}
	\label{CondLoopMispredEvent}
\end{figure}

The second experiment continuously executes an $N$-iteration loop, whose body is filled with 400 fixed-outcome conditional branches.
We pick this value so that it will not fit in a loop predictor (if present) or a reasonably-sized BHR.
Thus, the BHR value of the loop's backwards branch remains fixed, as it is fully-determined by the branch sequence within the loop body and does not record history of the loop branch.
We expect to see a misprediction rate of $\approx \dfrac{1}{N}$ for the loop's backward branch, as it should mispredict only on the loop's last iteration when its outcome flips.
\Cref{CondLoopMispredEvent} shows the misprediction rate reported by \texttt{PMC5} event \texttt{0xc5} as a function of $N$, which closely match the expected result.

Overall, we identify \texttt{PMC5} event \texttt{0xc5} as counting conditional branch mispredictions.
We remark that additional \texttt{PMC5} event encodings produced similar results (\texttt{0xcb}, \texttt{0xcd}),
but we chose to pick \texttt{0xc5} due to the other events being used for other purposes on past Apple processors~\cite{OldApplePMC}.

Using analogous experiments, we find that \texttt{PMC5} event \texttt{0xc6} counts indirect branch mispredictions.
Again, additional event encodings produced similar results (\texttt{0xcb}, \texttt{0xce}), but we choose to pick \texttt{0xc6} due to its similarity to event-encodings of past Apple processors.
We further verify that these events are mutually exclusive, i.e., \texttt{0xc5} does not increment when \texttt{0xc6} does and vice versa.
This finding enables us to perform experiments that use both indirect and conditional branches, while counting mispredictions of only one branch type.

\subsection{Determining the Saturating Prediction Counter Width} \label{sec:counter}

Recall that the saturating prediction counter in a TAGE component entry is a $c$-bit counter that increments (respectively, decrements) when the outcome of a branch mapped to the entry is taken (respectively, not-taken).
The prediction made by an entry is determined by the most significant bit of the counter (if set, predict taken; otherwise, predict not-taken).

\begin{figure}[t]
	\centering
    \vspace{-15pt}
	\resizebox{\columnwidth}{!}{\input{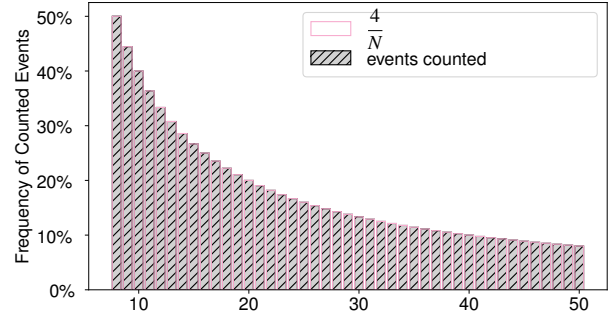}}
	\caption{Frequency of branch mispredictions for a branch whose outcome flips ${N}$ iterations (x-axis).}
	\label{PredContMispredEvent}
\end{figure}

To determine the counter size (in bits), we measure the branch misprediction rate of a branch whose outcome is flipped every $N$ iterations, for various values of $N$.
As in~\cref{sec:PMC Event Encodings}, we set this branch to have a fixed BHR value by prepending its execution with a branch slide of fixed-outcome branches, which guarantees that the prediction for each execution of the branch is provided by the same TAGE component entry.
Consequently, when $N > 2^c$, the branch is mispredicted for the first $2^{c-1}$ executions each time its outcome is flipped and thus its misprediction rate is $\dfrac{2^{c-1}}{N}$.
\Cref{PredContMispredEvent} shows the branch's measured misprediction rate is $\dfrac{4}{N}$, indicating a 3-bit counter.

\section{Figures Omitted From~\cref{sec:BHR Update Policy} and~\cref{sec:Conditional Branch Outcome Effect}} \label{sec:appendix-figures}

The figures appear on the next pages.

\begin{figure*}
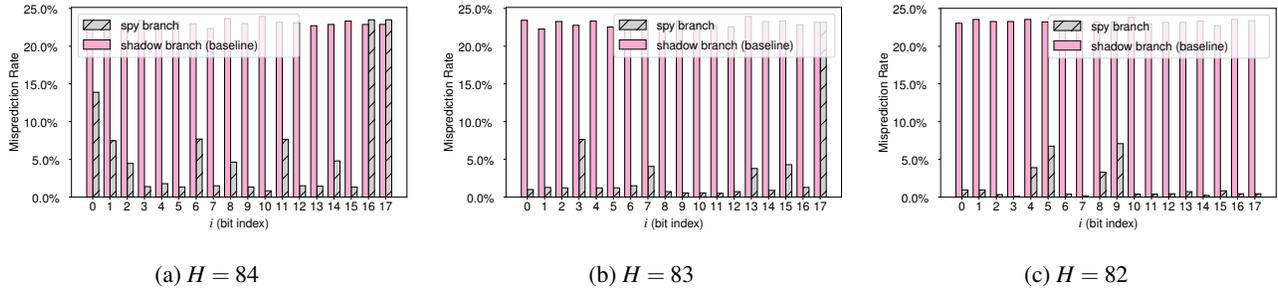

	\centering

	\begin{subfigure}[t]{0.32\textwidth}
		\centering
		\resizebox{\columnwidth}{!}{\input{./src/TAGE-SCL/BHR/conditional-branch-immediate/results/firestorm/H_84.pgf}}
		\caption{$H=84$}
	\end{subfigure}
	\begin{subfigure}[t]{0.32\textwidth}
		\centering
		\resizebox{\columnwidth}{!}{\input{./src/TAGE-SCL/BHR/conditional-branch-immediate/results/firestorm/H_83.pgf}}
		\caption{$H=83$}
	\end{subfigure}
	\begin{subfigure}[t]{0.32\textwidth}
		\centering
		\resizebox{\columnwidth}{!}{\input{./src/TAGE-SCL/BHR/conditional-branch-immediate/results/firestorm/H_82.pgf}}
		\caption{$H=82$}
	\end{subfigure}

	\caption{Effect of conditional branch immediate jump offset bit $i$ on the BHR is lost at $H=100-i$ on the Firestorm microarchitecture. For bits that have effect, the spy branch's misprediction rate is $\approx 0$\%. The figure shows results for bits 17 and 16. Omitted results for other bits are similarly consistent.}
	\label{CondImmediateBitsBHR-Update}
\end{figure*}

\begin{figure*}
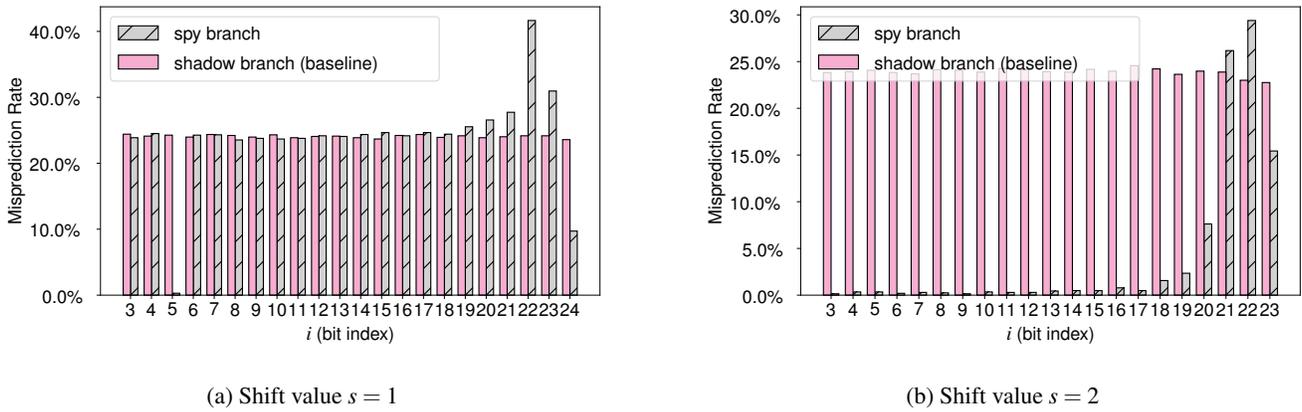

	\centering

	\begin{subfigure}[!]{\columnwidth}
		\centering
		\resizebox{\columnwidth}{!}{\input{./src/TAGE-SCL/BHR/update-policy/update-policy-cond-ip/results/firestorm/shift_1.pgf}}
		\caption{Shift value $s=1$}
	\end{subfigure}
	\hfill
	\begin{subfigure}[!]{\columnwidth}
		\centering
		\resizebox{\columnwidth}{!}{\input{./src/TAGE-SCL/BHR/update-policy/update-policy-cond-ip/results/firestorm/shift_2.pgf}}
		\caption{Shift value $s=2$}
	\end{subfigure}

	\caption{BHR update policy of taken conditional branch PC address bits is shift-and-XOR with a shift value of 1. High spy branch misprediction rate for bit $i$ and shift value $s$ indicates that when two conditional branches update the BHR consecutively, bit $i+s$ of the first branch's PC gets XORed with bit $i$ of the second branch's PC.}
	\label{BHRUpdatePolicyCondIP}
\end{figure*}

\begin{figure*}
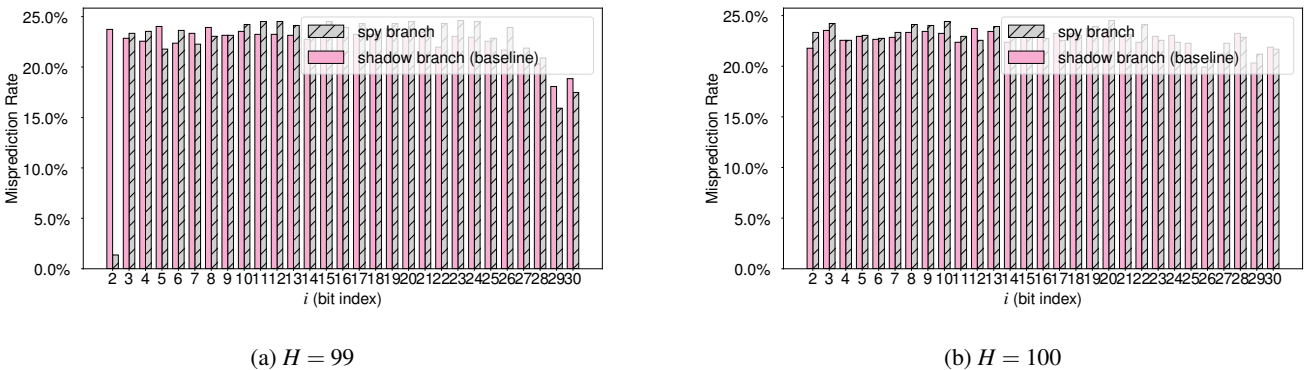

	\centering

	\begin{subfigure}[t]{\columnwidth}
		\centering
		\resizebox{\columnwidth}{!}{\input{./src/TAGE-SCL/BHR/conditional-branch-ip-bits/results/firestorm/H_99.pgf}}
		\caption{$H=99$}
	\end{subfigure}
	\hfill
	\begin{subfigure}[t]{\columnwidth}
		\centering
		\resizebox{\columnwidth}{!}{\input{./src/TAGE-SCL/BHR/conditional-branch-ip-bits/results/firestorm/H_100.pgf}}
		\caption{$H=100$}
	\end{subfigure}

	\caption{Effect of conditional branch PC address bits on the BHR on the Firestorm microarchitecture. For bits that have effect, the spy branch's misprediction rate is $\approx 0$\%. The figure shows that bit $i$'s effect is lost at $H=100-(i-2)$.
The figure shows results for bit 2. Omitted results for other bits are similarly consistent.}
	\label{CondBranchBitsBHR-Update}
\end{figure*}

\begin{figure*}
	\centering

	\begin{subfigure}[t]{\columnwidth}
		\centering
		\resizebox{\columnwidth}{!}{\input{./src/TAGE-SCL/BHR/indirect-branch-ip-bits/results/firestorm/H_27.pgf}}
		\caption{$H=27$}
	\end{subfigure}
	\hfill
	\begin{subfigure}[t]{\columnwidth}
		\centering
		\resizebox{\columnwidth}{!}{\input{./src/TAGE-SCL/BHR/indirect-branch-ip-bits/results/firestorm/H_28.pgf}}
		\caption{$H=28$}
	\end{subfigure}

	\caption{Effect of indirect branch PC address bit $i$ on the BHR is lost at distance $H=100-(72-(i-2))$ from the spy branch (Firestorm microarchitecture). The figure shows bit 2's effect; results for other bits are similarly consistent.}
	\label{IndirBranchBitsBHR-Update}
\end{figure*}

\begin{figure*}
	\centering

	\begin{subfigure}[!]{\columnwidth}
		\centering
		\resizebox{\columnwidth}{!}{\input{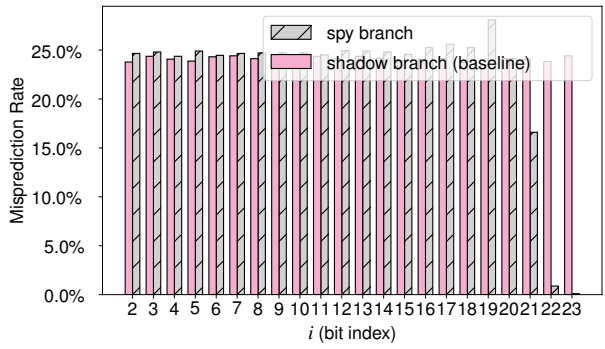}}
		\caption{Shift value $s=1$}
	\end{subfigure}
	\hfill
	\begin{subfigure}[!]{\columnwidth}
		\centering
		\resizebox{\columnwidth}{!}{\input{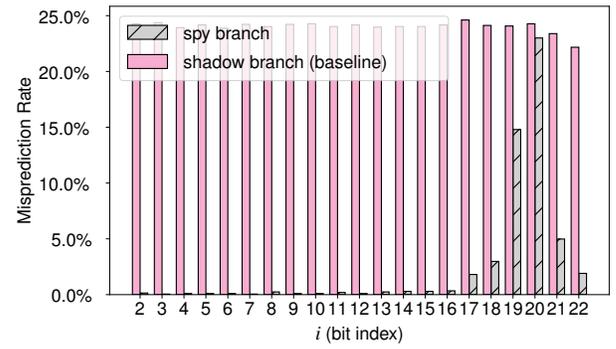}}
		\caption{Shift value $s=2$}
	\end{subfigure}

	\caption{BHR update policy of indirect branch target address bits is shift-and-XOR with a shift value of 1. High spy branch misprediction rate for bit $i$ and shift value $s$ indicates that when two indirect branches update the BHR consecutively, bit $i+s$ of the first branch's target gets XORed with bit $i$ of the second branch's target.}
	\label{BHRUpdatePolicyIndirectTarget}
\end{figure*}

\begin{figure*}
	\centering

	\begin{subfigure}[t]{\columnwidth}
		\centering
		\resizebox{\columnwidth}{!}{\input{./src/TAGE-SCL/BHR/indirect-branch-target/results/firestorm/H_77.pgf}}
		\caption{$H=77$}
	\end{subfigure}
    \hfill
	\begin{subfigure}[t]{\columnwidth}
		\centering
		\resizebox{\columnwidth}{!}{\input{./src/TAGE-SCL/BHR/indirect-branch-target/results/firestorm/H_78.pgf}}
		\caption{$H=78$}
	\end{subfigure}

	\caption{Effect of indirect branch target address bit $i$ on the BHR is lost at distance $H=100-(i-2)$ from the spy branch (Firestorm microarchitecture). The figure shows bit 24's effect; results for other bits are similarly consistent.}
	\label{IndirTargetBitsBHR-Update}
\end{figure*}

\end{document}

%% file: diagrams/tage.tex
\tikzset{every picture/.style={line width=0.75pt}} %

\begin{tikzpicture}[x=0.75pt,y=0.75pt,yscale=-1,xscale=1]

\draw  [color={rgb, 255:red, 162; green, 162; blue, 162 }  ,draw opacity=1 ][fill={rgb, 255:red, 226; green, 226; blue, 226 }  ,fill opacity=1 ] (149,328) -- (219,328) -- (219,368) -- (149,368) -- cycle ;
\draw    (185,209) -- (185,328) ;
\draw    (183.6,368) -- (184,437) ;
\draw  [color={rgb, 255:red, 162; green, 162; blue, 162 }  ,draw opacity=1 ][fill={rgb, 255:red, 226; green, 226; blue, 226 }  ,fill opacity=1 ] (300,232) -- (396.6,232) -- (396.6,314) -- (300,314) -- cycle ;
\draw  [color={rgb, 255:red, 162; green, 162; blue, 162 }  ,draw opacity=1 ][fill={rgb, 255:red, 226; green, 226; blue, 226 }  ,fill opacity=1 ] (320,252) -- (416.6,252) -- (416.6,334) -- (320,334) -- cycle ;
\draw  [color={rgb, 255:red, 162; green, 162; blue, 162 }  ,draw opacity=1 ][fill={rgb, 255:red, 226; green, 226; blue, 226 }  ,fill opacity=1 ] (340,272) -- (436.6,272) -- (436.6,354) -- (340,354) -- cycle ;
\draw  [color={rgb, 255:red, 162; green, 162; blue, 162 }  ,draw opacity=1 ][fill={rgb, 255:red, 226; green, 226; blue, 226 }  ,fill opacity=1 ] (360,292) -- (456.6,292) -- (456.6,374) -- (360,374) -- cycle ;
\draw  [color={rgb, 255:red, 162; green, 162; blue, 162 }  ,draw opacity=1 ][fill={rgb, 255:red, 226; green, 226; blue, 226 }  ,fill opacity=1 ] (380,312) -- (476.6,312) -- (476.6,394) -- (380,394) -- cycle ;
\draw  [color={rgb, 255:red, 162; green, 162; blue, 162 }  ,draw opacity=1 ][fill={rgb, 255:red, 226; green, 226; blue, 226 }  ,fill opacity=1 ] (400,332) -- (496.6,332) -- (496.6,414) -- (400,414) -- cycle ;
\draw    (325.3,232) -- (423.6,332) ;
\draw    (360.3,232) -- (458.6,332) ;
\draw    (361,220) -- (361,265) ;
\draw [shift={(361,268)}, rotate = 270] [fill={rgb, 255:red, 0; green, 0; blue, 0 }  ][line width=0.08]  [draw opacity=0] (8.93,-4.29) -- (0,0) -- (8.93,4.29) -- cycle    ;
\draw    (185,220) -- (361,220) ;
\draw    (405.6,208) -- (405.03,273) ;
\draw [shift={(405,276)}, rotate = 270.51] [fill={rgb, 255:red, 0; green, 0; blue, 0 }  ][line width=0.08]  [draw opacity=0] (8.93,-4.29) -- (0,0) -- (8.93,4.29) -- cycle    ;
\draw    (184,437) -- (294,437) ;
\draw    (294,437) -- (294,466) ;
\draw [shift={(294,469)}, rotate = 270] [fill={rgb, 255:red, 0; green, 0; blue, 0 }  ][line width=0.08]  [draw opacity=0] (8.93,-4.29) -- (0,0) -- (8.93,4.29) -- cycle    ;
\draw    (309,437) -- (309,466) ;
\draw [shift={(309,469)}, rotate = 270] [fill={rgb, 255:red, 0; green, 0; blue, 0 }  ][line width=0.08]  [draw opacity=0] (8.93,-4.29) -- (0,0) -- (8.93,4.29) -- cycle    ;
\draw    (309,437) -- (450,438) ;
\draw    (450,414) -- (450,438) ;
\draw   (321,469) -- (317.1,482) -- (287.9,482) -- (284,469) -- cycle ;
\draw    (303,482) -- (303,507) ;
\draw [shift={(303,510)}, rotate = 270] [fill={rgb, 255:red, 0; green, 0; blue, 0 }  ][line width=0.08]  [draw opacity=0] (8.93,-4.29) -- (0,0) -- (8.93,4.29) -- cycle    ;
\draw  [color={rgb, 255:red, 131; green, 131; blue, 131 }  ,draw opacity=1 ] (120,178) -- (545.13,178) -- (545.13,490.47) -- (120,490.47) -- cycle ;
\draw  [color={rgb, 255:red, 162; green, 162; blue, 162 }  ,draw opacity=1 ][fill={rgb, 255:red, 226; green, 226; blue, 226 }  ,fill opacity=1 ] (392,511) -- (473.13,511) -- (473.13,564.47) -- (392,564.47) -- cycle ;
\draw  [color={rgb, 255:red, 162; green, 162; blue, 162 }  ,draw opacity=1 ][fill={rgb, 255:red, 226; green, 226; blue, 226 }  ,fill opacity=1 ] (263,510) -- (343.13,510) -- (343.13,564.47) -- (263,564.47) -- cycle ;
\draw    (300,580) -- (360,580) ;
\draw    (360,580) -- (360,597.2) ;
\draw [shift={(360,600.2)}, rotate = 270] [fill={rgb, 255:red, 0; green, 0; blue, 0 }  ][line width=0.08]  [draw opacity=0] (8.93,-4.29) -- (0,0) -- (8.93,4.29) -- cycle    ;
\draw    (375,579.95) -- (375,597) ;
\draw [shift={(375,600)}, rotate = 270] [fill={rgb, 255:red, 0; green, 0; blue, 0 }  ][line width=0.08]  [draw opacity=0] (8.93,-4.29) -- (0,0) -- (8.93,4.29) -- cycle    ;
\draw    (375,579.95) -- (434,580.04) ;
\draw    (434,565) -- (434,580.04) ;

\draw    (300,564.93) -- (300,580) ;
\draw   (385,600) -- (381.1,613) -- (351.9,613) -- (348,600) -- cycle ;
\draw    (367,613) -- (367,640) ;
\draw [shift={(367,643)}, rotate = 270] [fill={rgb, 255:red, 0; green, 0; blue, 0 }  ][line width=0.08]  [draw opacity=0] (8.93,-4.29) -- (0,0) -- (8.93,4.29) -- cycle    ;

\draw (187,202) node [anchor=north west][inner sep=0.75pt]   [align=left] {\textbf{{\fontfamily{helvet}\selectfont PC}}};
\draw (408,201) node [anchor=north west][inner sep=0.75pt]   [align=left] {\textbf{{\fontfamily{helvet}\selectfont BHR}}};
\draw (369,616) node [anchor=north west][inner sep=0.75pt]   [align=left] {\textbf{{\fontfamily{helvet}\selectfont Prediction}}};
\draw (184,348) node   [align=left] {\textbf{{\fontfamily{helvet}\selectfont Bimodal}}};
\draw (303.07,537.23) node   [align=left] {\begin{minipage}[lt]{50.92pt}\setlength\topsep{0pt}
\begin{center}
\textbf{{\fontfamily{helvet}\selectfont Statistical}}\\\textbf{{\fontfamily{helvet}\selectfont Corrector}}
\end{center}

\end{minipage}};
\draw (432.57,537.73) node   [align=left] {\begin{minipage}[lt]{47.5pt}\setlength\topsep{0pt}
\begin{center}
\textbf{{\fontfamily{helvet}\selectfont Loop}}\\\textbf{{\fontfamily{helvet}\selectfont Predictor}}
\end{center}

\end{minipage}};
\draw (448.3,373) node   [align=left] {\begin{minipage}[lt]{34.28pt}\setlength\topsep{0pt}
\begin{center}
\textbf{{\fontfamily{helvet}\selectfont TAGE}}\\\textbf{{\fontfamily{helvet}\selectfont Tables}}
\end{center}

\end{minipage}};

\end{tikzpicture}

%% file: diagrams/bhr-baseline-experiment.tex
\tikzset{every picture/.style={line width=0.75pt}} %

\begin{tikzpicture}[x=0.75pt,y=0.75pt,yscale=-1,xscale=1]

\draw    (343.22,537) -- (343.22,570.19) ;
\draw [shift={(343.22,573.19)}, rotate = 270] [fill={rgb, 255:red, 0; green, 0; blue, 0 }  ][line width=0.08]  [draw opacity=0] (8.93,-4.29) -- (0,0) -- (8.93,4.29) -- cycle    ;
\draw    (246.35,342.29) -- (246.12,370.88) ;
\draw [shift={(246.09,373.88)}, rotate = 270.46] [fill={rgb, 255:red, 0; green, 0; blue, 0 }  ][line width=0.08]  [draw opacity=0] (8.93,-4.29) -- (0,0) -- (8.93,4.29) -- cycle    ;
\draw    (246.35,342.29) -- (333.9,342.73) ;
\draw    (333.9,332.21) -- (333.9,342.73) ;

\draw    (349.94,342.09) -- (437.32,342.09) ;
\draw    (437.32,342.09) -- (437.02,369.91) ;
\draw [shift={(436.99,372.91)}, rotate = 270.61] [fill={rgb, 255:red, 0; green, 0; blue, 0 }  ][line width=0.08]  [draw opacity=0] (8.93,-4.29) -- (0,0) -- (8.93,4.29) -- cycle    ;
\draw    (349.94,331.82) -- (349.94,342.09) ;

\draw    (351.32,423.18) -- (351.09,452.07) ;
\draw [shift={(351.07,455.07)}, rotate = 270.45] [fill={rgb, 255:red, 0; green, 0; blue, 0 }  ][line width=0.08]  [draw opacity=0] (8.93,-4.29) -- (0,0) -- (8.93,4.29) -- cycle    ;
\draw    (351.32,423.18) -- (436.86,423.62) ;
\draw    (436.86,413) -- (436.86,423.62) ;

\draw    (245,423.96) -- (330.37,423.96) ;
\draw    (330.37,423.96) -- (330.08,452.07) ;
\draw [shift={(330.05,455.07)}, rotate = 270.59] [fill={rgb, 255:red, 0; green, 0; blue, 0 }  ][line width=0.08]  [draw opacity=0] (8.93,-4.29) -- (0,0) -- (8.93,4.29) -- cycle    ;
\draw    (245,413.6) -- (245,423.96) ;

\draw  [color={rgb, 255:red, 155; green, 155; blue, 155 }  ,draw opacity=1 ][fill={rgb, 255:red, 226; green, 226; blue, 226 }  ,fill opacity=1 ] (291,281.8) .. controls (291,274.73) and (296.73,269) .. (303.8,269) -- (382.2,269) .. controls (389.27,269) and (395,274.73) .. (395,281.8) -- (395,320.2) .. controls (395,327.27) and (389.27,333) .. (382.2,333) -- (303.8,333) .. controls (296.73,333) and (291,327.27) .. (291,320.2) -- cycle ;
\draw  [color={rgb, 255:red, 155; green, 155; blue, 155 }  ,draw opacity=1 ][fill={rgb, 255:red, 226; green, 226; blue, 226 }  ,fill opacity=1 ] (369,381.2) .. controls (369,376.67) and (372.67,373) .. (377.2,373) -- (496.8,373) .. controls (501.33,373) and (505,376.67) .. (505,381.2) -- (505,405.8) .. controls (505,410.33) and (501.33,414) .. (496.8,414) -- (377.2,414) .. controls (372.67,414) and (369,410.33) .. (369,405.8) -- cycle ;
\draw  [color={rgb, 255:red, 155; green, 155; blue, 155 }  ,draw opacity=1 ][fill={rgb, 255:red, 226; green, 226; blue, 226 }  ,fill opacity=1 ] (257,473.4) .. controls (257,464.34) and (264.34,457) .. (273.4,457) -- (410.6,457) .. controls (419.66,457) and (427,464.34) .. (427,473.4) -- (427,522.6) .. controls (427,531.66) and (419.66,539) .. (410.6,539) -- (273.4,539) .. controls (264.34,539) and (257,531.66) .. (257,522.6) -- cycle ;
\draw  [color={rgb, 255:red, 155; green, 155; blue, 155 }  ,draw opacity=1 ][fill={rgb, 255:red, 226; green, 226; blue, 226 }  ,fill opacity=1 ] (305,584.2) .. controls (305,578.01) and (310.01,573) .. (316.2,573) -- (370.8,573) .. controls (376.99,573) and (382,578.01) .. (382,584.2) -- (382,617.8) .. controls (382,623.99) and (376.99,629) .. (370.8,629) -- (316.2,629) .. controls (310.01,629) and (305,623.99) .. (305,617.8) -- cycle ;
\draw  [color={rgb, 255:red, 155; green, 155; blue, 155 }  ,draw opacity=1 ][fill={rgb, 255:red, 226; green, 226; blue, 226 }  ,fill opacity=1 ] (178,382.2) .. controls (178,377.67) and (181.67,374) .. (186.2,374) -- (305.8,374) .. controls (310.33,374) and (314,377.67) .. (314,382.2) -- (314,406.8) .. controls (314,411.33) and (310.33,415) .. (305.8,415) -- (186.2,415) .. controls (181.67,415) and (178,411.33) .. (178,406.8) -- cycle ;
\draw    (159.81,643.83) -- (344.81,643.83) ;
\draw    (342.8,245) -- (343.08,265.35) ;
\draw [shift={(343.12,268.35)}, rotate = 269.22] [fill={rgb, 255:red, 0; green, 0; blue, 0 }  ][line width=0.08]  [draw opacity=0] (8.93,-4.29) -- (0,0) -- (8.93,4.29) -- cycle    ;
\draw    (160.43,245.28) -- (159.81,643.83) ;

\draw    (342.8,245) -- (160.43,245.28) ;

\draw    (344.78,629.08) -- (344.81,643.83) ;

\draw (211,301) node [anchor=north west][inner sep=0.75pt]   [align=left] {\textbf{{\fontfamily{helvet}\selectfont Path 1}}};
\draw (428,301) node [anchor=north west][inner sep=0.75pt]   [align=left] {\textbf{{\fontfamily{helvet}\selectfont Path 2}}};
\draw (250,585) node [anchor=north west][inner sep=0.75pt]   [align=left] {\begin{minipage}[lt]{29.37pt}\setlength\topsep{0pt}
\begin{center}
\textbf{{\fontfamily{helvet}\selectfont not}}\\\textbf{{\fontfamily{helvet}\selectfont taken}}
\end{center}

\end{minipage}};
\draw (392,594) node [anchor=north west][inner sep=0.75pt]   [align=left] {\textbf{{\fontfamily{helvet}\selectfont taken}}};
\draw (342,498) node   [align=left] {\begin{minipage}[lt]{106.44pt}\setlength\topsep{0pt}
\begin{center}
\textbf{{\fontfamily{helvet}\selectfont Buffer of H branches }}\\\textbf{{\fontfamily{helvet}\selectfont of the type of the}}\\\textbf{{\fontfamily{helvet}\selectfont  setup branches}}
\end{center}

\end{minipage}};
\draw (343,301) node   [align=left] {\begin{minipage}[lt]{40.7pt}\setlength\topsep{0pt}
\begin{center}
\textbf{{\fontfamily{helvet}\selectfont Branch }}\\\textbf{{\fontfamily{helvet}\selectfont Slide}}
\end{center}

\end{minipage}};
\draw (437,393.5) node   [align=left] {\begin{minipage}[lt]{77.54pt}\setlength\topsep{0pt}
\begin{center}
\textbf{{\fontfamily{helvet}\selectfont Setup Branch 2}}
\end{center}

\end{minipage}};
\draw (246,394.5) node   [align=left] {\begin{minipage}[lt]{77.54pt}\setlength\topsep{0pt}
\begin{center}
\textbf{{\fontfamily{helvet}\selectfont Setup Branch 1}}
\end{center}

\end{minipage}};
\draw (343.5,601) node   [align=left] {\begin{minipage}[lt]{37.87pt}\setlength\topsep{0pt}
\begin{center}
\textbf{{\fontfamily{helvet}\selectfont Spy }}\\\textbf{{\fontfamily{helvet}\selectfont Branch}}
\end{center}

\end{minipage}};

\end{tikzpicture}

%% file: diagrams/shift-and-xor.tex
\tikzset{every picture/.style={line width=0.75pt}} %

\begin{tikzpicture}[x=0.75pt,y=0.75pt,yscale=-1,xscale=1]

\draw  [fill={rgb, 255:red, 255; green, 255; blue, 255 }  ,fill opacity=1 ] (208.33,400) -- (256,400) -- (256,440) -- (208.33,440) -- cycle ;
\draw  [fill={rgb, 255:red, 255; green, 255; blue, 255 }  ,fill opacity=1 ] (256,400) -- (303.67,400) -- (303.67,440) -- (256,440) -- cycle ;
\draw  [fill={rgb, 255:red, 255; green, 255; blue, 255 }  ,fill opacity=1 ] (303.67,400) -- (351.33,400) -- (351.33,440) -- (303.67,440) -- cycle ;
\draw  [fill={rgb, 255:red, 255; green, 255; blue, 255 }  ,fill opacity=1 ] (351.33,400) -- (399,400) -- (399,440) -- (351.33,440) -- cycle ;
\draw  [fill={rgb, 255:red, 255; green, 255; blue, 255 }  ,fill opacity=1 ] (399,400) -- (446.67,400) -- (446.67,440) -- (399,440) -- cycle ;
\draw  [fill={rgb, 255:red, 255; green, 255; blue, 255 }  ,fill opacity=1 ] (114.67,408) .. controls (114.67,403.58) and (118.25,400) .. (122.67,400) -- (160.67,400) .. controls (165.08,400) and (168.67,403.58) .. (168.67,408) -- (168.67,432) .. controls (168.67,436.42) and (165.08,440) .. (160.67,440) -- (122.67,440) .. controls (118.25,440) and (114.67,436.42) .. (114.67,432) -- cycle ;
\draw  [fill={rgb, 255:red, 255; green, 255; blue, 255 }  ,fill opacity=1 ] (160.67,400) -- (208.33,400) -- (208.33,440) -- (160.67,440) -- cycle ;
\draw  [fill={rgb, 255:red, 255; green, 255; blue, 255 }  ,fill opacity=1 ] (486.33,408) .. controls (486.33,403.58) and (489.92,400) .. (494.33,400) -- (532.33,400) .. controls (536.75,400) and (540.33,403.58) .. (540.33,408) -- (540.33,432) .. controls (540.33,436.42) and (536.75,440) .. (532.33,440) -- (494.33,440) .. controls (489.92,440) and (486.33,436.42) .. (486.33,432) -- cycle ;
\draw  [fill={rgb, 255:red, 255; green, 255; blue, 255 }  ,fill opacity=1 ] (446.67,400) -- (494.33,400) -- (494.33,440) -- (446.67,440) -- cycle ;

\draw    (184,441) -- (137.65,472.87) ;
\draw [shift={(136,474)}, rotate = 325.49] [color={rgb, 255:red, 0; green, 0; blue, 0 }  ][line width=0.75]    (10.93,-3.29) .. controls (6.95,-1.4) and (3.31,-0.3) .. (0,0) .. controls (3.31,0.3) and (6.95,1.4) .. (10.93,3.29)   ;
\draw    (233,440) -- (186.65,471.87) ;
\draw [shift={(185,473)}, rotate = 325.49] [color={rgb, 255:red, 0; green, 0; blue, 0 }  ][line width=0.75]    (10.93,-3.29) .. controls (6.95,-1.4) and (3.31,-0.3) .. (0,0) .. controls (3.31,0.3) and (6.95,1.4) .. (10.93,3.29)   ;
\draw    (520,440) -- (473.65,471.87) ;
\draw [shift={(472,473)}, rotate = 325.49] [color={rgb, 255:red, 0; green, 0; blue, 0 }  ][line width=0.75]    (10.93,-3.29) .. controls (6.95,-1.4) and (3.31,-0.3) .. (0,0) .. controls (3.31,0.3) and (6.95,1.4) .. (10.93,3.29)   ;
\draw  [fill={rgb, 255:red, 255; green, 255; blue, 255 }  ,fill opacity=1 ] (208.33,473) -- (256,473) -- (256,513) -- (208.33,513) -- cycle ;
\draw  [fill={rgb, 255:red, 255; green, 255; blue, 255 }  ,fill opacity=1 ] (256,473) -- (303.67,473) -- (303.67,513) -- (256,513) -- cycle ;
\draw  [fill={rgb, 255:red, 255; green, 255; blue, 255 }  ,fill opacity=1 ] (303.67,473) -- (351.33,473) -- (351.33,513) -- (303.67,513) -- cycle ;
\draw  [fill={rgb, 255:red, 255; green, 255; blue, 255 }  ,fill opacity=1 ] (351.33,473) -- (399,473) -- (399,513) -- (351.33,513) -- cycle ;
\draw  [fill={rgb, 255:red, 255; green, 255; blue, 255 }  ,fill opacity=1 ] (399,473) -- (446.67,473) -- (446.67,513) -- (399,513) -- cycle ;
\draw  [fill={rgb, 255:red, 255; green, 255; blue, 255 }  ,fill opacity=1 ] (114.67,481) .. controls (114.67,476.58) and (118.25,473) .. (122.67,473) -- (160.67,473) .. controls (165.08,473) and (168.67,476.58) .. (168.67,481) -- (168.67,505) .. controls (168.67,509.42) and (165.08,513) .. (160.67,513) -- (122.67,513) .. controls (118.25,513) and (114.67,509.42) .. (114.67,505) -- cycle ;
\draw  [fill={rgb, 255:red, 255; green, 255; blue, 255 }  ,fill opacity=1 ] (160.67,473) -- (208.33,473) -- (208.33,513) -- (160.67,513) -- cycle ;
\draw  [fill={rgb, 255:red, 255; green, 255; blue, 255 }  ,fill opacity=1 ] (486.33,481) .. controls (486.33,476.58) and (489.92,473) .. (494.33,473) -- (532.33,473) .. controls (536.75,473) and (540.33,476.58) .. (540.33,481) -- (540.33,505) .. controls (540.33,509.42) and (536.75,513) .. (532.33,513) -- (494.33,513) .. controls (489.92,513) and (486.33,509.42) .. (486.33,505) -- cycle ;
\draw  [fill={rgb, 255:red, 255; green, 255; blue, 255 }  ,fill opacity=1 ] (446.67,473) -- (494.33,473) -- (494.33,513) -- (446.67,513) -- cycle ;
\draw  [fill={rgb, 255:red, 255; green, 255; blue, 255 }  ,fill opacity=1 ] (398,550) .. controls (398,545.58) and (401.58,542) .. (406,542) -- (445.67,542) .. controls (450.08,542) and (453.67,545.58) .. (453.67,550) -- (453.67,574) .. controls (453.67,578.42) and (450.08,582) .. (445.67,582) -- (406,582) .. controls (401.58,582) and (398,578.42) .. (398,574) -- cycle ;
\draw  [fill={rgb, 255:red, 255; green, 255; blue, 255 }  ,fill opacity=1 ] (485.33,550) .. controls (485.33,545.58) and (488.92,542) .. (493.33,542) -- (533,542) .. controls (537.42,542) and (541,545.58) .. (541,550) -- (541,574) .. controls (541,578.42) and (537.42,582) .. (533,582) -- (493.33,582) .. controls (488.92,582) and (485.33,578.42) .. (485.33,574) -- cycle ;
\draw  [fill={rgb, 255:red, 255; green, 255; blue, 255 }  ,fill opacity=1 ] (445.67,542) -- (493.33,542) -- (493.33,582) -- (445.67,582) -- cycle ;

\draw    (342,561) -- (386,561) ;
\draw [shift={(388,561)}, rotate = 180] [color={rgb, 255:red, 0; green, 0; blue, 0 }  ][line width=0.75]    (10.93,-3.29) .. controls (6.95,-1.4) and (3.31,-0.3) .. (0,0) .. controls (3.31,0.3) and (6.95,1.4) .. (10.93,3.29)   ;

\draw (369,451) node [anchor=north west][inner sep=0.75pt]   [align=left] {\textbf{{\large {\fontfamily{helvet}\selectfont .}}}};
\draw (323,451) node [anchor=north west][inner sep=0.75pt]   [align=left] {\textbf{{\large {\fontfamily{helvet}\selectfont .}}}};
\draw (278,451) node [anchor=north west][inner sep=0.75pt]   [align=left] {\textbf{{\large {\fontfamily{helvet}\selectfont .}}}};
\draw (369,378) node [anchor=north west][inner sep=0.75pt]   [align=left] {\textbf{{\large {\fontfamily{helvet}\selectfont .}}}};
\draw (323,378) node [anchor=north west][inner sep=0.75pt]   [align=left] {\textbf{{\large {\fontfamily{helvet}\selectfont .}}}};
\draw (278,378) node [anchor=north west][inner sep=0.75pt]   [align=left] {\textbf{{\large {\fontfamily{helvet}\selectfont .}}}};
\draw (469,376) node [anchor=north west][inner sep=0.75pt]   [align=left] {{\large \textbf{{\fontfamily{helvet}\selectfont 1}}}};
\draw (512,376) node [anchor=north west][inner sep=0.75pt]   [align=left] {{\large \textbf{{\fontfamily{helvet}\selectfont 0}}}};
\draw (130,375) node [anchor=north west][inner sep=0.75pt]   [align=left] {{\large \textbf{{\fontfamily{helvet}\selectfont 8}}}};
\draw (307,356) node [anchor=north west][inner sep=0.75pt]   [align=left] {\textbf{{\fontfamily{helvet}\selectfont {\Large BHR}}}};
\draw (251.63,561.11) node   [align=left] {\begin{minipage}[lt]{130.05pt}\setlength\topsep{0pt}
\begin{center}

\textbf{{\fontfamily{helvet}\selectfont {\Large attrs(new\_branch)}}}

\end{center}

\end{minipage}};
\draw (410,518.4) node [anchor=north west][inner sep=0.75pt]    {$\bigoplus $};
\draw (457,518.4) node [anchor=north west][inner sep=0.75pt]    {$\bigoplus $};
\draw (542.33,435) node [anchor=north west][inner sep=0.75pt]   [align=left] {\textbf{{\fontfamily{helvet}\selectfont {\Large shift}}}};
\draw (542.33,508) node [anchor=north west][inner sep=0.75pt]   [align=left] {\textbf{{\fontfamily{helvet}\selectfont {\Large xor}}}};
\draw (504,518.4) node [anchor=north west][inner sep=0.75pt]    {$\bigoplus $};
\draw (516.34,493) node   [align=left] {{\Large \textbf{{\fontfamily{helvet}\selectfont 0}}}};

\end{tikzpicture}

%% file: diagrams/outcome-effect.tex
\tikzset{every picture/.style={line width=0.75pt}} %

\begin{tikzpicture}[x=0.75pt,y=0.75pt,yscale=-1,xscale=1]

\draw  [color={rgb, 255:red, 155; green, 155; blue, 155 }  ,draw opacity=1 ][fill={rgb, 255:red, 226; green, 226; blue, 226 }  ,fill opacity=1 ] (237,579.2) .. controls (237,573.01) and (242.01,568) .. (248.2,568) -- (302.8,568) .. controls (308.99,568) and (314,573.01) .. (314,579.2) -- (314,612.8) .. controls (314,618.99) and (308.99,624) .. (302.8,624) -- (248.2,624) .. controls (242.01,624) and (237,618.99) .. (237,612.8) -- cycle ;
\draw    (276.22,446) -- (276.37,480.69) ;
\draw [shift={(276.39,483.69)}, rotate = 269.75] [fill={rgb, 255:red, 0; green, 0; blue, 0 }  ][line width=0.08]  [draw opacity=0] (8.93,-4.29) -- (0,0) -- (8.93,4.29) -- cycle    ;
\draw  [color={rgb, 255:red, 155; green, 155; blue, 155 }  ,draw opacity=1 ][fill={rgb, 255:red, 226; green, 226; blue, 226 }  ,fill opacity=1 ] (190.97,416) .. controls (190.97,410.48) and (195.44,406) .. (200.97,406) -- (355.97,406) .. controls (361.49,406) and (365.97,410.48) .. (365.97,416) -- (365.97,446) .. controls (365.97,451.52) and (361.49,456) .. (355.97,456) -- (200.97,456) .. controls (195.44,456) and (190.97,451.52) .. (190.97,446) -- cycle ;
\draw  [color={rgb, 255:red, 155; green, 155; blue, 155 }  ,draw opacity=1 ][fill={rgb, 255:red, 226; green, 226; blue, 226 }  ,fill opacity=1 ] (238,495.2) .. controls (238,489.01) and (243.01,484) .. (249.2,484) -- (303.8,484) .. controls (309.99,484) and (315,489.01) .. (315,495.2) -- (315,528.8) .. controls (315,534.99) and (309.99,540) .. (303.8,540) -- (249.2,540) .. controls (243.01,540) and (238,534.99) .. (238,528.8) -- cycle ;
\draw  [color={rgb, 255:red, 155; green, 155; blue, 155 }  ,draw opacity=1 ][fill={rgb, 255:red, 226; green, 226; blue, 226 }  ,fill opacity=1 ] (189.97,260) .. controls (189.97,254.48) and (194.44,250) .. (199.97,250) -- (355,250) .. controls (360.52,250) and (365,254.48) .. (365,260) -- (365,290) .. controls (365,295.52) and (360.52,300) .. (355,300) -- (199.97,300) .. controls (194.44,300) and (189.97,295.52) .. (189.97,290) -- cycle ;
\draw    (448,274.74) -- (368.31,274.99) ;
\draw [shift={(366.31,275)}, rotate = 359.82] [color={rgb, 255:red, 0; green, 0; blue, 0 }  ][line width=0.75]    (10.93,-3.29) .. controls (6.95,-1.4) and (3.31,-0.3) .. (0,0) .. controls (3.31,0.3) and (6.95,1.4) .. (10.93,3.29)   ;
\draw    (448,430.74) -- (367.97,430.06) ;
\draw [shift={(365.97,430.04)}, rotate = 0.49] [color={rgb, 255:red, 0; green, 0; blue, 0 }  ][line width=0.75]    (10.93,-3.29) .. controls (6.95,-1.4) and (3.31,-0.3) .. (0,0) .. controls (3.31,0.3) and (6.95,1.4) .. (10.93,3.29)   ;
\draw    (446.69,507.74) -- (367,507.99) ;
\draw [shift={(365,508)}, rotate = 359.82] [color={rgb, 255:red, 0; green, 0; blue, 0 }  ][line width=0.75]    (10.93,-3.29) .. controls (6.95,-1.4) and (3.31,-0.3) .. (0,0) .. controls (3.31,0.3) and (6.95,1.4) .. (10.93,3.29)   ;
\draw    (448,274.74) -- (446.69,507.74) ;
\draw    (126,644) -- (275.79,644) ;
\draw    (275.77,624.64) -- (275.79,644) ;

\draw    (278.6,192.67) -- (278.6,247.13) ;
\draw [shift={(278.6,250.13)}, rotate = 270] [fill={rgb, 255:red, 0; green, 0; blue, 0 }  ][line width=0.08]  [draw opacity=0] (8.93,-4.29) -- (0,0) -- (8.93,4.29) -- cycle    ;
\draw    (278.6,192.67) -- (126.51,192.82) ;

\draw    (126.51,192.82) -- (126,644) ;
\draw    (275.39,540.3) -- (275.39,564.69) ;
\draw [shift={(275.39,567.69)}, rotate = 270] [fill={rgb, 255:red, 0; green, 0; blue, 0 }  ][line width=0.08]  [draw opacity=0] (8.93,-4.29) -- (0,0) -- (8.93,4.29) -- cycle    ;
\draw    (277.33,382.67) -- (277.38,403.69) ;
\draw [shift={(277.39,406.69)}, rotate = 269.87] [fill={rgb, 255:red, 0; green, 0; blue, 0 }  ][line width=0.08]  [draw opacity=0] (8.93,-4.29) -- (0,0) -- (8.93,4.29) -- cycle    ;
\draw    (278.33,300) -- (278.33,321.67) ;
\draw [shift={(278.33,324.67)}, rotate = 270] [fill={rgb, 255:red, 0; green, 0; blue, 0 }  ][line width=0.08]  [draw opacity=0] (8.93,-4.29) -- (0,0) -- (8.93,4.29) -- cycle    ;

\draw (165,210) node [anchor=north west][inner sep=0.75pt]   [align=left] {\textbf{{\fontfamily{helvet}\selectfont Path 1}}};
\draw (354,210) node [anchor=north west][inner sep=0.75pt]   [align=left] {\textbf{{\fontfamily{helvet}\selectfont Path 2}}};
\draw (182,580) node [anchor=north west][inner sep=0.75pt]   [align=left] {\begin{minipage}[lt]{29.37pt}\setlength\topsep{0pt}
\begin{center}
\textbf{{\fontfamily{helvet}\selectfont not}}\\\textbf{{\fontfamily{helvet}\selectfont taken}}
\end{center}

\end{minipage}};
\draw (327,589) node [anchor=north west][inner sep=0.75pt]   [align=left] {\textbf{{\fontfamily{helvet}\selectfont taken}}};
\draw (275.5,596) node   [align=left] {\begin{minipage}[lt]{37.87pt}\setlength\topsep{0pt}
\begin{center}
\textbf{{\fontfamily{helvet}\selectfont Spy }}\\\textbf{{\fontfamily{helvet}\selectfont Branch}}
\end{center}

\end{minipage}};
\draw (277.48,275) node   [align=left] {\begin{minipage}[lt]{90.38pt}\setlength\topsep{0pt}
\begin{center}
\textbf{{\fontfamily{helvet}\selectfont Conditional Taken}}\\\textbf{{\fontfamily{helvet}\selectfont Branch}}
\end{center}

\end{minipage}};
\draw (452,270) node [anchor=north west][inner sep=0.75pt]   [align=left] {\textbf{{\fontfamily{helvet}\selectfont $+0$}}\\\textbf{{\fontfamily{helvet}\selectfont base of }}\\\textbf{{\fontfamily{helvet}\selectfont experiment}}};
\draw (452,437.69) node   [anchor=north west][align=left] {\textbf{{\fontfamily{helvet}\selectfont $+(2^{32} \times 99)$}}};
\draw (452,512.69) node   [anchor=north west][align=left] {\textbf{{\fontfamily{helvet}\selectfont $+(2^{32} \times 100)$}}};
\draw (284,333) node [anchor=north west][inner sep=0.75pt]   [align=left] {\begin{minipage}[lt]{86.6pt}\setlength\topsep{0pt}
\begin{center}
\textbf{{\fontfamily{helvet}\selectfont 100}}\\\textbf{{\fontfamily{helvet}\selectfont conditional taken}}\\\textbf{{\fontfamily{helvet}\selectfont branches}}
\end{center}

\end{minipage}};
\draw (278.47,431) node   [align=left] {\begin{minipage}[lt]{90.38pt}\setlength\topsep{0pt}
\begin{center}
\textbf{{\fontfamily{helvet}\selectfont Conditional Taken}}\\\textbf{{\fontfamily{helvet}\selectfont Branch}}
\end{center}

\end{minipage}};
\draw (274,335) node [anchor=north west][inner sep=0.75pt]   [align=left] {\textbf{.}\\\textbf{.}\\\textbf{.}};
\draw (182,497) node [anchor=north west][inner sep=0.75pt]   [align=left] {\begin{minipage}[lt]{29.37pt}\setlength\topsep{0pt}
\begin{center}
\textbf{{\fontfamily{helvet}\selectfont not}}\\\textbf{{\fontfamily{helvet}\selectfont taken}}
\end{center}

\end{minipage}};
\draw (326,503.2) node [anchor=north west][inner sep=0.75pt]   [align=left] {\textbf{{\fontfamily{helvet}\selectfont taken}}};
\draw (276.5,512) node   [align=left] {\begin{minipage}[lt]{38.43pt}\setlength\topsep{0pt}
\begin{center}
\textbf{{\fontfamily{helvet}\selectfont Switch }}\\\textbf{{\fontfamily{helvet}\selectfont Branch}}
\end{center}

\end{minipage}};

\end{tikzpicture}

%% file: src/TAGE-SCL/BHR/conditional-branch-outcome/results/firestorm.pgf
\begingroup%
\makeatletter%
\begin{pgfpicture}%
\pgfpathrectangle{\pgfpointorigin}{\pgfqpoint{8.000000in}{4.800000in}}%
\pgfusepath{use as bounding box, clip}%
\begin{pgfscope}%
\pgfsetbuttcap%
\pgfsetmiterjoin%
\definecolor{currentfill}{rgb}{1.000000,1.000000,1.000000}%
\pgfsetfillcolor{currentfill}%
\pgfsetlinewidth{0.000000pt}%
\definecolor{currentstroke}{rgb}{1.000000,1.000000,1.000000}%
\pgfsetstrokecolor{currentstroke}%
\pgfsetdash{}{0pt}%
\pgfpathmoveto{\pgfqpoint{0.000000in}{0.000000in}}%
\pgfpathlineto{\pgfqpoint{8.000000in}{0.000000in}}%
\pgfpathlineto{\pgfqpoint{8.000000in}{4.800000in}}%
\pgfpathlineto{\pgfqpoint{0.000000in}{4.800000in}}%
\pgfpathlineto{\pgfqpoint{0.000000in}{0.000000in}}%
\pgfpathclose%
\pgfusepath{fill}%
\end{pgfscope}%
\begin{pgfscope}%
\pgfsetbuttcap%
\pgfsetmiterjoin%
\definecolor{currentfill}{rgb}{1.000000,1.000000,1.000000}%
\pgfsetfillcolor{currentfill}%
\pgfsetlinewidth{0.000000pt}%
\definecolor{currentstroke}{rgb}{0.000000,0.000000,0.000000}%
\pgfsetstrokecolor{currentstroke}%
\pgfsetstrokeopacity{0.000000}%
\pgfsetdash{}{0pt}%
\pgfpathmoveto{\pgfqpoint{1.458910in}{0.906664in}}%
\pgfpathlineto{\pgfqpoint{7.730000in}{0.906664in}}%
\pgfpathlineto{\pgfqpoint{7.730000in}{4.497481in}}%
\pgfpathlineto{\pgfqpoint{1.458910in}{4.497481in}}%
\pgfpathlineto{\pgfqpoint{1.458910in}{0.906664in}}%
\pgfpathclose%
\pgfusepath{fill}%
\end{pgfscope}%
\begin{pgfscope}%
\pgfpathrectangle{\pgfqpoint{1.458910in}{0.906664in}}{\pgfqpoint{6.271090in}{3.590817in}}%
\pgfusepath{clip}%
\pgfsetbuttcap%
\pgfsetmiterjoin%
\definecolor{currentfill}{rgb}{0.819608,0.819608,0.819608}%
\pgfsetfillcolor{currentfill}%
\pgfsetlinewidth{1.003750pt}%
\definecolor{currentstroke}{rgb}{0.000000,0.000000,0.000000}%
\pgfsetstrokecolor{currentstroke}%
\pgfsetdash{}{0pt}%
\pgfpathmoveto{\pgfqpoint{2.041773in}{0.906664in}}%
\pgfpathlineto{\pgfqpoint{2.339586in}{0.906664in}}%
\pgfpathlineto{\pgfqpoint{2.339586in}{0.920454in}}%
\pgfpathlineto{\pgfqpoint{2.041773in}{0.920454in}}%
\pgfpathlineto{\pgfqpoint{2.041773in}{0.906664in}}%
\pgfpathclose%
\pgfusepath{stroke,fill}%
\end{pgfscope}%
\begin{pgfscope}%
\pgfsetbuttcap%
\pgfsetmiterjoin%
\definecolor{currentfill}{rgb}{0.819608,0.819608,0.819608}%
\pgfsetfillcolor{currentfill}%
\pgfsetlinewidth{1.003750pt}%
\definecolor{currentstroke}{rgb}{0.000000,0.000000,0.000000}%
\pgfsetstrokecolor{currentstroke}%
\pgfsetdash{}{0pt}%
\pgfpathrectangle{\pgfqpoint{1.458910in}{0.906664in}}{\pgfqpoint{6.271090in}{3.590817in}}%
\pgfusepath{clip}%
\pgfpathmoveto{\pgfqpoint{2.041773in}{0.906664in}}%
\pgfpathlineto{\pgfqpoint{2.339586in}{0.906664in}}%
\pgfpathlineto{\pgfqpoint{2.339586in}{0.920454in}}%
\pgfpathlineto{\pgfqpoint{2.041773in}{0.920454in}}%
\pgfpathlineto{\pgfqpoint{2.041773in}{0.906664in}}%
\pgfpathclose%
\pgfusepath{clip}%
\pgfsys@defobject{currentpattern}{\pgfqpoint{0in}{0in}}{\pgfqpoint{1in}{1in}}{%
\begin{pgfscope}%
\pgfpathrectangle{\pgfqpoint{0in}{0in}}{\pgfqpoint{1in}{1in}}%
\pgfusepath{clip}%
\pgfpathmoveto{\pgfqpoint{-0.500000in}{0.500000in}}%
\pgfpathlineto{\pgfqpoint{0.500000in}{1.500000in}}%
\pgfpathmoveto{\pgfqpoint{-0.333333in}{0.333333in}}%
\pgfpathlineto{\pgfqpoint{0.666667in}{1.333333in}}%
\pgfpathmoveto{\pgfqpoint{-0.166667in}{0.166667in}}%
\pgfpathlineto{\pgfqpoint{0.833333in}{1.166667in}}%
\pgfpathmoveto{\pgfqpoint{0.000000in}{0.000000in}}%
\pgfpathlineto{\pgfqpoint{1.000000in}{1.000000in}}%
\pgfpathmoveto{\pgfqpoint{0.166667in}{-0.166667in}}%
\pgfpathlineto{\pgfqpoint{1.166667in}{0.833333in}}%
\pgfpathmoveto{\pgfqpoint{0.333333in}{-0.333333in}}%
\pgfpathlineto{\pgfqpoint{1.333333in}{0.666667in}}%
\pgfpathmoveto{\pgfqpoint{0.500000in}{-0.500000in}}%
\pgfpathlineto{\pgfqpoint{1.500000in}{0.500000in}}%
\pgfusepath{stroke}%
\end{pgfscope}%
}%
\pgfsys@transformshift{2.041773in}{0.906664in}%
\pgfsys@useobject{currentpattern}{}%
\pgfsys@transformshift{1in}{0in}%
\pgfsys@transformshift{-1in}{0in}%
\pgfsys@transformshift{0in}{1in}%
\end{pgfscope}%
\begin{pgfscope}%
\pgfpathrectangle{\pgfqpoint{1.458910in}{0.906664in}}{\pgfqpoint{6.271090in}{3.590817in}}%
\pgfusepath{clip}%
\pgfsetbuttcap%
\pgfsetmiterjoin%
\definecolor{currentfill}{rgb}{0.819608,0.819608,0.819608}%
\pgfsetfillcolor{currentfill}%
\pgfsetlinewidth{1.003750pt}%
\definecolor{currentstroke}{rgb}{0.000000,0.000000,0.000000}%
\pgfsetstrokecolor{currentstroke}%
\pgfsetdash{}{0pt}%
\pgfpathmoveto{\pgfqpoint{2.892667in}{0.906664in}}%
\pgfpathlineto{\pgfqpoint{3.190480in}{0.906664in}}%
\pgfpathlineto{\pgfqpoint{3.190480in}{0.906664in}}%
\pgfpathlineto{\pgfqpoint{2.892667in}{0.906664in}}%
\pgfpathlineto{\pgfqpoint{2.892667in}{0.906664in}}%
\pgfpathclose%
\pgfusepath{stroke,fill}%
\end{pgfscope}%
\begin{pgfscope}%
\pgfsetbuttcap%
\pgfsetmiterjoin%
\definecolor{currentfill}{rgb}{0.819608,0.819608,0.819608}%
\pgfsetfillcolor{currentfill}%
\pgfsetlinewidth{1.003750pt}%
\definecolor{currentstroke}{rgb}{0.000000,0.000000,0.000000}%
\pgfsetstrokecolor{currentstroke}%
\pgfsetdash{}{0pt}%
\pgfpathrectangle{\pgfqpoint{1.458910in}{0.906664in}}{\pgfqpoint{6.271090in}{3.590817in}}%
\pgfusepath{clip}%
\pgfpathmoveto{\pgfqpoint{2.892667in}{0.906664in}}%
\pgfpathlineto{\pgfqpoint{3.190480in}{0.906664in}}%
\pgfpathlineto{\pgfqpoint{3.190480in}{0.906664in}}%
\pgfpathlineto{\pgfqpoint{2.892667in}{0.906664in}}%
\pgfpathlineto{\pgfqpoint{2.892667in}{0.906664in}}%
\pgfpathclose%
\pgfusepath{clip}%
\pgfsys@defobject{currentpattern}{\pgfqpoint{0in}{0in}}{\pgfqpoint{1in}{1in}}{%
\begin{pgfscope}%
\pgfpathrectangle{\pgfqpoint{0in}{0in}}{\pgfqpoint{1in}{1in}}%
\pgfusepath{clip}%
\pgfpathmoveto{\pgfqpoint{-0.500000in}{0.500000in}}%
\pgfpathlineto{\pgfqpoint{0.500000in}{1.500000in}}%
\pgfpathmoveto{\pgfqpoint{-0.333333in}{0.333333in}}%
\pgfpathlineto{\pgfqpoint{0.666667in}{1.333333in}}%
\pgfpathmoveto{\pgfqpoint{-0.166667in}{0.166667in}}%
\pgfpathlineto{\pgfqpoint{0.833333in}{1.166667in}}%
\pgfpathmoveto{\pgfqpoint{0.000000in}{0.000000in}}%
\pgfpathlineto{\pgfqpoint{1.000000in}{1.000000in}}%
\pgfpathmoveto{\pgfqpoint{0.166667in}{-0.166667in}}%
\pgfpathlineto{\pgfqpoint{1.166667in}{0.833333in}}%
\pgfpathmoveto{\pgfqpoint{0.333333in}{-0.333333in}}%
\pgfpathlineto{\pgfqpoint{1.333333in}{0.666667in}}%
\pgfpathmoveto{\pgfqpoint{0.500000in}{-0.500000in}}%
\pgfpathlineto{\pgfqpoint{1.500000in}{0.500000in}}%
\pgfusepath{stroke}%
\end{pgfscope}%
}%
\pgfsys@transformshift{2.892667in}{0.906664in}%
\end{pgfscope}%
\begin{pgfscope}%
\pgfpathrectangle{\pgfqpoint{1.458910in}{0.906664in}}{\pgfqpoint{6.271090in}{3.590817in}}%
\pgfusepath{clip}%
\pgfsetbuttcap%
\pgfsetmiterjoin%
\definecolor{currentfill}{rgb}{0.819608,0.819608,0.819608}%
\pgfsetfillcolor{currentfill}%
\pgfsetlinewidth{1.003750pt}%
\definecolor{currentstroke}{rgb}{0.000000,0.000000,0.000000}%
\pgfsetstrokecolor{currentstroke}%
\pgfsetdash{}{0pt}%
\pgfpathmoveto{\pgfqpoint{3.743561in}{0.906664in}}%
\pgfpathlineto{\pgfqpoint{4.041374in}{0.906664in}}%
\pgfpathlineto{\pgfqpoint{4.041374in}{0.948033in}}%
\pgfpathlineto{\pgfqpoint{3.743561in}{0.948033in}}%
\pgfpathlineto{\pgfqpoint{3.743561in}{0.906664in}}%
\pgfpathclose%
\pgfusepath{stroke,fill}%
\end{pgfscope}%
\begin{pgfscope}%
\pgfsetbuttcap%
\pgfsetmiterjoin%
\definecolor{currentfill}{rgb}{0.819608,0.819608,0.819608}%
\pgfsetfillcolor{currentfill}%
\pgfsetlinewidth{1.003750pt}%
\definecolor{currentstroke}{rgb}{0.000000,0.000000,0.000000}%
\pgfsetstrokecolor{currentstroke}%
\pgfsetdash{}{0pt}%
\pgfpathrectangle{\pgfqpoint{1.458910in}{0.906664in}}{\pgfqpoint{6.271090in}{3.590817in}}%
\pgfusepath{clip}%
\pgfpathmoveto{\pgfqpoint{3.743561in}{0.906664in}}%
\pgfpathlineto{\pgfqpoint{4.041374in}{0.906664in}}%
\pgfpathlineto{\pgfqpoint{4.041374in}{0.948033in}}%
\pgfpathlineto{\pgfqpoint{3.743561in}{0.948033in}}%
\pgfpathlineto{\pgfqpoint{3.743561in}{0.906664in}}%
\pgfpathclose%
\pgfusepath{clip}%
\pgfsys@defobject{currentpattern}{\pgfqpoint{0in}{0in}}{\pgfqpoint{1in}{1in}}{%
\begin{pgfscope}%
\pgfpathrectangle{\pgfqpoint{0in}{0in}}{\pgfqpoint{1in}{1in}}%
\pgfusepath{clip}%
\pgfpathmoveto{\pgfqpoint{-0.500000in}{0.500000in}}%
\pgfpathlineto{\pgfqpoint{0.500000in}{1.500000in}}%
\pgfpathmoveto{\pgfqpoint{-0.333333in}{0.333333in}}%
\pgfpathlineto{\pgfqpoint{0.666667in}{1.333333in}}%
\pgfpathmoveto{\pgfqpoint{-0.166667in}{0.166667in}}%
\pgfpathlineto{\pgfqpoint{0.833333in}{1.166667in}}%
\pgfpathmoveto{\pgfqpoint{0.000000in}{0.000000in}}%
\pgfpathlineto{\pgfqpoint{1.000000in}{1.000000in}}%
\pgfpathmoveto{\pgfqpoint{0.166667in}{-0.166667in}}%
\pgfpathlineto{\pgfqpoint{1.166667in}{0.833333in}}%
\pgfpathmoveto{\pgfqpoint{0.333333in}{-0.333333in}}%
\pgfpathlineto{\pgfqpoint{1.333333in}{0.666667in}}%
\pgfpathmoveto{\pgfqpoint{0.500000in}{-0.500000in}}%
\pgfpathlineto{\pgfqpoint{1.500000in}{0.500000in}}%
\pgfusepath{stroke}%
\end{pgfscope}%
}%
\pgfsys@transformshift{3.743561in}{0.906664in}%
\pgfsys@useobject{currentpattern}{}%
\pgfsys@transformshift{1in}{0in}%
\pgfsys@transformshift{-1in}{0in}%
\pgfsys@transformshift{0in}{1in}%
\end{pgfscope}%
\begin{pgfscope}%
\pgfpathrectangle{\pgfqpoint{1.458910in}{0.906664in}}{\pgfqpoint{6.271090in}{3.590817in}}%
\pgfusepath{clip}%
\pgfsetbuttcap%
\pgfsetmiterjoin%
\definecolor{currentfill}{rgb}{0.819608,0.819608,0.819608}%
\pgfsetfillcolor{currentfill}%
\pgfsetlinewidth{1.003750pt}%
\definecolor{currentstroke}{rgb}{0.000000,0.000000,0.000000}%
\pgfsetstrokecolor{currentstroke}%
\pgfsetdash{}{0pt}%
\pgfpathmoveto{\pgfqpoint{4.594455in}{0.906664in}}%
\pgfpathlineto{\pgfqpoint{4.892268in}{0.906664in}}%
\pgfpathlineto{\pgfqpoint{4.892268in}{0.934243in}}%
\pgfpathlineto{\pgfqpoint{4.594455in}{0.934243in}}%
\pgfpathlineto{\pgfqpoint{4.594455in}{0.906664in}}%
\pgfpathclose%
\pgfusepath{stroke,fill}%
\end{pgfscope}%
\begin{pgfscope}%
\pgfsetbuttcap%
\pgfsetmiterjoin%
\definecolor{currentfill}{rgb}{0.819608,0.819608,0.819608}%
\pgfsetfillcolor{currentfill}%
\pgfsetlinewidth{1.003750pt}%
\definecolor{currentstroke}{rgb}{0.000000,0.000000,0.000000}%
\pgfsetstrokecolor{currentstroke}%
\pgfsetdash{}{0pt}%
\pgfpathrectangle{\pgfqpoint{1.458910in}{0.906664in}}{\pgfqpoint{6.271090in}{3.590817in}}%
\pgfusepath{clip}%
\pgfpathmoveto{\pgfqpoint{4.594455in}{0.906664in}}%
\pgfpathlineto{\pgfqpoint{4.892268in}{0.906664in}}%
\pgfpathlineto{\pgfqpoint{4.892268in}{0.934243in}}%
\pgfpathlineto{\pgfqpoint{4.594455in}{0.934243in}}%
\pgfpathlineto{\pgfqpoint{4.594455in}{0.906664in}}%
\pgfpathclose%
\pgfusepath{clip}%
\pgfsys@defobject{currentpattern}{\pgfqpoint{0in}{0in}}{\pgfqpoint{1in}{1in}}{%
\begin{pgfscope}%
\pgfpathrectangle{\pgfqpoint{0in}{0in}}{\pgfqpoint{1in}{1in}}%
\pgfusepath{clip}%
\pgfpathmoveto{\pgfqpoint{-0.500000in}{0.500000in}}%
\pgfpathlineto{\pgfqpoint{0.500000in}{1.500000in}}%
\pgfpathmoveto{\pgfqpoint{-0.333333in}{0.333333in}}%
\pgfpathlineto{\pgfqpoint{0.666667in}{1.333333in}}%
\pgfpathmoveto{\pgfqpoint{-0.166667in}{0.166667in}}%
\pgfpathlineto{\pgfqpoint{0.833333in}{1.166667in}}%
\pgfpathmoveto{\pgfqpoint{0.000000in}{0.000000in}}%
\pgfpathlineto{\pgfqpoint{1.000000in}{1.000000in}}%
\pgfpathmoveto{\pgfqpoint{0.166667in}{-0.166667in}}%
\pgfpathlineto{\pgfqpoint{1.166667in}{0.833333in}}%
\pgfpathmoveto{\pgfqpoint{0.333333in}{-0.333333in}}%
\pgfpathlineto{\pgfqpoint{1.333333in}{0.666667in}}%
\pgfpathmoveto{\pgfqpoint{0.500000in}{-0.500000in}}%
\pgfpathlineto{\pgfqpoint{1.500000in}{0.500000in}}%
\pgfusepath{stroke}%
\end{pgfscope}%
}%
\pgfsys@transformshift{4.594455in}{0.906664in}%
\pgfsys@useobject{currentpattern}{}%
\pgfsys@transformshift{1in}{0in}%
\pgfsys@transformshift{-1in}{0in}%
\pgfsys@transformshift{0in}{1in}%
\end{pgfscope}%
\begin{pgfscope}%
\pgfpathrectangle{\pgfqpoint{1.458910in}{0.906664in}}{\pgfqpoint{6.271090in}{3.590817in}}%
\pgfusepath{clip}%
\pgfsetbuttcap%
\pgfsetmiterjoin%
\definecolor{currentfill}{rgb}{0.819608,0.819608,0.819608}%
\pgfsetfillcolor{currentfill}%
\pgfsetlinewidth{1.003750pt}%
\definecolor{currentstroke}{rgb}{0.000000,0.000000,0.000000}%
\pgfsetstrokecolor{currentstroke}%
\pgfsetdash{}{0pt}%
\pgfpathmoveto{\pgfqpoint{5.445349in}{0.906664in}}%
\pgfpathlineto{\pgfqpoint{5.743162in}{0.906664in}}%
\pgfpathlineto{\pgfqpoint{5.743162in}{0.934243in}}%
\pgfpathlineto{\pgfqpoint{5.445349in}{0.934243in}}%
\pgfpathlineto{\pgfqpoint{5.445349in}{0.906664in}}%
\pgfpathclose%
\pgfusepath{stroke,fill}%
\end{pgfscope}%
\begin{pgfscope}%
\pgfsetbuttcap%
\pgfsetmiterjoin%
\definecolor{currentfill}{rgb}{0.819608,0.819608,0.819608}%
\pgfsetfillcolor{currentfill}%
\pgfsetlinewidth{1.003750pt}%
\definecolor{currentstroke}{rgb}{0.000000,0.000000,0.000000}%
\pgfsetstrokecolor{currentstroke}%
\pgfsetdash{}{0pt}%
\pgfpathrectangle{\pgfqpoint{1.458910in}{0.906664in}}{\pgfqpoint{6.271090in}{3.590817in}}%
\pgfusepath{clip}%
\pgfpathmoveto{\pgfqpoint{5.445349in}{0.906664in}}%
\pgfpathlineto{\pgfqpoint{5.743162in}{0.906664in}}%
\pgfpathlineto{\pgfqpoint{5.743162in}{0.934243in}}%
\pgfpathlineto{\pgfqpoint{5.445349in}{0.934243in}}%
\pgfpathlineto{\pgfqpoint{5.445349in}{0.906664in}}%
\pgfpathclose%
\pgfusepath{clip}%
\pgfsys@defobject{currentpattern}{\pgfqpoint{0in}{0in}}{\pgfqpoint{1in}{1in}}{%
\begin{pgfscope}%
\pgfpathrectangle{\pgfqpoint{0in}{0in}}{\pgfqpoint{1in}{1in}}%
\pgfusepath{clip}%
\pgfpathmoveto{\pgfqpoint{-0.500000in}{0.500000in}}%
\pgfpathlineto{\pgfqpoint{0.500000in}{1.500000in}}%
\pgfpathmoveto{\pgfqpoint{-0.333333in}{0.333333in}}%
\pgfpathlineto{\pgfqpoint{0.666667in}{1.333333in}}%
\pgfpathmoveto{\pgfqpoint{-0.166667in}{0.166667in}}%
\pgfpathlineto{\pgfqpoint{0.833333in}{1.166667in}}%
\pgfpathmoveto{\pgfqpoint{0.000000in}{0.000000in}}%
\pgfpathlineto{\pgfqpoint{1.000000in}{1.000000in}}%
\pgfpathmoveto{\pgfqpoint{0.166667in}{-0.166667in}}%
\pgfpathlineto{\pgfqpoint{1.166667in}{0.833333in}}%
\pgfpathmoveto{\pgfqpoint{0.333333in}{-0.333333in}}%
\pgfpathlineto{\pgfqpoint{1.333333in}{0.666667in}}%
\pgfpathmoveto{\pgfqpoint{0.500000in}{-0.500000in}}%
\pgfpathlineto{\pgfqpoint{1.500000in}{0.500000in}}%
\pgfusepath{stroke}%
\end{pgfscope}%
}%
\pgfsys@transformshift{5.445349in}{0.906664in}%
\pgfsys@useobject{currentpattern}{}%
\pgfsys@transformshift{1in}{0in}%
\pgfsys@transformshift{-1in}{0in}%
\pgfsys@transformshift{0in}{1in}%
\end{pgfscope}%
\begin{pgfscope}%
\pgfpathrectangle{\pgfqpoint{1.458910in}{0.906664in}}{\pgfqpoint{6.271090in}{3.590817in}}%
\pgfusepath{clip}%
\pgfsetbuttcap%
\pgfsetmiterjoin%
\definecolor{currentfill}{rgb}{0.819608,0.819608,0.819608}%
\pgfsetfillcolor{currentfill}%
\pgfsetlinewidth{1.003750pt}%
\definecolor{currentstroke}{rgb}{0.000000,0.000000,0.000000}%
\pgfsetstrokecolor{currentstroke}%
\pgfsetdash{}{0pt}%
\pgfpathmoveto{\pgfqpoint{6.296243in}{0.906664in}}%
\pgfpathlineto{\pgfqpoint{6.594056in}{0.906664in}}%
\pgfpathlineto{\pgfqpoint{6.594056in}{0.934243in}}%
\pgfpathlineto{\pgfqpoint{6.296243in}{0.934243in}}%
\pgfpathlineto{\pgfqpoint{6.296243in}{0.906664in}}%
\pgfpathclose%
\pgfusepath{stroke,fill}%
\end{pgfscope}%
\begin{pgfscope}%
\pgfsetbuttcap%
\pgfsetmiterjoin%
\definecolor{currentfill}{rgb}{0.819608,0.819608,0.819608}%
\pgfsetfillcolor{currentfill}%
\pgfsetlinewidth{1.003750pt}%
\definecolor{currentstroke}{rgb}{0.000000,0.000000,0.000000}%
\pgfsetstrokecolor{currentstroke}%
\pgfsetdash{}{0pt}%
\pgfpathrectangle{\pgfqpoint{1.458910in}{0.906664in}}{\pgfqpoint{6.271090in}{3.590817in}}%
\pgfusepath{clip}%
\pgfpathmoveto{\pgfqpoint{6.296243in}{0.906664in}}%
\pgfpathlineto{\pgfqpoint{6.594056in}{0.906664in}}%
\pgfpathlineto{\pgfqpoint{6.594056in}{0.934243in}}%
\pgfpathlineto{\pgfqpoint{6.296243in}{0.934243in}}%
\pgfpathlineto{\pgfqpoint{6.296243in}{0.906664in}}%
\pgfpathclose%
\pgfusepath{clip}%
\pgfsys@defobject{currentpattern}{\pgfqpoint{0in}{0in}}{\pgfqpoint{1in}{1in}}{%
\begin{pgfscope}%
\pgfpathrectangle{\pgfqpoint{0in}{0in}}{\pgfqpoint{1in}{1in}}%
\pgfusepath{clip}%
\pgfpathmoveto{\pgfqpoint{-0.500000in}{0.500000in}}%
\pgfpathlineto{\pgfqpoint{0.500000in}{1.500000in}}%
\pgfpathmoveto{\pgfqpoint{-0.333333in}{0.333333in}}%
\pgfpathlineto{\pgfqpoint{0.666667in}{1.333333in}}%
\pgfpathmoveto{\pgfqpoint{-0.166667in}{0.166667in}}%
\pgfpathlineto{\pgfqpoint{0.833333in}{1.166667in}}%
\pgfpathmoveto{\pgfqpoint{0.000000in}{0.000000in}}%
\pgfpathlineto{\pgfqpoint{1.000000in}{1.000000in}}%
\pgfpathmoveto{\pgfqpoint{0.166667in}{-0.166667in}}%
\pgfpathlineto{\pgfqpoint{1.166667in}{0.833333in}}%
\pgfpathmoveto{\pgfqpoint{0.333333in}{-0.333333in}}%
\pgfpathlineto{\pgfqpoint{1.333333in}{0.666667in}}%
\pgfpathmoveto{\pgfqpoint{0.500000in}{-0.500000in}}%
\pgfpathlineto{\pgfqpoint{1.500000in}{0.500000in}}%
\pgfusepath{stroke}%
\end{pgfscope}%
}%
\pgfsys@transformshift{6.296243in}{0.906664in}%
\pgfsys@useobject{currentpattern}{}%
\pgfsys@transformshift{1in}{0in}%
\pgfsys@transformshift{-1in}{0in}%
\pgfsys@transformshift{0in}{1in}%
\end{pgfscope}%
\begin{pgfscope}%
\pgfpathrectangle{\pgfqpoint{1.458910in}{0.906664in}}{\pgfqpoint{6.271090in}{3.590817in}}%
\pgfusepath{clip}%
\pgfsetbuttcap%
\pgfsetmiterjoin%
\definecolor{currentfill}{rgb}{0.819608,0.819608,0.819608}%
\pgfsetfillcolor{currentfill}%
\pgfsetlinewidth{1.003750pt}%
\definecolor{currentstroke}{rgb}{0.000000,0.000000,0.000000}%
\pgfsetstrokecolor{currentstroke}%
\pgfsetdash{}{0pt}%
\pgfpathmoveto{\pgfqpoint{7.147138in}{0.906664in}}%
\pgfpathlineto{\pgfqpoint{7.444950in}{0.906664in}}%
\pgfpathlineto{\pgfqpoint{7.444950in}{0.948033in}}%
\pgfpathlineto{\pgfqpoint{7.147138in}{0.948033in}}%
\pgfpathlineto{\pgfqpoint{7.147138in}{0.906664in}}%
\pgfpathclose%
\pgfusepath{stroke,fill}%
\end{pgfscope}%
\begin{pgfscope}%
\pgfsetbuttcap%
\pgfsetmiterjoin%
\definecolor{currentfill}{rgb}{0.819608,0.819608,0.819608}%
\pgfsetfillcolor{currentfill}%
\pgfsetlinewidth{1.003750pt}%
\definecolor{currentstroke}{rgb}{0.000000,0.000000,0.000000}%
\pgfsetstrokecolor{currentstroke}%
\pgfsetdash{}{0pt}%
\pgfpathrectangle{\pgfqpoint{1.458910in}{0.906664in}}{\pgfqpoint{6.271090in}{3.590817in}}%
\pgfusepath{clip}%
\pgfpathmoveto{\pgfqpoint{7.147138in}{0.906664in}}%
\pgfpathlineto{\pgfqpoint{7.444950in}{0.906664in}}%
\pgfpathlineto{\pgfqpoint{7.444950in}{0.948033in}}%
\pgfpathlineto{\pgfqpoint{7.147138in}{0.948033in}}%
\pgfpathlineto{\pgfqpoint{7.147138in}{0.906664in}}%
\pgfpathclose%
\pgfusepath{clip}%
\pgfsys@defobject{currentpattern}{\pgfqpoint{0in}{0in}}{\pgfqpoint{1in}{1in}}{%
\begin{pgfscope}%
\pgfpathrectangle{\pgfqpoint{0in}{0in}}{\pgfqpoint{1in}{1in}}%
\pgfusepath{clip}%
\pgfpathmoveto{\pgfqpoint{-0.500000in}{0.500000in}}%
\pgfpathlineto{\pgfqpoint{0.500000in}{1.500000in}}%
\pgfpathmoveto{\pgfqpoint{-0.333333in}{0.333333in}}%
\pgfpathlineto{\pgfqpoint{0.666667in}{1.333333in}}%
\pgfpathmoveto{\pgfqpoint{-0.166667in}{0.166667in}}%
\pgfpathlineto{\pgfqpoint{0.833333in}{1.166667in}}%
\pgfpathmoveto{\pgfqpoint{0.000000in}{0.000000in}}%
\pgfpathlineto{\pgfqpoint{1.000000in}{1.000000in}}%
\pgfpathmoveto{\pgfqpoint{0.166667in}{-0.166667in}}%
\pgfpathlineto{\pgfqpoint{1.166667in}{0.833333in}}%
\pgfpathmoveto{\pgfqpoint{0.333333in}{-0.333333in}}%
\pgfpathlineto{\pgfqpoint{1.333333in}{0.666667in}}%
\pgfpathmoveto{\pgfqpoint{0.500000in}{-0.500000in}}%
\pgfpathlineto{\pgfqpoint{1.500000in}{0.500000in}}%
\pgfusepath{stroke}%
\end{pgfscope}%
}%
\pgfsys@transformshift{7.147138in}{0.906664in}%
\pgfsys@useobject{currentpattern}{}%
\pgfsys@transformshift{1in}{0in}%
\pgfsys@transformshift{-1in}{0in}%
\pgfsys@transformshift{0in}{1in}%
\end{pgfscope}%
\begin{pgfscope}%
\pgfpathrectangle{\pgfqpoint{1.458910in}{0.906664in}}{\pgfqpoint{6.271090in}{3.590817in}}%
\pgfusepath{clip}%
\pgfsetbuttcap%
\pgfsetmiterjoin%
\definecolor{currentfill}{rgb}{0.968627,0.678431,0.811765}%
\pgfsetfillcolor{currentfill}%
\pgfsetlinewidth{1.003750pt}%
\definecolor{currentstroke}{rgb}{0.000000,0.000000,0.000000}%
\pgfsetstrokecolor{currentstroke}%
\pgfsetdash{}{0pt}%
\pgfpathmoveto{\pgfqpoint{1.743960in}{0.906664in}}%
\pgfpathlineto{\pgfqpoint{2.041773in}{0.906664in}}%
\pgfpathlineto{\pgfqpoint{2.041773in}{4.174804in}}%
\pgfpathlineto{\pgfqpoint{1.743960in}{4.174804in}}%
\pgfpathlineto{\pgfqpoint{1.743960in}{0.906664in}}%
\pgfpathclose%
\pgfusepath{stroke,fill}%
\end{pgfscope}%
\begin{pgfscope}%
\pgfpathrectangle{\pgfqpoint{1.458910in}{0.906664in}}{\pgfqpoint{6.271090in}{3.590817in}}%
\pgfusepath{clip}%
\pgfsetbuttcap%
\pgfsetmiterjoin%
\definecolor{currentfill}{rgb}{0.968627,0.678431,0.811765}%
\pgfsetfillcolor{currentfill}%
\pgfsetlinewidth{1.003750pt}%
\definecolor{currentstroke}{rgb}{0.000000,0.000000,0.000000}%
\pgfsetstrokecolor{currentstroke}%
\pgfsetdash{}{0pt}%
\pgfpathmoveto{\pgfqpoint{2.594854in}{0.906664in}}%
\pgfpathlineto{\pgfqpoint{2.892667in}{0.906664in}}%
\pgfpathlineto{\pgfqpoint{2.892667in}{4.243752in}}%
\pgfpathlineto{\pgfqpoint{2.594854in}{4.243752in}}%
\pgfpathlineto{\pgfqpoint{2.594854in}{0.906664in}}%
\pgfpathclose%
\pgfusepath{stroke,fill}%
\end{pgfscope}%
\begin{pgfscope}%
\pgfpathrectangle{\pgfqpoint{1.458910in}{0.906664in}}{\pgfqpoint{6.271090in}{3.590817in}}%
\pgfusepath{clip}%
\pgfsetbuttcap%
\pgfsetmiterjoin%
\definecolor{currentfill}{rgb}{0.968627,0.678431,0.811765}%
\pgfsetfillcolor{currentfill}%
\pgfsetlinewidth{1.003750pt}%
\definecolor{currentstroke}{rgb}{0.000000,0.000000,0.000000}%
\pgfsetstrokecolor{currentstroke}%
\pgfsetdash{}{0pt}%
\pgfpathmoveto{\pgfqpoint{3.445748in}{0.906664in}}%
\pgfpathlineto{\pgfqpoint{3.743561in}{0.906664in}}%
\pgfpathlineto{\pgfqpoint{3.743561in}{4.188593in}}%
\pgfpathlineto{\pgfqpoint{3.445748in}{4.188593in}}%
\pgfpathlineto{\pgfqpoint{3.445748in}{0.906664in}}%
\pgfpathclose%
\pgfusepath{stroke,fill}%
\end{pgfscope}%
\begin{pgfscope}%
\pgfpathrectangle{\pgfqpoint{1.458910in}{0.906664in}}{\pgfqpoint{6.271090in}{3.590817in}}%
\pgfusepath{clip}%
\pgfsetbuttcap%
\pgfsetmiterjoin%
\definecolor{currentfill}{rgb}{0.968627,0.678431,0.811765}%
\pgfsetfillcolor{currentfill}%
\pgfsetlinewidth{1.003750pt}%
\definecolor{currentstroke}{rgb}{0.000000,0.000000,0.000000}%
\pgfsetstrokecolor{currentstroke}%
\pgfsetdash{}{0pt}%
\pgfpathmoveto{\pgfqpoint{4.296642in}{0.906664in}}%
\pgfpathlineto{\pgfqpoint{4.594455in}{0.906664in}}%
\pgfpathlineto{\pgfqpoint{4.594455in}{4.243752in}}%
\pgfpathlineto{\pgfqpoint{4.296642in}{4.243752in}}%
\pgfpathlineto{\pgfqpoint{4.296642in}{0.906664in}}%
\pgfpathclose%
\pgfusepath{stroke,fill}%
\end{pgfscope}%
\begin{pgfscope}%
\pgfpathrectangle{\pgfqpoint{1.458910in}{0.906664in}}{\pgfqpoint{6.271090in}{3.590817in}}%
\pgfusepath{clip}%
\pgfsetbuttcap%
\pgfsetmiterjoin%
\definecolor{currentfill}{rgb}{0.968627,0.678431,0.811765}%
\pgfsetfillcolor{currentfill}%
\pgfsetlinewidth{1.003750pt}%
\definecolor{currentstroke}{rgb}{0.000000,0.000000,0.000000}%
\pgfsetstrokecolor{currentstroke}%
\pgfsetdash{}{0pt}%
\pgfpathmoveto{\pgfqpoint{5.147536in}{0.906664in}}%
\pgfpathlineto{\pgfqpoint{5.445349in}{0.906664in}}%
\pgfpathlineto{\pgfqpoint{5.445349in}{4.147225in}}%
\pgfpathlineto{\pgfqpoint{5.147536in}{4.147225in}}%
\pgfpathlineto{\pgfqpoint{5.147536in}{0.906664in}}%
\pgfpathclose%
\pgfusepath{stroke,fill}%
\end{pgfscope}%
\begin{pgfscope}%
\pgfpathrectangle{\pgfqpoint{1.458910in}{0.906664in}}{\pgfqpoint{6.271090in}{3.590817in}}%
\pgfusepath{clip}%
\pgfsetbuttcap%
\pgfsetmiterjoin%
\definecolor{currentfill}{rgb}{0.968627,0.678431,0.811765}%
\pgfsetfillcolor{currentfill}%
\pgfsetlinewidth{1.003750pt}%
\definecolor{currentstroke}{rgb}{0.000000,0.000000,0.000000}%
\pgfsetstrokecolor{currentstroke}%
\pgfsetdash{}{0pt}%
\pgfpathmoveto{\pgfqpoint{5.998430in}{0.906664in}}%
\pgfpathlineto{\pgfqpoint{6.296243in}{0.906664in}}%
\pgfpathlineto{\pgfqpoint{6.296243in}{4.285121in}}%
\pgfpathlineto{\pgfqpoint{5.998430in}{4.285121in}}%
\pgfpathlineto{\pgfqpoint{5.998430in}{0.906664in}}%
\pgfpathclose%
\pgfusepath{stroke,fill}%
\end{pgfscope}%
\begin{pgfscope}%
\pgfpathrectangle{\pgfqpoint{1.458910in}{0.906664in}}{\pgfqpoint{6.271090in}{3.590817in}}%
\pgfusepath{clip}%
\pgfsetbuttcap%
\pgfsetmiterjoin%
\definecolor{currentfill}{rgb}{0.968627,0.678431,0.811765}%
\pgfsetfillcolor{currentfill}%
\pgfsetlinewidth{1.003750pt}%
\definecolor{currentstroke}{rgb}{0.000000,0.000000,0.000000}%
\pgfsetstrokecolor{currentstroke}%
\pgfsetdash{}{0pt}%
\pgfpathmoveto{\pgfqpoint{6.849325in}{0.906664in}}%
\pgfpathlineto{\pgfqpoint{7.147138in}{0.906664in}}%
\pgfpathlineto{\pgfqpoint{7.147138in}{4.326490in}}%
\pgfpathlineto{\pgfqpoint{6.849325in}{4.326490in}}%
\pgfpathlineto{\pgfqpoint{6.849325in}{0.906664in}}%
\pgfpathclose%
\pgfusepath{stroke,fill}%
\end{pgfscope}%
\begin{pgfscope}%
\pgfsetbuttcap%
\pgfsetroundjoin%
\definecolor{currentfill}{rgb}{0.000000,0.000000,0.000000}%
\pgfsetfillcolor{currentfill}%
\pgfsetlinewidth{0.803000pt}%
\definecolor{currentstroke}{rgb}{0.000000,0.000000,0.000000}%
\pgfsetstrokecolor{currentstroke}%
\pgfsetdash{}{0pt}%
\pgfsys@defobject{currentmarker}{\pgfqpoint{0.000000in}{-0.048611in}}{\pgfqpoint{0.000000in}{0.000000in}}{%
\pgfpathmoveto{\pgfqpoint{0.000000in}{0.000000in}}%
\pgfpathlineto{\pgfqpoint{0.000000in}{-0.048611in}}%
\pgfusepath{stroke,fill}%
}%
\begin{pgfscope}%
\pgfsys@transformshift{2.041773in}{0.906664in}%
\pgfsys@useobject{currentmarker}{}%
\end{pgfscope}%
\end{pgfscope}%
\begin{pgfscope}%
\definecolor{textcolor}{rgb}{0.000000,0.000000,0.000000}%
\pgfsetstrokecolor{textcolor}%
\pgfsetfillcolor{textcolor}%
\pgftext[x=2.041773in,y=0.809442in,,top]{\color{textcolor}\sffamily\fontsize{18.000000}{21.600000}\selectfont 32}%
\end{pgfscope}%
\begin{pgfscope}%
\pgfsetbuttcap%
\pgfsetroundjoin%
\definecolor{currentfill}{rgb}{0.000000,0.000000,0.000000}%
\pgfsetfillcolor{currentfill}%
\pgfsetlinewidth{0.803000pt}%
\definecolor{currentstroke}{rgb}{0.000000,0.000000,0.000000}%
\pgfsetstrokecolor{currentstroke}%
\pgfsetdash{}{0pt}%
\pgfsys@defobject{currentmarker}{\pgfqpoint{0.000000in}{-0.048611in}}{\pgfqpoint{0.000000in}{0.000000in}}{%
\pgfpathmoveto{\pgfqpoint{0.000000in}{0.000000in}}%
\pgfpathlineto{\pgfqpoint{0.000000in}{-0.048611in}}%
\pgfusepath{stroke,fill}%
}%
\begin{pgfscope}%
\pgfsys@transformshift{2.892667in}{0.906664in}%
\pgfsys@useobject{currentmarker}{}%
\end{pgfscope}%
\end{pgfscope}%
\begin{pgfscope}%
\definecolor{textcolor}{rgb}{0.000000,0.000000,0.000000}%
\pgfsetstrokecolor{textcolor}%
\pgfsetfillcolor{textcolor}%
\pgftext[x=2.892667in,y=0.809442in,,top]{\color{textcolor}\sffamily\fontsize{18.000000}{21.600000}\selectfont 33}%
\end{pgfscope}%
\begin{pgfscope}%
\pgfsetbuttcap%
\pgfsetroundjoin%
\definecolor{currentfill}{rgb}{0.000000,0.000000,0.000000}%
\pgfsetfillcolor{currentfill}%
\pgfsetlinewidth{0.803000pt}%
\definecolor{currentstroke}{rgb}{0.000000,0.000000,0.000000}%
\pgfsetstrokecolor{currentstroke}%
\pgfsetdash{}{0pt}%
\pgfsys@defobject{currentmarker}{\pgfqpoint{0.000000in}{-0.048611in}}{\pgfqpoint{0.000000in}{0.000000in}}{%
\pgfpathmoveto{\pgfqpoint{0.000000in}{0.000000in}}%
\pgfpathlineto{\pgfqpoint{0.000000in}{-0.048611in}}%
\pgfusepath{stroke,fill}%
}%
\begin{pgfscope}%
\pgfsys@transformshift{3.743561in}{0.906664in}%
\pgfsys@useobject{currentmarker}{}%
\end{pgfscope}%
\end{pgfscope}%
\begin{pgfscope}%
\definecolor{textcolor}{rgb}{0.000000,0.000000,0.000000}%
\pgfsetstrokecolor{textcolor}%
\pgfsetfillcolor{textcolor}%
\pgftext[x=3.743561in,y=0.809442in,,top]{\color{textcolor}\sffamily\fontsize{18.000000}{21.600000}\selectfont 34}%
\end{pgfscope}%
\begin{pgfscope}%
\pgfsetbuttcap%
\pgfsetroundjoin%
\definecolor{currentfill}{rgb}{0.000000,0.000000,0.000000}%
\pgfsetfillcolor{currentfill}%
\pgfsetlinewidth{0.803000pt}%
\definecolor{currentstroke}{rgb}{0.000000,0.000000,0.000000}%
\pgfsetstrokecolor{currentstroke}%
\pgfsetdash{}{0pt}%
\pgfsys@defobject{currentmarker}{\pgfqpoint{0.000000in}{-0.048611in}}{\pgfqpoint{0.000000in}{0.000000in}}{%
\pgfpathmoveto{\pgfqpoint{0.000000in}{0.000000in}}%
\pgfpathlineto{\pgfqpoint{0.000000in}{-0.048611in}}%
\pgfusepath{stroke,fill}%
}%
\begin{pgfscope}%
\pgfsys@transformshift{4.594455in}{0.906664in}%
\pgfsys@useobject{currentmarker}{}%
\end{pgfscope}%
\end{pgfscope}%
\begin{pgfscope}%
\definecolor{textcolor}{rgb}{0.000000,0.000000,0.000000}%
\pgfsetstrokecolor{textcolor}%
\pgfsetfillcolor{textcolor}%
\pgftext[x=4.594455in,y=0.809442in,,top]{\color{textcolor}\sffamily\fontsize{18.000000}{21.600000}\selectfont 35}%
\end{pgfscope}%
\begin{pgfscope}%
\pgfsetbuttcap%
\pgfsetroundjoin%
\definecolor{currentfill}{rgb}{0.000000,0.000000,0.000000}%
\pgfsetfillcolor{currentfill}%
\pgfsetlinewidth{0.803000pt}%
\definecolor{currentstroke}{rgb}{0.000000,0.000000,0.000000}%
\pgfsetstrokecolor{currentstroke}%
\pgfsetdash{}{0pt}%
\pgfsys@defobject{currentmarker}{\pgfqpoint{0.000000in}{-0.048611in}}{\pgfqpoint{0.000000in}{0.000000in}}{%
\pgfpathmoveto{\pgfqpoint{0.000000in}{0.000000in}}%
\pgfpathlineto{\pgfqpoint{0.000000in}{-0.048611in}}%
\pgfusepath{stroke,fill}%
}%
\begin{pgfscope}%
\pgfsys@transformshift{5.445349in}{0.906664in}%
\pgfsys@useobject{currentmarker}{}%
\end{pgfscope}%
\end{pgfscope}%
\begin{pgfscope}%
\definecolor{textcolor}{rgb}{0.000000,0.000000,0.000000}%
\pgfsetstrokecolor{textcolor}%
\pgfsetfillcolor{textcolor}%
\pgftext[x=5.445349in,y=0.809442in,,top]{\color{textcolor}\sffamily\fontsize{18.000000}{21.600000}\selectfont 36}%
\end{pgfscope}%
\begin{pgfscope}%
\pgfsetbuttcap%
\pgfsetroundjoin%
\definecolor{currentfill}{rgb}{0.000000,0.000000,0.000000}%
\pgfsetfillcolor{currentfill}%
\pgfsetlinewidth{0.803000pt}%
\definecolor{currentstroke}{rgb}{0.000000,0.000000,0.000000}%
\pgfsetstrokecolor{currentstroke}%
\pgfsetdash{}{0pt}%
\pgfsys@defobject{currentmarker}{\pgfqpoint{0.000000in}{-0.048611in}}{\pgfqpoint{0.000000in}{0.000000in}}{%
\pgfpathmoveto{\pgfqpoint{0.000000in}{0.000000in}}%
\pgfpathlineto{\pgfqpoint{0.000000in}{-0.048611in}}%
\pgfusepath{stroke,fill}%
}%
\begin{pgfscope}%
\pgfsys@transformshift{6.296243in}{0.906664in}%
\pgfsys@useobject{currentmarker}{}%
\end{pgfscope}%
\end{pgfscope}%
\begin{pgfscope}%
\definecolor{textcolor}{rgb}{0.000000,0.000000,0.000000}%
\pgfsetstrokecolor{textcolor}%
\pgfsetfillcolor{textcolor}%
\pgftext[x=6.296243in,y=0.809442in,,top]{\color{textcolor}\sffamily\fontsize{18.000000}{21.600000}\selectfont 37}%
\end{pgfscope}%
\begin{pgfscope}%
\pgfsetbuttcap%
\pgfsetroundjoin%
\definecolor{currentfill}{rgb}{0.000000,0.000000,0.000000}%
\pgfsetfillcolor{currentfill}%
\pgfsetlinewidth{0.803000pt}%
\definecolor{currentstroke}{rgb}{0.000000,0.000000,0.000000}%
\pgfsetstrokecolor{currentstroke}%
\pgfsetdash{}{0pt}%
\pgfsys@defobject{currentmarker}{\pgfqpoint{0.000000in}{-0.048611in}}{\pgfqpoint{0.000000in}{0.000000in}}{%
\pgfpathmoveto{\pgfqpoint{0.000000in}{0.000000in}}%
\pgfpathlineto{\pgfqpoint{0.000000in}{-0.048611in}}%
\pgfusepath{stroke,fill}%
}%
\begin{pgfscope}%
\pgfsys@transformshift{7.147138in}{0.906664in}%
\pgfsys@useobject{currentmarker}{}%
\end{pgfscope}%
\end{pgfscope}%
\begin{pgfscope}%
\definecolor{textcolor}{rgb}{0.000000,0.000000,0.000000}%
\pgfsetstrokecolor{textcolor}%
\pgfsetfillcolor{textcolor}%
\pgftext[x=7.147138in,y=0.809442in,,top]{\color{textcolor}\sffamily\fontsize{18.000000}{21.600000}\selectfont 38}%
\end{pgfscope}%
\begin{pgfscope}%
\definecolor{textcolor}{rgb}{0.000000,0.000000,0.000000}%
\pgfsetstrokecolor{textcolor}%
\pgfsetfillcolor{textcolor}%
\pgftext[x=4.594455in,y=0.511943in,,top]{\color{textcolor}\sffamily\fontsize{18.000000}{21.600000}\selectfont Offset bit}%
\end{pgfscope}%
\begin{pgfscope}%
\pgfsetbuttcap%
\pgfsetroundjoin%
\definecolor{currentfill}{rgb}{0.000000,0.000000,0.000000}%
\pgfsetfillcolor{currentfill}%
\pgfsetlinewidth{0.803000pt}%
\definecolor{currentstroke}{rgb}{0.000000,0.000000,0.000000}%
\pgfsetstrokecolor{currentstroke}%
\pgfsetdash{}{0pt}%
\pgfsys@defobject{currentmarker}{\pgfqpoint{-0.048611in}{0.000000in}}{\pgfqpoint{-0.000000in}{0.000000in}}{%
\pgfpathmoveto{\pgfqpoint{-0.000000in}{0.000000in}}%
\pgfpathlineto{\pgfqpoint{-0.048611in}{0.000000in}}%
\pgfusepath{stroke,fill}%
}%
\begin{pgfscope}%
\pgfsys@transformshift{1.458910in}{0.906664in}%
\pgfsys@useobject{currentmarker}{}%
\end{pgfscope}%
\end{pgfscope}%
\begin{pgfscope}%
\definecolor{textcolor}{rgb}{0.000000,0.000000,0.000000}%
\pgfsetstrokecolor{textcolor}%
\pgfsetfillcolor{textcolor}%
\pgftext[x=0.726556in, y=0.811693in, left, base]{\color{textcolor}\sffamily\fontsize{18.000000}{21.600000}\selectfont 0.0\%}%
\end{pgfscope}%
\begin{pgfscope}%
\pgfsetbuttcap%
\pgfsetroundjoin%
\definecolor{currentfill}{rgb}{0.000000,0.000000,0.000000}%
\pgfsetfillcolor{currentfill}%
\pgfsetlinewidth{0.803000pt}%
\definecolor{currentstroke}{rgb}{0.000000,0.000000,0.000000}%
\pgfsetstrokecolor{currentstroke}%
\pgfsetdash{}{0pt}%
\pgfsys@defobject{currentmarker}{\pgfqpoint{-0.048611in}{0.000000in}}{\pgfqpoint{-0.000000in}{0.000000in}}{%
\pgfpathmoveto{\pgfqpoint{-0.000000in}{0.000000in}}%
\pgfpathlineto{\pgfqpoint{-0.048611in}{0.000000in}}%
\pgfusepath{stroke,fill}%
}%
\begin{pgfscope}%
\pgfsys@transformshift{1.458910in}{1.612693in}%
\pgfsys@useobject{currentmarker}{}%
\end{pgfscope}%
\end{pgfscope}%
\begin{pgfscope}%
\definecolor{textcolor}{rgb}{0.000000,0.000000,0.000000}%
\pgfsetstrokecolor{textcolor}%
\pgfsetfillcolor{textcolor}%
\pgftext[x=0.726556in, y=1.517722in, left, base]{\color{textcolor}\sffamily\fontsize{18.000000}{21.600000}\selectfont 5.0\%}%
\end{pgfscope}%
\begin{pgfscope}%
\pgfsetbuttcap%
\pgfsetroundjoin%
\definecolor{currentfill}{rgb}{0.000000,0.000000,0.000000}%
\pgfsetfillcolor{currentfill}%
\pgfsetlinewidth{0.803000pt}%
\definecolor{currentstroke}{rgb}{0.000000,0.000000,0.000000}%
\pgfsetstrokecolor{currentstroke}%
\pgfsetdash{}{0pt}%
\pgfsys@defobject{currentmarker}{\pgfqpoint{-0.048611in}{0.000000in}}{\pgfqpoint{-0.000000in}{0.000000in}}{%
\pgfpathmoveto{\pgfqpoint{-0.000000in}{0.000000in}}%
\pgfpathlineto{\pgfqpoint{-0.048611in}{0.000000in}}%
\pgfusepath{stroke,fill}%
}%
\begin{pgfscope}%
\pgfsys@transformshift{1.458910in}{2.318721in}%
\pgfsys@useobject{currentmarker}{}%
\end{pgfscope}%
\end{pgfscope}%
\begin{pgfscope}%
\definecolor{textcolor}{rgb}{0.000000,0.000000,0.000000}%
\pgfsetstrokecolor{textcolor}%
\pgfsetfillcolor{textcolor}%
\pgftext[x=0.567499in, y=2.223750in, left, base]{\color{textcolor}\sffamily\fontsize{18.000000}{21.600000}\selectfont 10.0\%}%
\end{pgfscope}%
\begin{pgfscope}%
\pgfsetbuttcap%
\pgfsetroundjoin%
\definecolor{currentfill}{rgb}{0.000000,0.000000,0.000000}%
\pgfsetfillcolor{currentfill}%
\pgfsetlinewidth{0.803000pt}%
\definecolor{currentstroke}{rgb}{0.000000,0.000000,0.000000}%
\pgfsetstrokecolor{currentstroke}%
\pgfsetdash{}{0pt}%
\pgfsys@defobject{currentmarker}{\pgfqpoint{-0.048611in}{0.000000in}}{\pgfqpoint{-0.000000in}{0.000000in}}{%
\pgfpathmoveto{\pgfqpoint{-0.000000in}{0.000000in}}%
\pgfpathlineto{\pgfqpoint{-0.048611in}{0.000000in}}%
\pgfusepath{stroke,fill}%
}%
\begin{pgfscope}%
\pgfsys@transformshift{1.458910in}{3.024750in}%
\pgfsys@useobject{currentmarker}{}%
\end{pgfscope}%
\end{pgfscope}%
\begin{pgfscope}%
\definecolor{textcolor}{rgb}{0.000000,0.000000,0.000000}%
\pgfsetstrokecolor{textcolor}%
\pgfsetfillcolor{textcolor}%
\pgftext[x=0.567499in, y=2.929779in, left, base]{\color{textcolor}\sffamily\fontsize{18.000000}{21.600000}\selectfont 15.0\%}%
\end{pgfscope}%
\begin{pgfscope}%
\pgfsetbuttcap%
\pgfsetroundjoin%
\definecolor{currentfill}{rgb}{0.000000,0.000000,0.000000}%
\pgfsetfillcolor{currentfill}%
\pgfsetlinewidth{0.803000pt}%
\definecolor{currentstroke}{rgb}{0.000000,0.000000,0.000000}%
\pgfsetstrokecolor{currentstroke}%
\pgfsetdash{}{0pt}%
\pgfsys@defobject{currentmarker}{\pgfqpoint{-0.048611in}{0.000000in}}{\pgfqpoint{-0.000000in}{0.000000in}}{%
\pgfpathmoveto{\pgfqpoint{-0.000000in}{0.000000in}}%
\pgfpathlineto{\pgfqpoint{-0.048611in}{0.000000in}}%
\pgfusepath{stroke,fill}%
}%
\begin{pgfscope}%
\pgfsys@transformshift{1.458910in}{3.730778in}%
\pgfsys@useobject{currentmarker}{}%
\end{pgfscope}%
\end{pgfscope}%
\begin{pgfscope}%
\definecolor{textcolor}{rgb}{0.000000,0.000000,0.000000}%
\pgfsetstrokecolor{textcolor}%
\pgfsetfillcolor{textcolor}%
\pgftext[x=0.567499in, y=3.635807in, left, base]{\color{textcolor}\sffamily\fontsize{18.000000}{21.600000}\selectfont 20.0\%}%
\end{pgfscope}%
\begin{pgfscope}%
\pgfsetbuttcap%
\pgfsetroundjoin%
\definecolor{currentfill}{rgb}{0.000000,0.000000,0.000000}%
\pgfsetfillcolor{currentfill}%
\pgfsetlinewidth{0.803000pt}%
\definecolor{currentstroke}{rgb}{0.000000,0.000000,0.000000}%
\pgfsetstrokecolor{currentstroke}%
\pgfsetdash{}{0pt}%
\pgfsys@defobject{currentmarker}{\pgfqpoint{-0.048611in}{0.000000in}}{\pgfqpoint{-0.000000in}{0.000000in}}{%
\pgfpathmoveto{\pgfqpoint{-0.000000in}{0.000000in}}%
\pgfpathlineto{\pgfqpoint{-0.048611in}{0.000000in}}%
\pgfusepath{stroke,fill}%
}%
\begin{pgfscope}%
\pgfsys@transformshift{1.458910in}{4.436807in}%
\pgfsys@useobject{currentmarker}{}%
\end{pgfscope}%
\end{pgfscope}%
\begin{pgfscope}%
\definecolor{textcolor}{rgb}{0.000000,0.000000,0.000000}%
\pgfsetstrokecolor{textcolor}%
\pgfsetfillcolor{textcolor}%
\pgftext[x=0.567499in, y=4.341836in, left, base]{\color{textcolor}\sffamily\fontsize{18.000000}{21.600000}\selectfont 25.0\%}%
\end{pgfscope}%
\begin{pgfscope}%
\definecolor{textcolor}{rgb}{0.000000,0.000000,0.000000}%
\pgfsetstrokecolor{textcolor}%
\pgfsetfillcolor{textcolor}%
\pgftext[x=0.511943in,y=2.702073in,,bottom,rotate=90.000000]{\color{textcolor}\sffamily\fontsize{18.000000}{21.600000}\selectfont Misprediction Rate}%
\end{pgfscope}%
\begin{pgfscope}%
\pgfsetrectcap%
\pgfsetmiterjoin%
\pgfsetlinewidth{0.803000pt}%
\definecolor{currentstroke}{rgb}{0.000000,0.000000,0.000000}%
\pgfsetstrokecolor{currentstroke}%
\pgfsetdash{}{0pt}%
\pgfpathmoveto{\pgfqpoint{1.458910in}{0.906664in}}%
\pgfpathlineto{\pgfqpoint{1.458910in}{4.497481in}}%
\pgfusepath{stroke}%
\end{pgfscope}%
\begin{pgfscope}%
\pgfsetrectcap%
\pgfsetmiterjoin%
\pgfsetlinewidth{0.803000pt}%
\definecolor{currentstroke}{rgb}{0.000000,0.000000,0.000000}%
\pgfsetstrokecolor{currentstroke}%
\pgfsetdash{}{0pt}%
\pgfpathmoveto{\pgfqpoint{7.730000in}{0.906664in}}%
\pgfpathlineto{\pgfqpoint{7.730000in}{4.497481in}}%
\pgfusepath{stroke}%
\end{pgfscope}%
\begin{pgfscope}%
\pgfsetrectcap%
\pgfsetmiterjoin%
\pgfsetlinewidth{0.803000pt}%
\definecolor{currentstroke}{rgb}{0.000000,0.000000,0.000000}%
\pgfsetstrokecolor{currentstroke}%
\pgfsetdash{}{0pt}%
\pgfpathmoveto{\pgfqpoint{1.458910in}{0.906664in}}%
\pgfpathlineto{\pgfqpoint{7.730000in}{0.906664in}}%
\pgfusepath{stroke}%
\end{pgfscope}%
\begin{pgfscope}%
\pgfsetrectcap%
\pgfsetmiterjoin%
\pgfsetlinewidth{0.803000pt}%
\definecolor{currentstroke}{rgb}{0.000000,0.000000,0.000000}%
\pgfsetstrokecolor{currentstroke}%
\pgfsetdash{}{0pt}%
\pgfpathmoveto{\pgfqpoint{1.458910in}{4.497481in}}%
\pgfpathlineto{\pgfqpoint{7.730000in}{4.497481in}}%
\pgfusepath{stroke}%
\end{pgfscope}%
\begin{pgfscope}%
\pgfsetbuttcap%
\pgfsetmiterjoin%
\definecolor{currentfill}{rgb}{1.000000,1.000000,1.000000}%
\pgfsetfillcolor{currentfill}%
\pgfsetfillopacity{0.800000}%
\pgfsetlinewidth{1.003750pt}%
\definecolor{currentstroke}{rgb}{0.800000,0.800000,0.800000}%
\pgfsetstrokecolor{currentstroke}%
\pgfsetstrokeopacity{0.800000}%
\pgfsetdash{}{0pt}%
\pgfpathmoveto{\pgfqpoint{3.526851in}{3.563595in}}%
\pgfpathlineto{\pgfqpoint{7.555000in}{3.563595in}}%
\pgfpathquadraticcurveto{\pgfqpoint{7.605000in}{3.563595in}}{\pgfqpoint{7.605000in}{3.613595in}}%
\pgfpathlineto{\pgfqpoint{7.605000in}{4.322481in}}%
\pgfpathquadraticcurveto{\pgfqpoint{7.605000in}{4.372481in}}{\pgfqpoint{7.555000in}{4.372481in}}%
\pgfpathlineto{\pgfqpoint{3.526851in}{4.372481in}}%
\pgfpathquadraticcurveto{\pgfqpoint{3.476851in}{4.372481in}}{\pgfqpoint{3.476851in}{4.322481in}}%
\pgfpathlineto{\pgfqpoint{3.476851in}{3.613595in}}%
\pgfpathquadraticcurveto{\pgfqpoint{3.476851in}{3.563595in}}{\pgfqpoint{3.526851in}{3.563595in}}%
\pgfpathlineto{\pgfqpoint{3.526851in}{3.563595in}}%
\pgfpathclose%
\pgfusepath{stroke,fill}%
\end{pgfscope}%
\begin{pgfscope}%
\pgfsetbuttcap%
\pgfsetmiterjoin%
\definecolor{currentfill}{rgb}{0.819608,0.819608,0.819608}%
\pgfsetfillcolor{currentfill}%
\pgfsetlinewidth{1.003750pt}%
\definecolor{currentstroke}{rgb}{0.000000,0.000000,0.000000}%
\pgfsetstrokecolor{currentstroke}%
\pgfsetdash{}{0pt}%
\pgfpathmoveto{\pgfqpoint{3.576851in}{4.082540in}}%
\pgfpathlineto{\pgfqpoint{4.076851in}{4.082540in}}%
\pgfpathlineto{\pgfqpoint{4.076851in}{4.257540in}}%
\pgfpathlineto{\pgfqpoint{3.576851in}{4.257540in}}%
\pgfpathlineto{\pgfqpoint{3.576851in}{4.082540in}}%
\pgfpathclose%
\pgfusepath{stroke,fill}%
\end{pgfscope}%
\begin{pgfscope}%
\pgfsetbuttcap%
\pgfsetmiterjoin%
\definecolor{currentfill}{rgb}{0.819608,0.819608,0.819608}%
\pgfsetfillcolor{currentfill}%
\pgfsetlinewidth{1.003750pt}%
\definecolor{currentstroke}{rgb}{0.000000,0.000000,0.000000}%
\pgfsetstrokecolor{currentstroke}%
\pgfsetdash{}{0pt}%
\pgfpathmoveto{\pgfqpoint{3.576851in}{4.082540in}}%
\pgfpathlineto{\pgfqpoint{4.076851in}{4.082540in}}%
\pgfpathlineto{\pgfqpoint{4.076851in}{4.257540in}}%
\pgfpathlineto{\pgfqpoint{3.576851in}{4.257540in}}%
\pgfpathlineto{\pgfqpoint{3.576851in}{4.082540in}}%
\pgfpathclose%
\pgfusepath{clip}%
\pgfsys@defobject{currentpattern}{\pgfqpoint{0in}{0in}}{\pgfqpoint{1in}{1in}}{%
\begin{pgfscope}%
\pgfpathrectangle{\pgfqpoint{0in}{0in}}{\pgfqpoint{1in}{1in}}%
\pgfusepath{clip}%
\pgfpathmoveto{\pgfqpoint{-0.500000in}{0.500000in}}%
\pgfpathlineto{\pgfqpoint{0.500000in}{1.500000in}}%
\pgfpathmoveto{\pgfqpoint{-0.333333in}{0.333333in}}%
\pgfpathlineto{\pgfqpoint{0.666667in}{1.333333in}}%
\pgfpathmoveto{\pgfqpoint{-0.166667in}{0.166667in}}%
\pgfpathlineto{\pgfqpoint{0.833333in}{1.166667in}}%
\pgfpathmoveto{\pgfqpoint{0.000000in}{0.000000in}}%
\pgfpathlineto{\pgfqpoint{1.000000in}{1.000000in}}%
\pgfpathmoveto{\pgfqpoint{0.166667in}{-0.166667in}}%
\pgfpathlineto{\pgfqpoint{1.166667in}{0.833333in}}%
\pgfpathmoveto{\pgfqpoint{0.333333in}{-0.333333in}}%
\pgfpathlineto{\pgfqpoint{1.333333in}{0.666667in}}%
\pgfpathmoveto{\pgfqpoint{0.500000in}{-0.500000in}}%
\pgfpathlineto{\pgfqpoint{1.500000in}{0.500000in}}%
\pgfusepath{stroke}%
\end{pgfscope}%
}%
\pgfsys@transformshift{3.576851in}{4.082540in}%
\pgfsys@useobject{currentpattern}{}%
\pgfsys@transformshift{1in}{0in}%
\pgfsys@transformshift{-1in}{0in}%
\pgfsys@transformshift{0in}{1in}%
\end{pgfscope}%
\begin{pgfscope}%
\definecolor{textcolor}{rgb}{0.000000,0.000000,0.000000}%
\pgfsetstrokecolor{textcolor}%
\pgfsetfillcolor{textcolor}%
\pgftext[x=4.276851in,y=4.082540in,left,base]{\color{textcolor}\sffamily\fontsize{18.000000}{21.600000}\selectfont spy branch}%
\end{pgfscope}%
\begin{pgfscope}%
\pgfsetbuttcap%
\pgfsetmiterjoin%
\definecolor{currentfill}{rgb}{0.968627,0.678431,0.811765}%
\pgfsetfillcolor{currentfill}%
\pgfsetlinewidth{1.003750pt}%
\definecolor{currentstroke}{rgb}{0.000000,0.000000,0.000000}%
\pgfsetstrokecolor{currentstroke}%
\pgfsetdash{}{0pt}%
\pgfpathmoveto{\pgfqpoint{3.576851in}{3.715596in}}%
\pgfpathlineto{\pgfqpoint{4.076851in}{3.715596in}}%
\pgfpathlineto{\pgfqpoint{4.076851in}{3.890596in}}%
\pgfpathlineto{\pgfqpoint{3.576851in}{3.890596in}}%
\pgfpathlineto{\pgfqpoint{3.576851in}{3.715596in}}%
\pgfpathclose%
\pgfusepath{stroke,fill}%
\end{pgfscope}%
\begin{pgfscope}%
\definecolor{textcolor}{rgb}{0.000000,0.000000,0.000000}%
\pgfsetstrokecolor{textcolor}%
\pgfsetfillcolor{textcolor}%
\pgftext[x=4.276851in,y=3.715596in,left,base]{\color{textcolor}\sffamily\fontsize{18.000000}{21.600000}\selectfont shadow branch (baseline)}%
\end{pgfscope}%
\end{pgfpicture}%
\makeatother%
\endgroup%

%% file: src/TAGE-SCL/Security-Tags/EL0_EL1-Security-Tag/results.pgf
\begingroup%
\makeatletter%
\begin{pgfpicture}%
\pgfpathrectangle{\pgfpointorigin}{\pgfqpoint{8.000000in}{4.000000in}}%
\pgfusepath{use as bounding box, clip}%
\begin{pgfscope}%
\pgfsetbuttcap%
\pgfsetmiterjoin%
\definecolor{currentfill}{rgb}{1.000000,1.000000,1.000000}%
\pgfsetfillcolor{currentfill}%
\pgfsetlinewidth{0.000000pt}%
\definecolor{currentstroke}{rgb}{1.000000,1.000000,1.000000}%
\pgfsetstrokecolor{currentstroke}%
\pgfsetdash{}{0pt}%
\pgfpathmoveto{\pgfqpoint{0.000000in}{0.000000in}}%
\pgfpathlineto{\pgfqpoint{8.000000in}{0.000000in}}%
\pgfpathlineto{\pgfqpoint{8.000000in}{4.000000in}}%
\pgfpathlineto{\pgfqpoint{0.000000in}{4.000000in}}%
\pgfpathlineto{\pgfqpoint{0.000000in}{0.000000in}}%
\pgfpathclose%
\pgfusepath{fill}%
\end{pgfscope}%
\begin{pgfscope}%
\pgfsetbuttcap%
\pgfsetmiterjoin%
\definecolor{currentfill}{rgb}{1.000000,1.000000,1.000000}%
\pgfsetfillcolor{currentfill}%
\pgfsetlinewidth{0.000000pt}%
\definecolor{currentstroke}{rgb}{0.000000,0.000000,0.000000}%
\pgfsetstrokecolor{currentstroke}%
\pgfsetstrokeopacity{0.000000}%
\pgfsetdash{}{0pt}%
\pgfpathmoveto{\pgfqpoint{1.458910in}{0.609165in}}%
\pgfpathlineto{\pgfqpoint{7.730000in}{0.609165in}}%
\pgfpathlineto{\pgfqpoint{7.730000in}{3.730000in}}%
\pgfpathlineto{\pgfqpoint{1.458910in}{3.730000in}}%
\pgfpathlineto{\pgfqpoint{1.458910in}{0.609165in}}%
\pgfpathclose%
\pgfusepath{fill}%
\end{pgfscope}%
\begin{pgfscope}%
\pgfpathrectangle{\pgfqpoint{1.458910in}{0.609165in}}{\pgfqpoint{6.271090in}{3.120835in}}%
\pgfusepath{clip}%
\pgfsetbuttcap%
\pgfsetmiterjoin%
\definecolor{currentfill}{rgb}{0.819608,0.819608,0.819608}%
\pgfsetfillcolor{currentfill}%
\pgfsetlinewidth{1.003750pt}%
\definecolor{currentstroke}{rgb}{0.000000,0.000000,0.000000}%
\pgfsetstrokecolor{currentstroke}%
\pgfsetdash{}{0pt}%
\pgfpathmoveto{\pgfqpoint{1.743960in}{0.609165in}}%
\pgfpathlineto{\pgfqpoint{4.277734in}{0.609165in}}%
\pgfpathlineto{\pgfqpoint{4.277734in}{3.581389in}}%
\pgfpathlineto{\pgfqpoint{1.743960in}{3.581389in}}%
\pgfpathlineto{\pgfqpoint{1.743960in}{0.609165in}}%
\pgfpathclose%
\pgfusepath{stroke,fill}%
\end{pgfscope}%
\begin{pgfscope}%
\pgfsetbuttcap%
\pgfsetmiterjoin%
\definecolor{currentfill}{rgb}{0.819608,0.819608,0.819608}%
\pgfsetfillcolor{currentfill}%
\pgfsetlinewidth{1.003750pt}%
\definecolor{currentstroke}{rgb}{0.000000,0.000000,0.000000}%
\pgfsetstrokecolor{currentstroke}%
\pgfsetdash{}{0pt}%
\pgfpathrectangle{\pgfqpoint{1.458910in}{0.609165in}}{\pgfqpoint{6.271090in}{3.120835in}}%
\pgfusepath{clip}%
\pgfpathmoveto{\pgfqpoint{1.743960in}{0.609165in}}%
\pgfpathlineto{\pgfqpoint{4.277734in}{0.609165in}}%
\pgfpathlineto{\pgfqpoint{4.277734in}{3.581389in}}%
\pgfpathlineto{\pgfqpoint{1.743960in}{3.581389in}}%
\pgfpathlineto{\pgfqpoint{1.743960in}{0.609165in}}%
\pgfpathclose%
\pgfusepath{clip}%
\pgfsys@defobject{currentpattern}{\pgfqpoint{0in}{0in}}{\pgfqpoint{1in}{1in}}{%
\begin{pgfscope}%
\pgfpathrectangle{\pgfqpoint{0in}{0in}}{\pgfqpoint{1in}{1in}}%
\pgfusepath{clip}%
\pgfpathmoveto{\pgfqpoint{-0.500000in}{0.500000in}}%
\pgfpathlineto{\pgfqpoint{0.500000in}{1.500000in}}%
\pgfpathmoveto{\pgfqpoint{-0.333333in}{0.333333in}}%
\pgfpathlineto{\pgfqpoint{0.666667in}{1.333333in}}%
\pgfpathmoveto{\pgfqpoint{-0.166667in}{0.166667in}}%
\pgfpathlineto{\pgfqpoint{0.833333in}{1.166667in}}%
\pgfpathmoveto{\pgfqpoint{0.000000in}{0.000000in}}%
\pgfpathlineto{\pgfqpoint{1.000000in}{1.000000in}}%
\pgfpathmoveto{\pgfqpoint{0.166667in}{-0.166667in}}%
\pgfpathlineto{\pgfqpoint{1.166667in}{0.833333in}}%
\pgfpathmoveto{\pgfqpoint{0.333333in}{-0.333333in}}%
\pgfpathlineto{\pgfqpoint{1.333333in}{0.666667in}}%
\pgfpathmoveto{\pgfqpoint{0.500000in}{-0.500000in}}%
\pgfpathlineto{\pgfqpoint{1.500000in}{0.500000in}}%
\pgfusepath{stroke}%
\end{pgfscope}%
}%
\pgfsys@transformshift{1.743960in}{0.609165in}%
\pgfsys@useobject{currentpattern}{}%
\pgfsys@transformshift{1in}{0in}%
\pgfsys@useobject{currentpattern}{}%
\pgfsys@transformshift{1in}{0in}%
\pgfsys@useobject{currentpattern}{}%
\pgfsys@transformshift{1in}{0in}%
\pgfsys@transformshift{-3in}{0in}%
\pgfsys@transformshift{0in}{1in}%
\pgfsys@useobject{currentpattern}{}%
\pgfsys@transformshift{1in}{0in}%
\pgfsys@useobject{currentpattern}{}%
\pgfsys@transformshift{1in}{0in}%
\pgfsys@useobject{currentpattern}{}%
\pgfsys@transformshift{1in}{0in}%
\pgfsys@transformshift{-3in}{0in}%
\pgfsys@transformshift{0in}{1in}%
\pgfsys@useobject{currentpattern}{}%
\pgfsys@transformshift{1in}{0in}%
\pgfsys@useobject{currentpattern}{}%
\pgfsys@transformshift{1in}{0in}%
\pgfsys@useobject{currentpattern}{}%
\pgfsys@transformshift{1in}{0in}%
\pgfsys@transformshift{-3in}{0in}%
\pgfsys@transformshift{0in}{1in}%
\end{pgfscope}%
\begin{pgfscope}%
\pgfpathrectangle{\pgfqpoint{1.458910in}{0.609165in}}{\pgfqpoint{6.271090in}{3.120835in}}%
\pgfusepath{clip}%
\pgfsetbuttcap%
\pgfsetmiterjoin%
\definecolor{currentfill}{rgb}{0.819608,0.819608,0.819608}%
\pgfsetfillcolor{currentfill}%
\pgfsetlinewidth{1.003750pt}%
\definecolor{currentstroke}{rgb}{0.000000,0.000000,0.000000}%
\pgfsetstrokecolor{currentstroke}%
\pgfsetdash{}{0pt}%
\pgfpathmoveto{\pgfqpoint{4.911177in}{0.609165in}}%
\pgfpathlineto{\pgfqpoint{7.444950in}{0.609165in}}%
\pgfpathlineto{\pgfqpoint{7.444950in}{0.697101in}}%
\pgfpathlineto{\pgfqpoint{4.911177in}{0.697101in}}%
\pgfpathlineto{\pgfqpoint{4.911177in}{0.609165in}}%
\pgfpathclose%
\pgfusepath{stroke,fill}%
\end{pgfscope}%
\begin{pgfscope}%
\pgfsetbuttcap%
\pgfsetmiterjoin%
\definecolor{currentfill}{rgb}{0.819608,0.819608,0.819608}%
\pgfsetfillcolor{currentfill}%
\pgfsetlinewidth{1.003750pt}%
\definecolor{currentstroke}{rgb}{0.000000,0.000000,0.000000}%
\pgfsetstrokecolor{currentstroke}%
\pgfsetdash{}{0pt}%
\pgfpathrectangle{\pgfqpoint{1.458910in}{0.609165in}}{\pgfqpoint{6.271090in}{3.120835in}}%
\pgfusepath{clip}%
\pgfpathmoveto{\pgfqpoint{4.911177in}{0.609165in}}%
\pgfpathlineto{\pgfqpoint{7.444950in}{0.609165in}}%
\pgfpathlineto{\pgfqpoint{7.444950in}{0.697101in}}%
\pgfpathlineto{\pgfqpoint{4.911177in}{0.697101in}}%
\pgfpathlineto{\pgfqpoint{4.911177in}{0.609165in}}%
\pgfpathclose%
\pgfusepath{clip}%
\pgfsys@defobject{currentpattern}{\pgfqpoint{0in}{0in}}{\pgfqpoint{1in}{1in}}{%
\begin{pgfscope}%
\pgfpathrectangle{\pgfqpoint{0in}{0in}}{\pgfqpoint{1in}{1in}}%
\pgfusepath{clip}%
\pgfpathmoveto{\pgfqpoint{-0.500000in}{0.500000in}}%
\pgfpathlineto{\pgfqpoint{0.500000in}{1.500000in}}%
\pgfpathmoveto{\pgfqpoint{-0.333333in}{0.333333in}}%
\pgfpathlineto{\pgfqpoint{0.666667in}{1.333333in}}%
\pgfpathmoveto{\pgfqpoint{-0.166667in}{0.166667in}}%
\pgfpathlineto{\pgfqpoint{0.833333in}{1.166667in}}%
\pgfpathmoveto{\pgfqpoint{0.000000in}{0.000000in}}%
\pgfpathlineto{\pgfqpoint{1.000000in}{1.000000in}}%
\pgfpathmoveto{\pgfqpoint{0.166667in}{-0.166667in}}%
\pgfpathlineto{\pgfqpoint{1.166667in}{0.833333in}}%
\pgfpathmoveto{\pgfqpoint{0.333333in}{-0.333333in}}%
\pgfpathlineto{\pgfqpoint{1.333333in}{0.666667in}}%
\pgfpathmoveto{\pgfqpoint{0.500000in}{-0.500000in}}%
\pgfpathlineto{\pgfqpoint{1.500000in}{0.500000in}}%
\pgfusepath{stroke}%
\end{pgfscope}%
}%
\pgfsys@transformshift{4.911177in}{0.609165in}%
\pgfsys@useobject{currentpattern}{}%
\pgfsys@transformshift{1in}{0in}%
\pgfsys@useobject{currentpattern}{}%
\pgfsys@transformshift{1in}{0in}%
\pgfsys@useobject{currentpattern}{}%
\pgfsys@transformshift{1in}{0in}%
\pgfsys@transformshift{-3in}{0in}%
\pgfsys@transformshift{0in}{1in}%
\end{pgfscope}%
\begin{pgfscope}%
\pgfsetbuttcap%
\pgfsetroundjoin%
\definecolor{currentfill}{rgb}{0.000000,0.000000,0.000000}%
\pgfsetfillcolor{currentfill}%
\pgfsetlinewidth{0.803000pt}%
\definecolor{currentstroke}{rgb}{0.000000,0.000000,0.000000}%
\pgfsetstrokecolor{currentstroke}%
\pgfsetdash{}{0pt}%
\pgfsys@defobject{currentmarker}{\pgfqpoint{0.000000in}{-0.048611in}}{\pgfqpoint{0.000000in}{0.000000in}}{%
\pgfpathmoveto{\pgfqpoint{0.000000in}{0.000000in}}%
\pgfpathlineto{\pgfqpoint{0.000000in}{-0.048611in}}%
\pgfusepath{stroke,fill}%
}%
\begin{pgfscope}%
\pgfsys@transformshift{3.010847in}{0.609165in}%
\pgfsys@useobject{currentmarker}{}%
\end{pgfscope}%
\end{pgfscope}%
\begin{pgfscope}%
\definecolor{textcolor}{rgb}{0.000000,0.000000,0.000000}%
\pgfsetstrokecolor{textcolor}%
\pgfsetfillcolor{textcolor}%
\pgftext[x=3.010847in,y=0.511943in,,top]{\color{textcolor}\sffamily\fontsize{18.000000}{21.600000}\selectfont FIRESTORM}%
\end{pgfscope}%
\begin{pgfscope}%
\pgfsetbuttcap%
\pgfsetroundjoin%
\definecolor{currentfill}{rgb}{0.000000,0.000000,0.000000}%
\pgfsetfillcolor{currentfill}%
\pgfsetlinewidth{0.803000pt}%
\definecolor{currentstroke}{rgb}{0.000000,0.000000,0.000000}%
\pgfsetstrokecolor{currentstroke}%
\pgfsetdash{}{0pt}%
\pgfsys@defobject{currentmarker}{\pgfqpoint{0.000000in}{-0.048611in}}{\pgfqpoint{0.000000in}{0.000000in}}{%
\pgfpathmoveto{\pgfqpoint{0.000000in}{0.000000in}}%
\pgfpathlineto{\pgfqpoint{0.000000in}{-0.048611in}}%
\pgfusepath{stroke,fill}%
}%
\begin{pgfscope}%
\pgfsys@transformshift{6.178064in}{0.609165in}%
\pgfsys@useobject{currentmarker}{}%
\end{pgfscope}%
\end{pgfscope}%
\begin{pgfscope}%
\definecolor{textcolor}{rgb}{0.000000,0.000000,0.000000}%
\pgfsetstrokecolor{textcolor}%
\pgfsetfillcolor{textcolor}%
\pgftext[x=6.178064in,y=0.511943in,,top]{\color{textcolor}\sffamily\fontsize{18.000000}{21.600000}\selectfont ICESTORM}%
\end{pgfscope}%
\begin{pgfscope}%
\pgfsetbuttcap%
\pgfsetroundjoin%
\definecolor{currentfill}{rgb}{0.000000,0.000000,0.000000}%
\pgfsetfillcolor{currentfill}%
\pgfsetlinewidth{0.803000pt}%
\definecolor{currentstroke}{rgb}{0.000000,0.000000,0.000000}%
\pgfsetstrokecolor{currentstroke}%
\pgfsetdash{}{0pt}%
\pgfsys@defobject{currentmarker}{\pgfqpoint{-0.048611in}{0.000000in}}{\pgfqpoint{-0.000000in}{0.000000in}}{%
\pgfpathmoveto{\pgfqpoint{-0.000000in}{0.000000in}}%
\pgfpathlineto{\pgfqpoint{-0.048611in}{0.000000in}}%
\pgfusepath{stroke,fill}%
}%
\begin{pgfscope}%
\pgfsys@transformshift{1.458910in}{0.609165in}%
\pgfsys@useobject{currentmarker}{}%
\end{pgfscope}%
\end{pgfscope}%
\begin{pgfscope}%
\definecolor{textcolor}{rgb}{0.000000,0.000000,0.000000}%
\pgfsetstrokecolor{textcolor}%
\pgfsetfillcolor{textcolor}%
\pgftext[x=0.726556in, y=0.514195in, left, base]{\color{textcolor}\sffamily\fontsize{18.000000}{21.600000}\selectfont 0.0\%}%
\end{pgfscope}%
\begin{pgfscope}%
\pgfsetbuttcap%
\pgfsetroundjoin%
\definecolor{currentfill}{rgb}{0.000000,0.000000,0.000000}%
\pgfsetfillcolor{currentfill}%
\pgfsetlinewidth{0.803000pt}%
\definecolor{currentstroke}{rgb}{0.000000,0.000000,0.000000}%
\pgfsetstrokecolor{currentstroke}%
\pgfsetdash{}{0pt}%
\pgfsys@defobject{currentmarker}{\pgfqpoint{-0.048611in}{0.000000in}}{\pgfqpoint{-0.000000in}{0.000000in}}{%
\pgfpathmoveto{\pgfqpoint{-0.000000in}{0.000000in}}%
\pgfpathlineto{\pgfqpoint{-0.048611in}{0.000000in}}%
\pgfusepath{stroke,fill}%
}%
\begin{pgfscope}%
\pgfsys@transformshift{1.458910in}{1.209472in}%
\pgfsys@useobject{currentmarker}{}%
\end{pgfscope}%
\end{pgfscope}%
\begin{pgfscope}%
\definecolor{textcolor}{rgb}{0.000000,0.000000,0.000000}%
\pgfsetstrokecolor{textcolor}%
\pgfsetfillcolor{textcolor}%
\pgftext[x=0.726556in, y=1.114502in, left, base]{\color{textcolor}\sffamily\fontsize{18.000000}{21.600000}\selectfont 5.0\%}%
\end{pgfscope}%
\begin{pgfscope}%
\pgfsetbuttcap%
\pgfsetroundjoin%
\definecolor{currentfill}{rgb}{0.000000,0.000000,0.000000}%
\pgfsetfillcolor{currentfill}%
\pgfsetlinewidth{0.803000pt}%
\definecolor{currentstroke}{rgb}{0.000000,0.000000,0.000000}%
\pgfsetstrokecolor{currentstroke}%
\pgfsetdash{}{0pt}%
\pgfsys@defobject{currentmarker}{\pgfqpoint{-0.048611in}{0.000000in}}{\pgfqpoint{-0.000000in}{0.000000in}}{%
\pgfpathmoveto{\pgfqpoint{-0.000000in}{0.000000in}}%
\pgfpathlineto{\pgfqpoint{-0.048611in}{0.000000in}}%
\pgfusepath{stroke,fill}%
}%
\begin{pgfscope}%
\pgfsys@transformshift{1.458910in}{1.809780in}%
\pgfsys@useobject{currentmarker}{}%
\end{pgfscope}%
\end{pgfscope}%
\begin{pgfscope}%
\definecolor{textcolor}{rgb}{0.000000,0.000000,0.000000}%
\pgfsetstrokecolor{textcolor}%
\pgfsetfillcolor{textcolor}%
\pgftext[x=0.567499in, y=1.714809in, left, base]{\color{textcolor}\sffamily\fontsize{18.000000}{21.600000}\selectfont 10.0\%}%
\end{pgfscope}%
\begin{pgfscope}%
\pgfsetbuttcap%
\pgfsetroundjoin%
\definecolor{currentfill}{rgb}{0.000000,0.000000,0.000000}%
\pgfsetfillcolor{currentfill}%
\pgfsetlinewidth{0.803000pt}%
\definecolor{currentstroke}{rgb}{0.000000,0.000000,0.000000}%
\pgfsetstrokecolor{currentstroke}%
\pgfsetdash{}{0pt}%
\pgfsys@defobject{currentmarker}{\pgfqpoint{-0.048611in}{0.000000in}}{\pgfqpoint{-0.000000in}{0.000000in}}{%
\pgfpathmoveto{\pgfqpoint{-0.000000in}{0.000000in}}%
\pgfpathlineto{\pgfqpoint{-0.048611in}{0.000000in}}%
\pgfusepath{stroke,fill}%
}%
\begin{pgfscope}%
\pgfsys@transformshift{1.458910in}{2.410087in}%
\pgfsys@useobject{currentmarker}{}%
\end{pgfscope}%
\end{pgfscope}%
\begin{pgfscope}%
\definecolor{textcolor}{rgb}{0.000000,0.000000,0.000000}%
\pgfsetstrokecolor{textcolor}%
\pgfsetfillcolor{textcolor}%
\pgftext[x=0.567499in, y=2.315116in, left, base]{\color{textcolor}\sffamily\fontsize{18.000000}{21.600000}\selectfont 15.0\%}%
\end{pgfscope}%
\begin{pgfscope}%
\pgfsetbuttcap%
\pgfsetroundjoin%
\definecolor{currentfill}{rgb}{0.000000,0.000000,0.000000}%
\pgfsetfillcolor{currentfill}%
\pgfsetlinewidth{0.803000pt}%
\definecolor{currentstroke}{rgb}{0.000000,0.000000,0.000000}%
\pgfsetstrokecolor{currentstroke}%
\pgfsetdash{}{0pt}%
\pgfsys@defobject{currentmarker}{\pgfqpoint{-0.048611in}{0.000000in}}{\pgfqpoint{-0.000000in}{0.000000in}}{%
\pgfpathmoveto{\pgfqpoint{-0.000000in}{0.000000in}}%
\pgfpathlineto{\pgfqpoint{-0.048611in}{0.000000in}}%
\pgfusepath{stroke,fill}%
}%
\begin{pgfscope}%
\pgfsys@transformshift{1.458910in}{3.010394in}%
\pgfsys@useobject{currentmarker}{}%
\end{pgfscope}%
\end{pgfscope}%
\begin{pgfscope}%
\definecolor{textcolor}{rgb}{0.000000,0.000000,0.000000}%
\pgfsetstrokecolor{textcolor}%
\pgfsetfillcolor{textcolor}%
\pgftext[x=0.567499in, y=2.915423in, left, base]{\color{textcolor}\sffamily\fontsize{18.000000}{21.600000}\selectfont 20.0\%}%
\end{pgfscope}%
\begin{pgfscope}%
\pgfsetbuttcap%
\pgfsetroundjoin%
\definecolor{currentfill}{rgb}{0.000000,0.000000,0.000000}%
\pgfsetfillcolor{currentfill}%
\pgfsetlinewidth{0.803000pt}%
\definecolor{currentstroke}{rgb}{0.000000,0.000000,0.000000}%
\pgfsetstrokecolor{currentstroke}%
\pgfsetdash{}{0pt}%
\pgfsys@defobject{currentmarker}{\pgfqpoint{-0.048611in}{0.000000in}}{\pgfqpoint{-0.000000in}{0.000000in}}{%
\pgfpathmoveto{\pgfqpoint{-0.000000in}{0.000000in}}%
\pgfpathlineto{\pgfqpoint{-0.048611in}{0.000000in}}%
\pgfusepath{stroke,fill}%
}%
\begin{pgfscope}%
\pgfsys@transformshift{1.458910in}{3.610701in}%
\pgfsys@useobject{currentmarker}{}%
\end{pgfscope}%
\end{pgfscope}%
\begin{pgfscope}%
\definecolor{textcolor}{rgb}{0.000000,0.000000,0.000000}%
\pgfsetstrokecolor{textcolor}%
\pgfsetfillcolor{textcolor}%
\pgftext[x=0.567499in, y=3.515730in, left, base]{\color{textcolor}\sffamily\fontsize{18.000000}{21.600000}\selectfont 25.0\%}%
\end{pgfscope}%
\begin{pgfscope}%
\definecolor{textcolor}{rgb}{0.000000,0.000000,0.000000}%
\pgfsetstrokecolor{textcolor}%
\pgfsetfillcolor{textcolor}%
\pgftext[x=0.511943in,y=2.169583in,,bottom,rotate=90.000000]{\color{textcolor}\sffamily\fontsize{18.000000}{21.600000}\selectfont Misprediction Rate}%
\end{pgfscope}%
\begin{pgfscope}%
\pgfsetrectcap%
\pgfsetmiterjoin%
\pgfsetlinewidth{0.803000pt}%
\definecolor{currentstroke}{rgb}{0.000000,0.000000,0.000000}%
\pgfsetstrokecolor{currentstroke}%
\pgfsetdash{}{0pt}%
\pgfpathmoveto{\pgfqpoint{1.458910in}{0.609165in}}%
\pgfpathlineto{\pgfqpoint{1.458910in}{3.730000in}}%
\pgfusepath{stroke}%
\end{pgfscope}%
\begin{pgfscope}%
\pgfsetrectcap%
\pgfsetmiterjoin%
\pgfsetlinewidth{0.803000pt}%
\definecolor{currentstroke}{rgb}{0.000000,0.000000,0.000000}%
\pgfsetstrokecolor{currentstroke}%
\pgfsetdash{}{0pt}%
\pgfpathmoveto{\pgfqpoint{7.730000in}{0.609165in}}%
\pgfpathlineto{\pgfqpoint{7.730000in}{3.730000in}}%
\pgfusepath{stroke}%
\end{pgfscope}%
\begin{pgfscope}%
\pgfsetrectcap%
\pgfsetmiterjoin%
\pgfsetlinewidth{0.803000pt}%
\definecolor{currentstroke}{rgb}{0.000000,0.000000,0.000000}%
\pgfsetstrokecolor{currentstroke}%
\pgfsetdash{}{0pt}%
\pgfpathmoveto{\pgfqpoint{1.458910in}{0.609165in}}%
\pgfpathlineto{\pgfqpoint{7.730000in}{0.609165in}}%
\pgfusepath{stroke}%
\end{pgfscope}%
\begin{pgfscope}%
\pgfsetrectcap%
\pgfsetmiterjoin%
\pgfsetlinewidth{0.803000pt}%
\definecolor{currentstroke}{rgb}{0.000000,0.000000,0.000000}%
\pgfsetstrokecolor{currentstroke}%
\pgfsetdash{}{0pt}%
\pgfpathmoveto{\pgfqpoint{1.458910in}{3.730000in}}%
\pgfpathlineto{\pgfqpoint{7.730000in}{3.730000in}}%
\pgfusepath{stroke}%
\end{pgfscope}%
\begin{pgfscope}%
\pgfsetbuttcap%
\pgfsetmiterjoin%
\definecolor{currentfill}{rgb}{1.000000,1.000000,1.000000}%
\pgfsetfillcolor{currentfill}%
\pgfsetfillopacity{0.800000}%
\pgfsetlinewidth{1.003750pt}%
\definecolor{currentstroke}{rgb}{0.800000,0.800000,0.800000}%
\pgfsetstrokecolor{currentstroke}%
\pgfsetstrokeopacity{0.800000}%
\pgfsetdash{}{0pt}%
\pgfpathmoveto{\pgfqpoint{4.749995in}{3.163057in}}%
\pgfpathlineto{\pgfqpoint{7.555000in}{3.163057in}}%
\pgfpathquadraticcurveto{\pgfqpoint{7.605000in}{3.163057in}}{\pgfqpoint{7.605000in}{3.213057in}}%
\pgfpathlineto{\pgfqpoint{7.605000in}{3.555000in}}%
\pgfpathquadraticcurveto{\pgfqpoint{7.605000in}{3.605000in}}{\pgfqpoint{7.555000in}{3.605000in}}%
\pgfpathlineto{\pgfqpoint{4.749995in}{3.605000in}}%
\pgfpathquadraticcurveto{\pgfqpoint{4.699995in}{3.605000in}}{\pgfqpoint{4.699995in}{3.555000in}}%
\pgfpathlineto{\pgfqpoint{4.699995in}{3.213057in}}%
\pgfpathquadraticcurveto{\pgfqpoint{4.699995in}{3.163057in}}{\pgfqpoint{4.749995in}{3.163057in}}%
\pgfpathlineto{\pgfqpoint{4.749995in}{3.163057in}}%
\pgfpathclose%
\pgfusepath{stroke,fill}%
\end{pgfscope}%
\begin{pgfscope}%
\pgfsetbuttcap%
\pgfsetmiterjoin%
\definecolor{currentfill}{rgb}{0.819608,0.819608,0.819608}%
\pgfsetfillcolor{currentfill}%
\pgfsetlinewidth{1.003750pt}%
\definecolor{currentstroke}{rgb}{0.000000,0.000000,0.000000}%
\pgfsetstrokecolor{currentstroke}%
\pgfsetdash{}{0pt}%
\pgfpathmoveto{\pgfqpoint{4.799995in}{3.315059in}}%
\pgfpathlineto{\pgfqpoint{5.299995in}{3.315059in}}%
\pgfpathlineto{\pgfqpoint{5.299995in}{3.490059in}}%
\pgfpathlineto{\pgfqpoint{4.799995in}{3.490059in}}%
\pgfpathlineto{\pgfqpoint{4.799995in}{3.315059in}}%
\pgfpathclose%
\pgfusepath{stroke,fill}%
\end{pgfscope}%
\begin{pgfscope}%
\pgfsetbuttcap%
\pgfsetmiterjoin%
\definecolor{currentfill}{rgb}{0.819608,0.819608,0.819608}%
\pgfsetfillcolor{currentfill}%
\pgfsetlinewidth{1.003750pt}%
\definecolor{currentstroke}{rgb}{0.000000,0.000000,0.000000}%
\pgfsetstrokecolor{currentstroke}%
\pgfsetdash{}{0pt}%
\pgfpathmoveto{\pgfqpoint{4.799995in}{3.315059in}}%
\pgfpathlineto{\pgfqpoint{5.299995in}{3.315059in}}%
\pgfpathlineto{\pgfqpoint{5.299995in}{3.490059in}}%
\pgfpathlineto{\pgfqpoint{4.799995in}{3.490059in}}%
\pgfpathlineto{\pgfqpoint{4.799995in}{3.315059in}}%
\pgfpathclose%
\pgfusepath{clip}%
\pgfsys@defobject{currentpattern}{\pgfqpoint{0in}{0in}}{\pgfqpoint{1in}{1in}}{%
\begin{pgfscope}%
\pgfpathrectangle{\pgfqpoint{0in}{0in}}{\pgfqpoint{1in}{1in}}%
\pgfusepath{clip}%
\pgfpathmoveto{\pgfqpoint{-0.500000in}{0.500000in}}%
\pgfpathlineto{\pgfqpoint{0.500000in}{1.500000in}}%
\pgfpathmoveto{\pgfqpoint{-0.333333in}{0.333333in}}%
\pgfpathlineto{\pgfqpoint{0.666667in}{1.333333in}}%
\pgfpathmoveto{\pgfqpoint{-0.166667in}{0.166667in}}%
\pgfpathlineto{\pgfqpoint{0.833333in}{1.166667in}}%
\pgfpathmoveto{\pgfqpoint{0.000000in}{0.000000in}}%
\pgfpathlineto{\pgfqpoint{1.000000in}{1.000000in}}%
\pgfpathmoveto{\pgfqpoint{0.166667in}{-0.166667in}}%
\pgfpathlineto{\pgfqpoint{1.166667in}{0.833333in}}%
\pgfpathmoveto{\pgfqpoint{0.333333in}{-0.333333in}}%
\pgfpathlineto{\pgfqpoint{1.333333in}{0.666667in}}%
\pgfpathmoveto{\pgfqpoint{0.500000in}{-0.500000in}}%
\pgfpathlineto{\pgfqpoint{1.500000in}{0.500000in}}%
\pgfusepath{stroke}%
\end{pgfscope}%
}%
\pgfsys@transformshift{4.799995in}{3.315059in}%
\pgfsys@useobject{currentpattern}{}%
\pgfsys@transformshift{1in}{0in}%
\pgfsys@transformshift{-1in}{0in}%
\pgfsys@transformshift{0in}{1in}%
\end{pgfscope}%
\begin{pgfscope}%
\definecolor{textcolor}{rgb}{0.000000,0.000000,0.000000}%
\pgfsetstrokecolor{textcolor}%
\pgfsetfillcolor{textcolor}%
\pgftext[x=5.499995in,y=3.315059in,left,base]{\color{textcolor}\sffamily\fontsize{18.000000}{21.600000}\selectfont victim MPR}%
\end{pgfscope}%
\end{pgfpicture}%
\makeatother%
\endgroup%

%% file: src/TAGE-SCL/Security-Tags/PID-Security-Tag/results.pgf
\begingroup%
\makeatletter%
\begin{pgfpicture}%
\pgfpathrectangle{\pgfpointorigin}{\pgfqpoint{8.000000in}{4.000000in}}%
\pgfusepath{use as bounding box, clip}%
\begin{pgfscope}%
\pgfsetbuttcap%
\pgfsetmiterjoin%
\definecolor{currentfill}{rgb}{1.000000,1.000000,1.000000}%
\pgfsetfillcolor{currentfill}%
\pgfsetlinewidth{0.000000pt}%
\definecolor{currentstroke}{rgb}{1.000000,1.000000,1.000000}%
\pgfsetstrokecolor{currentstroke}%
\pgfsetdash{}{0pt}%
\pgfpathmoveto{\pgfqpoint{0.000000in}{0.000000in}}%
\pgfpathlineto{\pgfqpoint{8.000000in}{0.000000in}}%
\pgfpathlineto{\pgfqpoint{8.000000in}{4.000000in}}%
\pgfpathlineto{\pgfqpoint{0.000000in}{4.000000in}}%
\pgfpathlineto{\pgfqpoint{0.000000in}{0.000000in}}%
\pgfpathclose%
\pgfusepath{fill}%
\end{pgfscope}%
\begin{pgfscope}%
\pgfsetbuttcap%
\pgfsetmiterjoin%
\definecolor{currentfill}{rgb}{1.000000,1.000000,1.000000}%
\pgfsetfillcolor{currentfill}%
\pgfsetlinewidth{0.000000pt}%
\definecolor{currentstroke}{rgb}{0.000000,0.000000,0.000000}%
\pgfsetstrokecolor{currentstroke}%
\pgfsetstrokeopacity{0.000000}%
\pgfsetdash{}{0pt}%
\pgfpathmoveto{\pgfqpoint{1.458910in}{0.609165in}}%
\pgfpathlineto{\pgfqpoint{7.730000in}{0.609165in}}%
\pgfpathlineto{\pgfqpoint{7.730000in}{3.729211in}}%
\pgfpathlineto{\pgfqpoint{1.458910in}{3.729211in}}%
\pgfpathlineto{\pgfqpoint{1.458910in}{0.609165in}}%
\pgfpathclose%
\pgfusepath{fill}%
\end{pgfscope}%
\begin{pgfscope}%
\pgfpathrectangle{\pgfqpoint{1.458910in}{0.609165in}}{\pgfqpoint{6.271090in}{3.120045in}}%
\pgfusepath{clip}%
\pgfsetbuttcap%
\pgfsetmiterjoin%
\definecolor{currentfill}{rgb}{0.819608,0.819608,0.819608}%
\pgfsetfillcolor{currentfill}%
\pgfsetlinewidth{1.003750pt}%
\definecolor{currentstroke}{rgb}{0.000000,0.000000,0.000000}%
\pgfsetstrokecolor{currentstroke}%
\pgfsetdash{}{0pt}%
\pgfpathmoveto{\pgfqpoint{1.743960in}{0.609165in}}%
\pgfpathlineto{\pgfqpoint{4.277734in}{0.609165in}}%
\pgfpathlineto{\pgfqpoint{4.277734in}{3.568822in}}%
\pgfpathlineto{\pgfqpoint{1.743960in}{3.568822in}}%
\pgfpathlineto{\pgfqpoint{1.743960in}{0.609165in}}%
\pgfpathclose%
\pgfusepath{stroke,fill}%
\end{pgfscope}%
\begin{pgfscope}%
\pgfsetbuttcap%
\pgfsetmiterjoin%
\definecolor{currentfill}{rgb}{0.819608,0.819608,0.819608}%
\pgfsetfillcolor{currentfill}%
\pgfsetlinewidth{1.003750pt}%
\definecolor{currentstroke}{rgb}{0.000000,0.000000,0.000000}%
\pgfsetstrokecolor{currentstroke}%
\pgfsetdash{}{0pt}%
\pgfpathrectangle{\pgfqpoint{1.458910in}{0.609165in}}{\pgfqpoint{6.271090in}{3.120045in}}%
\pgfusepath{clip}%
\pgfpathmoveto{\pgfqpoint{1.743960in}{0.609165in}}%
\pgfpathlineto{\pgfqpoint{4.277734in}{0.609165in}}%
\pgfpathlineto{\pgfqpoint{4.277734in}{3.568822in}}%
\pgfpathlineto{\pgfqpoint{1.743960in}{3.568822in}}%
\pgfpathlineto{\pgfqpoint{1.743960in}{0.609165in}}%
\pgfpathclose%
\pgfusepath{clip}%
\pgfsys@defobject{currentpattern}{\pgfqpoint{0in}{0in}}{\pgfqpoint{1in}{1in}}{%
\begin{pgfscope}%
\pgfpathrectangle{\pgfqpoint{0in}{0in}}{\pgfqpoint{1in}{1in}}%
\pgfusepath{clip}%
\pgfpathmoveto{\pgfqpoint{-0.500000in}{0.500000in}}%
\pgfpathlineto{\pgfqpoint{0.500000in}{1.500000in}}%
\pgfpathmoveto{\pgfqpoint{-0.333333in}{0.333333in}}%
\pgfpathlineto{\pgfqpoint{0.666667in}{1.333333in}}%
\pgfpathmoveto{\pgfqpoint{-0.166667in}{0.166667in}}%
\pgfpathlineto{\pgfqpoint{0.833333in}{1.166667in}}%
\pgfpathmoveto{\pgfqpoint{0.000000in}{0.000000in}}%
\pgfpathlineto{\pgfqpoint{1.000000in}{1.000000in}}%
\pgfpathmoveto{\pgfqpoint{0.166667in}{-0.166667in}}%
\pgfpathlineto{\pgfqpoint{1.166667in}{0.833333in}}%
\pgfpathmoveto{\pgfqpoint{0.333333in}{-0.333333in}}%
\pgfpathlineto{\pgfqpoint{1.333333in}{0.666667in}}%
\pgfpathmoveto{\pgfqpoint{0.500000in}{-0.500000in}}%
\pgfpathlineto{\pgfqpoint{1.500000in}{0.500000in}}%
\pgfusepath{stroke}%
\end{pgfscope}%
}%
\pgfsys@transformshift{1.743960in}{0.609165in}%
\pgfsys@useobject{currentpattern}{}%
\pgfsys@transformshift{1in}{0in}%
\pgfsys@useobject{currentpattern}{}%
\pgfsys@transformshift{1in}{0in}%
\pgfsys@useobject{currentpattern}{}%
\pgfsys@transformshift{1in}{0in}%
\pgfsys@transformshift{-3in}{0in}%
\pgfsys@transformshift{0in}{1in}%
\pgfsys@useobject{currentpattern}{}%
\pgfsys@transformshift{1in}{0in}%
\pgfsys@useobject{currentpattern}{}%
\pgfsys@transformshift{1in}{0in}%
\pgfsys@useobject{currentpattern}{}%
\pgfsys@transformshift{1in}{0in}%
\pgfsys@transformshift{-3in}{0in}%
\pgfsys@transformshift{0in}{1in}%
\pgfsys@useobject{currentpattern}{}%
\pgfsys@transformshift{1in}{0in}%
\pgfsys@useobject{currentpattern}{}%
\pgfsys@transformshift{1in}{0in}%
\pgfsys@useobject{currentpattern}{}%
\pgfsys@transformshift{1in}{0in}%
\pgfsys@transformshift{-3in}{0in}%
\pgfsys@transformshift{0in}{1in}%
\end{pgfscope}%
\begin{pgfscope}%
\pgfpathrectangle{\pgfqpoint{1.458910in}{0.609165in}}{\pgfqpoint{6.271090in}{3.120045in}}%
\pgfusepath{clip}%
\pgfsetbuttcap%
\pgfsetmiterjoin%
\definecolor{currentfill}{rgb}{0.819608,0.819608,0.819608}%
\pgfsetfillcolor{currentfill}%
\pgfsetlinewidth{1.003750pt}%
\definecolor{currentstroke}{rgb}{0.000000,0.000000,0.000000}%
\pgfsetstrokecolor{currentstroke}%
\pgfsetdash{}{0pt}%
\pgfpathmoveto{\pgfqpoint{4.911177in}{0.609165in}}%
\pgfpathlineto{\pgfqpoint{7.444950in}{0.609165in}}%
\pgfpathlineto{\pgfqpoint{7.444950in}{3.580637in}}%
\pgfpathlineto{\pgfqpoint{4.911177in}{3.580637in}}%
\pgfpathlineto{\pgfqpoint{4.911177in}{0.609165in}}%
\pgfpathclose%
\pgfusepath{stroke,fill}%
\end{pgfscope}%
\begin{pgfscope}%
\pgfsetbuttcap%
\pgfsetmiterjoin%
\definecolor{currentfill}{rgb}{0.819608,0.819608,0.819608}%
\pgfsetfillcolor{currentfill}%
\pgfsetlinewidth{1.003750pt}%
\definecolor{currentstroke}{rgb}{0.000000,0.000000,0.000000}%
\pgfsetstrokecolor{currentstroke}%
\pgfsetdash{}{0pt}%
\pgfpathrectangle{\pgfqpoint{1.458910in}{0.609165in}}{\pgfqpoint{6.271090in}{3.120045in}}%
\pgfusepath{clip}%
\pgfpathmoveto{\pgfqpoint{4.911177in}{0.609165in}}%
\pgfpathlineto{\pgfqpoint{7.444950in}{0.609165in}}%
\pgfpathlineto{\pgfqpoint{7.444950in}{3.580637in}}%
\pgfpathlineto{\pgfqpoint{4.911177in}{3.580637in}}%
\pgfpathlineto{\pgfqpoint{4.911177in}{0.609165in}}%
\pgfpathclose%
\pgfusepath{clip}%
\pgfsys@defobject{currentpattern}{\pgfqpoint{0in}{0in}}{\pgfqpoint{1in}{1in}}{%
\begin{pgfscope}%
\pgfpathrectangle{\pgfqpoint{0in}{0in}}{\pgfqpoint{1in}{1in}}%
\pgfusepath{clip}%
\pgfpathmoveto{\pgfqpoint{-0.500000in}{0.500000in}}%
\pgfpathlineto{\pgfqpoint{0.500000in}{1.500000in}}%
\pgfpathmoveto{\pgfqpoint{-0.333333in}{0.333333in}}%
\pgfpathlineto{\pgfqpoint{0.666667in}{1.333333in}}%
\pgfpathmoveto{\pgfqpoint{-0.166667in}{0.166667in}}%
\pgfpathlineto{\pgfqpoint{0.833333in}{1.166667in}}%
\pgfpathmoveto{\pgfqpoint{0.000000in}{0.000000in}}%
\pgfpathlineto{\pgfqpoint{1.000000in}{1.000000in}}%
\pgfpathmoveto{\pgfqpoint{0.166667in}{-0.166667in}}%
\pgfpathlineto{\pgfqpoint{1.166667in}{0.833333in}}%
\pgfpathmoveto{\pgfqpoint{0.333333in}{-0.333333in}}%
\pgfpathlineto{\pgfqpoint{1.333333in}{0.666667in}}%
\pgfpathmoveto{\pgfqpoint{0.500000in}{-0.500000in}}%
\pgfpathlineto{\pgfqpoint{1.500000in}{0.500000in}}%
\pgfusepath{stroke}%
\end{pgfscope}%
}%
\pgfsys@transformshift{4.911177in}{0.609165in}%
\pgfsys@useobject{currentpattern}{}%
\pgfsys@transformshift{1in}{0in}%
\pgfsys@useobject{currentpattern}{}%
\pgfsys@transformshift{1in}{0in}%
\pgfsys@useobject{currentpattern}{}%
\pgfsys@transformshift{1in}{0in}%
\pgfsys@transformshift{-3in}{0in}%
\pgfsys@transformshift{0in}{1in}%
\pgfsys@useobject{currentpattern}{}%
\pgfsys@transformshift{1in}{0in}%
\pgfsys@useobject{currentpattern}{}%
\pgfsys@transformshift{1in}{0in}%
\pgfsys@useobject{currentpattern}{}%
\pgfsys@transformshift{1in}{0in}%
\pgfsys@transformshift{-3in}{0in}%
\pgfsys@transformshift{0in}{1in}%
\pgfsys@useobject{currentpattern}{}%
\pgfsys@transformshift{1in}{0in}%
\pgfsys@useobject{currentpattern}{}%
\pgfsys@transformshift{1in}{0in}%
\pgfsys@useobject{currentpattern}{}%
\pgfsys@transformshift{1in}{0in}%
\pgfsys@transformshift{-3in}{0in}%
\pgfsys@transformshift{0in}{1in}%
\end{pgfscope}%
\begin{pgfscope}%
\pgfsetbuttcap%
\pgfsetroundjoin%
\definecolor{currentfill}{rgb}{0.000000,0.000000,0.000000}%
\pgfsetfillcolor{currentfill}%
\pgfsetlinewidth{0.803000pt}%
\definecolor{currentstroke}{rgb}{0.000000,0.000000,0.000000}%
\pgfsetstrokecolor{currentstroke}%
\pgfsetdash{}{0pt}%
\pgfsys@defobject{currentmarker}{\pgfqpoint{0.000000in}{-0.048611in}}{\pgfqpoint{0.000000in}{0.000000in}}{%
\pgfpathmoveto{\pgfqpoint{0.000000in}{0.000000in}}%
\pgfpathlineto{\pgfqpoint{0.000000in}{-0.048611in}}%
\pgfusepath{stroke,fill}%
}%
\begin{pgfscope}%
\pgfsys@transformshift{3.010847in}{0.609165in}%
\pgfsys@useobject{currentmarker}{}%
\end{pgfscope}%
\end{pgfscope}%
\begin{pgfscope}%
\definecolor{textcolor}{rgb}{0.000000,0.000000,0.000000}%
\pgfsetstrokecolor{textcolor}%
\pgfsetfillcolor{textcolor}%
\pgftext[x=3.010847in,y=0.511943in,,top]{\color{textcolor}\sffamily\fontsize{18.000000}{21.600000}\selectfont FIRESTORM}%
\end{pgfscope}%
\begin{pgfscope}%
\pgfsetbuttcap%
\pgfsetroundjoin%
\definecolor{currentfill}{rgb}{0.000000,0.000000,0.000000}%
\pgfsetfillcolor{currentfill}%
\pgfsetlinewidth{0.803000pt}%
\definecolor{currentstroke}{rgb}{0.000000,0.000000,0.000000}%
\pgfsetstrokecolor{currentstroke}%
\pgfsetdash{}{0pt}%
\pgfsys@defobject{currentmarker}{\pgfqpoint{0.000000in}{-0.048611in}}{\pgfqpoint{0.000000in}{0.000000in}}{%
\pgfpathmoveto{\pgfqpoint{0.000000in}{0.000000in}}%
\pgfpathlineto{\pgfqpoint{0.000000in}{-0.048611in}}%
\pgfusepath{stroke,fill}%
}%
\begin{pgfscope}%
\pgfsys@transformshift{6.178064in}{0.609165in}%
\pgfsys@useobject{currentmarker}{}%
\end{pgfscope}%
\end{pgfscope}%
\begin{pgfscope}%
\definecolor{textcolor}{rgb}{0.000000,0.000000,0.000000}%
\pgfsetstrokecolor{textcolor}%
\pgfsetfillcolor{textcolor}%
\pgftext[x=6.178064in,y=0.511943in,,top]{\color{textcolor}\sffamily\fontsize{18.000000}{21.600000}\selectfont ICESTORM}%
\end{pgfscope}%
\begin{pgfscope}%
\pgfsetbuttcap%
\pgfsetroundjoin%
\definecolor{currentfill}{rgb}{0.000000,0.000000,0.000000}%
\pgfsetfillcolor{currentfill}%
\pgfsetlinewidth{0.803000pt}%
\definecolor{currentstroke}{rgb}{0.000000,0.000000,0.000000}%
\pgfsetstrokecolor{currentstroke}%
\pgfsetdash{}{0pt}%
\pgfsys@defobject{currentmarker}{\pgfqpoint{-0.048611in}{0.000000in}}{\pgfqpoint{-0.000000in}{0.000000in}}{%
\pgfpathmoveto{\pgfqpoint{-0.000000in}{0.000000in}}%
\pgfpathlineto{\pgfqpoint{-0.048611in}{0.000000in}}%
\pgfusepath{stroke,fill}%
}%
\begin{pgfscope}%
\pgfsys@transformshift{1.458910in}{0.609165in}%
\pgfsys@useobject{currentmarker}{}%
\end{pgfscope}%
\end{pgfscope}%
\begin{pgfscope}%
\definecolor{textcolor}{rgb}{0.000000,0.000000,0.000000}%
\pgfsetstrokecolor{textcolor}%
\pgfsetfillcolor{textcolor}%
\pgftext[x=0.726556in, y=0.514195in, left, base]{\color{textcolor}\sffamily\fontsize{18.000000}{21.600000}\selectfont 0.0\%}%
\end{pgfscope}%
\begin{pgfscope}%
\pgfsetbuttcap%
\pgfsetroundjoin%
\definecolor{currentfill}{rgb}{0.000000,0.000000,0.000000}%
\pgfsetfillcolor{currentfill}%
\pgfsetlinewidth{0.803000pt}%
\definecolor{currentstroke}{rgb}{0.000000,0.000000,0.000000}%
\pgfsetstrokecolor{currentstroke}%
\pgfsetdash{}{0pt}%
\pgfsys@defobject{currentmarker}{\pgfqpoint{-0.048611in}{0.000000in}}{\pgfqpoint{-0.000000in}{0.000000in}}{%
\pgfpathmoveto{\pgfqpoint{-0.000000in}{0.000000in}}%
\pgfpathlineto{\pgfqpoint{-0.048611in}{0.000000in}}%
\pgfusepath{stroke,fill}%
}%
\begin{pgfscope}%
\pgfsys@transformshift{1.458910in}{1.214093in}%
\pgfsys@useobject{currentmarker}{}%
\end{pgfscope}%
\end{pgfscope}%
\begin{pgfscope}%
\definecolor{textcolor}{rgb}{0.000000,0.000000,0.000000}%
\pgfsetstrokecolor{textcolor}%
\pgfsetfillcolor{textcolor}%
\pgftext[x=0.726556in, y=1.119123in, left, base]{\color{textcolor}\sffamily\fontsize{18.000000}{21.600000}\selectfont 5.0\%}%
\end{pgfscope}%
\begin{pgfscope}%
\pgfsetbuttcap%
\pgfsetroundjoin%
\definecolor{currentfill}{rgb}{0.000000,0.000000,0.000000}%
\pgfsetfillcolor{currentfill}%
\pgfsetlinewidth{0.803000pt}%
\definecolor{currentstroke}{rgb}{0.000000,0.000000,0.000000}%
\pgfsetstrokecolor{currentstroke}%
\pgfsetdash{}{0pt}%
\pgfsys@defobject{currentmarker}{\pgfqpoint{-0.048611in}{0.000000in}}{\pgfqpoint{-0.000000in}{0.000000in}}{%
\pgfpathmoveto{\pgfqpoint{-0.000000in}{0.000000in}}%
\pgfpathlineto{\pgfqpoint{-0.048611in}{0.000000in}}%
\pgfusepath{stroke,fill}%
}%
\begin{pgfscope}%
\pgfsys@transformshift{1.458910in}{1.819021in}%
\pgfsys@useobject{currentmarker}{}%
\end{pgfscope}%
\end{pgfscope}%
\begin{pgfscope}%
\definecolor{textcolor}{rgb}{0.000000,0.000000,0.000000}%
\pgfsetstrokecolor{textcolor}%
\pgfsetfillcolor{textcolor}%
\pgftext[x=0.567499in, y=1.724050in, left, base]{\color{textcolor}\sffamily\fontsize{18.000000}{21.600000}\selectfont 10.0\%}%
\end{pgfscope}%
\begin{pgfscope}%
\pgfsetbuttcap%
\pgfsetroundjoin%
\definecolor{currentfill}{rgb}{0.000000,0.000000,0.000000}%
\pgfsetfillcolor{currentfill}%
\pgfsetlinewidth{0.803000pt}%
\definecolor{currentstroke}{rgb}{0.000000,0.000000,0.000000}%
\pgfsetstrokecolor{currentstroke}%
\pgfsetdash{}{0pt}%
\pgfsys@defobject{currentmarker}{\pgfqpoint{-0.048611in}{0.000000in}}{\pgfqpoint{-0.000000in}{0.000000in}}{%
\pgfpathmoveto{\pgfqpoint{-0.000000in}{0.000000in}}%
\pgfpathlineto{\pgfqpoint{-0.048611in}{0.000000in}}%
\pgfusepath{stroke,fill}%
}%
\begin{pgfscope}%
\pgfsys@transformshift{1.458910in}{2.423949in}%
\pgfsys@useobject{currentmarker}{}%
\end{pgfscope}%
\end{pgfscope}%
\begin{pgfscope}%
\definecolor{textcolor}{rgb}{0.000000,0.000000,0.000000}%
\pgfsetstrokecolor{textcolor}%
\pgfsetfillcolor{textcolor}%
\pgftext[x=0.567499in, y=2.328978in, left, base]{\color{textcolor}\sffamily\fontsize{18.000000}{21.600000}\selectfont 15.0\%}%
\end{pgfscope}%
\begin{pgfscope}%
\pgfsetbuttcap%
\pgfsetroundjoin%
\definecolor{currentfill}{rgb}{0.000000,0.000000,0.000000}%
\pgfsetfillcolor{currentfill}%
\pgfsetlinewidth{0.803000pt}%
\definecolor{currentstroke}{rgb}{0.000000,0.000000,0.000000}%
\pgfsetstrokecolor{currentstroke}%
\pgfsetdash{}{0pt}%
\pgfsys@defobject{currentmarker}{\pgfqpoint{-0.048611in}{0.000000in}}{\pgfqpoint{-0.000000in}{0.000000in}}{%
\pgfpathmoveto{\pgfqpoint{-0.000000in}{0.000000in}}%
\pgfpathlineto{\pgfqpoint{-0.048611in}{0.000000in}}%
\pgfusepath{stroke,fill}%
}%
\begin{pgfscope}%
\pgfsys@transformshift{1.458910in}{3.028877in}%
\pgfsys@useobject{currentmarker}{}%
\end{pgfscope}%
\end{pgfscope}%
\begin{pgfscope}%
\definecolor{textcolor}{rgb}{0.000000,0.000000,0.000000}%
\pgfsetstrokecolor{textcolor}%
\pgfsetfillcolor{textcolor}%
\pgftext[x=0.567499in, y=2.933906in, left, base]{\color{textcolor}\sffamily\fontsize{18.000000}{21.600000}\selectfont 20.0\%}%
\end{pgfscope}%
\begin{pgfscope}%
\pgfsetbuttcap%
\pgfsetroundjoin%
\definecolor{currentfill}{rgb}{0.000000,0.000000,0.000000}%
\pgfsetfillcolor{currentfill}%
\pgfsetlinewidth{0.803000pt}%
\definecolor{currentstroke}{rgb}{0.000000,0.000000,0.000000}%
\pgfsetstrokecolor{currentstroke}%
\pgfsetdash{}{0pt}%
\pgfsys@defobject{currentmarker}{\pgfqpoint{-0.048611in}{0.000000in}}{\pgfqpoint{-0.000000in}{0.000000in}}{%
\pgfpathmoveto{\pgfqpoint{-0.000000in}{0.000000in}}%
\pgfpathlineto{\pgfqpoint{-0.048611in}{0.000000in}}%
\pgfusepath{stroke,fill}%
}%
\begin{pgfscope}%
\pgfsys@transformshift{1.458910in}{3.633805in}%
\pgfsys@useobject{currentmarker}{}%
\end{pgfscope}%
\end{pgfscope}%
\begin{pgfscope}%
\definecolor{textcolor}{rgb}{0.000000,0.000000,0.000000}%
\pgfsetstrokecolor{textcolor}%
\pgfsetfillcolor{textcolor}%
\pgftext[x=0.567499in, y=3.538834in, left, base]{\color{textcolor}\sffamily\fontsize{18.000000}{21.600000}\selectfont 25.0\%}%
\end{pgfscope}%
\begin{pgfscope}%
\definecolor{textcolor}{rgb}{0.000000,0.000000,0.000000}%
\pgfsetstrokecolor{textcolor}%
\pgfsetfillcolor{textcolor}%
\pgftext[x=0.511943in,y=2.169188in,,bottom,rotate=90.000000]{\color{textcolor}\sffamily\fontsize{18.000000}{21.600000}\selectfont Misprediction Rate}%
\end{pgfscope}%
\begin{pgfscope}%
\pgfsetrectcap%
\pgfsetmiterjoin%
\pgfsetlinewidth{0.803000pt}%
\definecolor{currentstroke}{rgb}{0.000000,0.000000,0.000000}%
\pgfsetstrokecolor{currentstroke}%
\pgfsetdash{}{0pt}%
\pgfpathmoveto{\pgfqpoint{1.458910in}{0.609165in}}%
\pgfpathlineto{\pgfqpoint{1.458910in}{3.729211in}}%
\pgfusepath{stroke}%
\end{pgfscope}%
\begin{pgfscope}%
\pgfsetrectcap%
\pgfsetmiterjoin%
\pgfsetlinewidth{0.803000pt}%
\definecolor{currentstroke}{rgb}{0.000000,0.000000,0.000000}%
\pgfsetstrokecolor{currentstroke}%
\pgfsetdash{}{0pt}%
\pgfpathmoveto{\pgfqpoint{7.730000in}{0.609165in}}%
\pgfpathlineto{\pgfqpoint{7.730000in}{3.729211in}}%
\pgfusepath{stroke}%
\end{pgfscope}%
\begin{pgfscope}%
\pgfsetrectcap%
\pgfsetmiterjoin%
\pgfsetlinewidth{0.803000pt}%
\definecolor{currentstroke}{rgb}{0.000000,0.000000,0.000000}%
\pgfsetstrokecolor{currentstroke}%
\pgfsetdash{}{0pt}%
\pgfpathmoveto{\pgfqpoint{1.458910in}{0.609165in}}%
\pgfpathlineto{\pgfqpoint{7.730000in}{0.609165in}}%
\pgfusepath{stroke}%
\end{pgfscope}%
\begin{pgfscope}%
\pgfsetrectcap%
\pgfsetmiterjoin%
\pgfsetlinewidth{0.803000pt}%
\definecolor{currentstroke}{rgb}{0.000000,0.000000,0.000000}%
\pgfsetstrokecolor{currentstroke}%
\pgfsetdash{}{0pt}%
\pgfpathmoveto{\pgfqpoint{1.458910in}{3.729211in}}%
\pgfpathlineto{\pgfqpoint{7.730000in}{3.729211in}}%
\pgfusepath{stroke}%
\end{pgfscope}%
\begin{pgfscope}%
\pgfsetbuttcap%
\pgfsetmiterjoin%
\definecolor{currentfill}{rgb}{1.000000,1.000000,1.000000}%
\pgfsetfillcolor{currentfill}%
\pgfsetfillopacity{0.800000}%
\pgfsetlinewidth{1.003750pt}%
\definecolor{currentstroke}{rgb}{0.800000,0.800000,0.800000}%
\pgfsetstrokecolor{currentstroke}%
\pgfsetstrokeopacity{0.800000}%
\pgfsetdash{}{0pt}%
\pgfpathmoveto{\pgfqpoint{4.749995in}{3.162268in}}%
\pgfpathlineto{\pgfqpoint{7.555000in}{3.162268in}}%
\pgfpathquadraticcurveto{\pgfqpoint{7.605000in}{3.162268in}}{\pgfqpoint{7.605000in}{3.212268in}}%
\pgfpathlineto{\pgfqpoint{7.605000in}{3.554211in}}%
\pgfpathquadraticcurveto{\pgfqpoint{7.605000in}{3.604211in}}{\pgfqpoint{7.555000in}{3.604211in}}%
\pgfpathlineto{\pgfqpoint{4.749995in}{3.604211in}}%
\pgfpathquadraticcurveto{\pgfqpoint{4.699995in}{3.604211in}}{\pgfqpoint{4.699995in}{3.554211in}}%
\pgfpathlineto{\pgfqpoint{4.699995in}{3.212268in}}%
\pgfpathquadraticcurveto{\pgfqpoint{4.699995in}{3.162268in}}{\pgfqpoint{4.749995in}{3.162268in}}%
\pgfpathlineto{\pgfqpoint{4.749995in}{3.162268in}}%
\pgfpathclose%
\pgfusepath{stroke,fill}%
\end{pgfscope}%
\begin{pgfscope}%
\pgfsetbuttcap%
\pgfsetmiterjoin%
\definecolor{currentfill}{rgb}{0.819608,0.819608,0.819608}%
\pgfsetfillcolor{currentfill}%
\pgfsetlinewidth{1.003750pt}%
\definecolor{currentstroke}{rgb}{0.000000,0.000000,0.000000}%
\pgfsetstrokecolor{currentstroke}%
\pgfsetdash{}{0pt}%
\pgfpathmoveto{\pgfqpoint{4.799995in}{3.314269in}}%
\pgfpathlineto{\pgfqpoint{5.299995in}{3.314269in}}%
\pgfpathlineto{\pgfqpoint{5.299995in}{3.489269in}}%
\pgfpathlineto{\pgfqpoint{4.799995in}{3.489269in}}%
\pgfpathlineto{\pgfqpoint{4.799995in}{3.314269in}}%
\pgfpathclose%
\pgfusepath{stroke,fill}%
\end{pgfscope}%
\begin{pgfscope}%
\pgfsetbuttcap%
\pgfsetmiterjoin%
\definecolor{currentfill}{rgb}{0.819608,0.819608,0.819608}%
\pgfsetfillcolor{currentfill}%
\pgfsetlinewidth{1.003750pt}%
\definecolor{currentstroke}{rgb}{0.000000,0.000000,0.000000}%
\pgfsetstrokecolor{currentstroke}%
\pgfsetdash{}{0pt}%
\pgfpathmoveto{\pgfqpoint{4.799995in}{3.314269in}}%
\pgfpathlineto{\pgfqpoint{5.299995in}{3.314269in}}%
\pgfpathlineto{\pgfqpoint{5.299995in}{3.489269in}}%
\pgfpathlineto{\pgfqpoint{4.799995in}{3.489269in}}%
\pgfpathlineto{\pgfqpoint{4.799995in}{3.314269in}}%
\pgfpathclose%
\pgfusepath{clip}%
\pgfsys@defobject{currentpattern}{\pgfqpoint{0in}{0in}}{\pgfqpoint{1in}{1in}}{%
\begin{pgfscope}%
\pgfpathrectangle{\pgfqpoint{0in}{0in}}{\pgfqpoint{1in}{1in}}%
\pgfusepath{clip}%
\pgfpathmoveto{\pgfqpoint{-0.500000in}{0.500000in}}%
\pgfpathlineto{\pgfqpoint{0.500000in}{1.500000in}}%
\pgfpathmoveto{\pgfqpoint{-0.333333in}{0.333333in}}%
\pgfpathlineto{\pgfqpoint{0.666667in}{1.333333in}}%
\pgfpathmoveto{\pgfqpoint{-0.166667in}{0.166667in}}%
\pgfpathlineto{\pgfqpoint{0.833333in}{1.166667in}}%
\pgfpathmoveto{\pgfqpoint{0.000000in}{0.000000in}}%
\pgfpathlineto{\pgfqpoint{1.000000in}{1.000000in}}%
\pgfpathmoveto{\pgfqpoint{0.166667in}{-0.166667in}}%
\pgfpathlineto{\pgfqpoint{1.166667in}{0.833333in}}%
\pgfpathmoveto{\pgfqpoint{0.333333in}{-0.333333in}}%
\pgfpathlineto{\pgfqpoint{1.333333in}{0.666667in}}%
\pgfpathmoveto{\pgfqpoint{0.500000in}{-0.500000in}}%
\pgfpathlineto{\pgfqpoint{1.500000in}{0.500000in}}%
\pgfusepath{stroke}%
\end{pgfscope}%
}%
\pgfsys@transformshift{4.799995in}{3.314269in}%
\pgfsys@useobject{currentpattern}{}%
\pgfsys@transformshift{1in}{0in}%
\pgfsys@transformshift{-1in}{0in}%
\pgfsys@transformshift{0in}{1in}%
\end{pgfscope}%
\begin{pgfscope}%
\definecolor{textcolor}{rgb}{0.000000,0.000000,0.000000}%
\pgfsetstrokecolor{textcolor}%
\pgfsetfillcolor{textcolor}%
\pgftext[x=5.499995in,y=3.314269in,left,base]{\color{textcolor}\sffamily\fontsize{18.000000}{21.600000}\selectfont victim MPR}%
\end{pgfscope}%
\end{pgfpicture}%
\makeatother%
\endgroup%

%% file: src/MacOS-Environment-Setup/Event-Encoding/cond-branch-mispredictions/PMC5/Random-Condition/results_firestorm/result0xc5.pgf
\begingroup%
\makeatletter%
\begin{pgfpicture}%
\pgfpathrectangle{\pgfpointorigin}{\pgfqpoint{6.400000in}{4.800000in}}%
\pgfusepath{use as bounding box, clip}%
\begin{pgfscope}%
\pgfsetbuttcap%
\pgfsetmiterjoin%
\definecolor{currentfill}{rgb}{1.000000,1.000000,1.000000}%
\pgfsetfillcolor{currentfill}%
\pgfsetlinewidth{0.000000pt}%
\definecolor{currentstroke}{rgb}{1.000000,1.000000,1.000000}%
\pgfsetstrokecolor{currentstroke}%
\pgfsetdash{}{0pt}%
\pgfpathmoveto{\pgfqpoint{0.000000in}{0.000000in}}%
\pgfpathlineto{\pgfqpoint{6.400000in}{0.000000in}}%
\pgfpathlineto{\pgfqpoint{6.400000in}{4.800000in}}%
\pgfpathlineto{\pgfqpoint{0.000000in}{4.800000in}}%
\pgfpathlineto{\pgfqpoint{0.000000in}{0.000000in}}%
\pgfpathclose%
\pgfusepath{fill}%
\end{pgfscope}%
\begin{pgfscope}%
\pgfsetbuttcap%
\pgfsetmiterjoin%
\definecolor{currentfill}{rgb}{1.000000,1.000000,1.000000}%
\pgfsetfillcolor{currentfill}%
\pgfsetlinewidth{0.000000pt}%
\definecolor{currentstroke}{rgb}{0.000000,0.000000,0.000000}%
\pgfsetstrokecolor{currentstroke}%
\pgfsetstrokeopacity{0.000000}%
\pgfsetdash{}{0pt}%
\pgfpathmoveto{\pgfqpoint{1.220385in}{0.906664in}}%
\pgfpathlineto{\pgfqpoint{6.130000in}{0.906664in}}%
\pgfpathlineto{\pgfqpoint{6.130000in}{4.218737in}}%
\pgfpathlineto{\pgfqpoint{1.220385in}{4.218737in}}%
\pgfpathlineto{\pgfqpoint{1.220385in}{0.906664in}}%
\pgfpathclose%
\pgfusepath{fill}%
\end{pgfscope}%
\begin{pgfscope}%
\pgfpathrectangle{\pgfqpoint{1.220385in}{0.906664in}}{\pgfqpoint{4.909615in}{3.312073in}}%
\pgfusepath{clip}%
\pgfsetbuttcap%
\pgfsetmiterjoin%
\definecolor{currentfill}{rgb}{0.819608,0.819608,0.819608}%
\pgfsetfillcolor{currentfill}%
\pgfsetlinewidth{1.003750pt}%
\definecolor{currentstroke}{rgb}{0.000000,0.000000,0.000000}%
\pgfsetstrokecolor{currentstroke}%
\pgfsetdash{}{0pt}%
\pgfpathmoveto{\pgfqpoint{1.443549in}{0.906664in}}%
\pgfpathlineto{\pgfqpoint{2.718774in}{0.906664in}}%
\pgfpathlineto{\pgfqpoint{2.718774in}{4.061019in}}%
\pgfpathlineto{\pgfqpoint{1.443549in}{4.061019in}}%
\pgfpathlineto{\pgfqpoint{1.443549in}{0.906664in}}%
\pgfpathclose%
\pgfusepath{stroke,fill}%
\end{pgfscope}%
\begin{pgfscope}%
\pgfsetbuttcap%
\pgfsetmiterjoin%
\definecolor{currentfill}{rgb}{0.819608,0.819608,0.819608}%
\pgfsetfillcolor{currentfill}%
\pgfsetlinewidth{1.003750pt}%
\definecolor{currentstroke}{rgb}{0.000000,0.000000,0.000000}%
\pgfsetstrokecolor{currentstroke}%
\pgfsetdash{}{0pt}%
\pgfpathrectangle{\pgfqpoint{1.220385in}{0.906664in}}{\pgfqpoint{4.909615in}{3.312073in}}%
\pgfusepath{clip}%
\pgfpathmoveto{\pgfqpoint{1.443549in}{0.906664in}}%
\pgfpathlineto{\pgfqpoint{2.718774in}{0.906664in}}%
\pgfpathlineto{\pgfqpoint{2.718774in}{4.061019in}}%
\pgfpathlineto{\pgfqpoint{1.443549in}{4.061019in}}%
\pgfpathlineto{\pgfqpoint{1.443549in}{0.906664in}}%
\pgfpathclose%
\pgfusepath{clip}%
\pgfsys@defobject{currentpattern}{\pgfqpoint{0in}{0in}}{\pgfqpoint{1in}{1in}}{%
\begin{pgfscope}%
\pgfpathrectangle{\pgfqpoint{0in}{0in}}{\pgfqpoint{1in}{1in}}%
\pgfusepath{clip}%
\pgfpathmoveto{\pgfqpoint{-0.500000in}{0.500000in}}%
\pgfpathlineto{\pgfqpoint{0.500000in}{1.500000in}}%
\pgfpathmoveto{\pgfqpoint{-0.333333in}{0.333333in}}%
\pgfpathlineto{\pgfqpoint{0.666667in}{1.333333in}}%
\pgfpathmoveto{\pgfqpoint{-0.166667in}{0.166667in}}%
\pgfpathlineto{\pgfqpoint{0.833333in}{1.166667in}}%
\pgfpathmoveto{\pgfqpoint{0.000000in}{0.000000in}}%
\pgfpathlineto{\pgfqpoint{1.000000in}{1.000000in}}%
\pgfpathmoveto{\pgfqpoint{0.166667in}{-0.166667in}}%
\pgfpathlineto{\pgfqpoint{1.166667in}{0.833333in}}%
\pgfpathmoveto{\pgfqpoint{0.333333in}{-0.333333in}}%
\pgfpathlineto{\pgfqpoint{1.333333in}{0.666667in}}%
\pgfpathmoveto{\pgfqpoint{0.500000in}{-0.500000in}}%
\pgfpathlineto{\pgfqpoint{1.500000in}{0.500000in}}%
\pgfusepath{stroke}%
\end{pgfscope}%
}%
\pgfsys@transformshift{1.443549in}{0.906664in}%
\pgfsys@useobject{currentpattern}{}%
\pgfsys@transformshift{1in}{0in}%
\pgfsys@useobject{currentpattern}{}%
\pgfsys@transformshift{1in}{0in}%
\pgfsys@transformshift{-2in}{0in}%
\pgfsys@transformshift{0in}{1in}%
\pgfsys@useobject{currentpattern}{}%
\pgfsys@transformshift{1in}{0in}%
\pgfsys@useobject{currentpattern}{}%
\pgfsys@transformshift{1in}{0in}%
\pgfsys@transformshift{-2in}{0in}%
\pgfsys@transformshift{0in}{1in}%
\pgfsys@useobject{currentpattern}{}%
\pgfsys@transformshift{1in}{0in}%
\pgfsys@useobject{currentpattern}{}%
\pgfsys@transformshift{1in}{0in}%
\pgfsys@transformshift{-2in}{0in}%
\pgfsys@transformshift{0in}{1in}%
\pgfsys@useobject{currentpattern}{}%
\pgfsys@transformshift{1in}{0in}%
\pgfsys@useobject{currentpattern}{}%
\pgfsys@transformshift{1in}{0in}%
\pgfsys@transformshift{-2in}{0in}%
\pgfsys@transformshift{0in}{1in}%
\end{pgfscope}%
\begin{pgfscope}%
\pgfpathrectangle{\pgfqpoint{1.220385in}{0.906664in}}{\pgfqpoint{4.909615in}{3.312073in}}%
\pgfusepath{clip}%
\pgfsetbuttcap%
\pgfsetmiterjoin%
\definecolor{currentfill}{rgb}{0.819608,0.819608,0.819608}%
\pgfsetfillcolor{currentfill}%
\pgfsetlinewidth{1.003750pt}%
\definecolor{currentstroke}{rgb}{0.000000,0.000000,0.000000}%
\pgfsetstrokecolor{currentstroke}%
\pgfsetdash{}{0pt}%
\pgfpathmoveto{\pgfqpoint{3.037580in}{0.906664in}}%
\pgfpathlineto{\pgfqpoint{4.312805in}{0.906664in}}%
\pgfpathlineto{\pgfqpoint{4.312805in}{3.913159in}}%
\pgfpathlineto{\pgfqpoint{3.037580in}{3.913159in}}%
\pgfpathlineto{\pgfqpoint{3.037580in}{0.906664in}}%
\pgfpathclose%
\pgfusepath{stroke,fill}%
\end{pgfscope}%
\begin{pgfscope}%
\pgfsetbuttcap%
\pgfsetmiterjoin%
\definecolor{currentfill}{rgb}{0.819608,0.819608,0.819608}%
\pgfsetfillcolor{currentfill}%
\pgfsetlinewidth{1.003750pt}%
\definecolor{currentstroke}{rgb}{0.000000,0.000000,0.000000}%
\pgfsetstrokecolor{currentstroke}%
\pgfsetdash{}{0pt}%
\pgfpathrectangle{\pgfqpoint{1.220385in}{0.906664in}}{\pgfqpoint{4.909615in}{3.312073in}}%
\pgfusepath{clip}%
\pgfpathmoveto{\pgfqpoint{3.037580in}{0.906664in}}%
\pgfpathlineto{\pgfqpoint{4.312805in}{0.906664in}}%
\pgfpathlineto{\pgfqpoint{4.312805in}{3.913159in}}%
\pgfpathlineto{\pgfqpoint{3.037580in}{3.913159in}}%
\pgfpathlineto{\pgfqpoint{3.037580in}{0.906664in}}%
\pgfpathclose%
\pgfusepath{clip}%
\pgfsys@defobject{currentpattern}{\pgfqpoint{0in}{0in}}{\pgfqpoint{1in}{1in}}{%
\begin{pgfscope}%
\pgfpathrectangle{\pgfqpoint{0in}{0in}}{\pgfqpoint{1in}{1in}}%
\pgfusepath{clip}%
\pgfpathmoveto{\pgfqpoint{-0.500000in}{0.500000in}}%
\pgfpathlineto{\pgfqpoint{0.500000in}{1.500000in}}%
\pgfpathmoveto{\pgfqpoint{-0.333333in}{0.333333in}}%
\pgfpathlineto{\pgfqpoint{0.666667in}{1.333333in}}%
\pgfpathmoveto{\pgfqpoint{-0.166667in}{0.166667in}}%
\pgfpathlineto{\pgfqpoint{0.833333in}{1.166667in}}%
\pgfpathmoveto{\pgfqpoint{0.000000in}{0.000000in}}%
\pgfpathlineto{\pgfqpoint{1.000000in}{1.000000in}}%
\pgfpathmoveto{\pgfqpoint{0.166667in}{-0.166667in}}%
\pgfpathlineto{\pgfqpoint{1.166667in}{0.833333in}}%
\pgfpathmoveto{\pgfqpoint{0.333333in}{-0.333333in}}%
\pgfpathlineto{\pgfqpoint{1.333333in}{0.666667in}}%
\pgfpathmoveto{\pgfqpoint{0.500000in}{-0.500000in}}%
\pgfpathlineto{\pgfqpoint{1.500000in}{0.500000in}}%
\pgfusepath{stroke}%
\end{pgfscope}%
}%
\pgfsys@transformshift{3.037580in}{0.906664in}%
\pgfsys@useobject{currentpattern}{}%
\pgfsys@transformshift{1in}{0in}%
\pgfsys@useobject{currentpattern}{}%
\pgfsys@transformshift{1in}{0in}%
\pgfsys@transformshift{-2in}{0in}%
\pgfsys@transformshift{0in}{1in}%
\pgfsys@useobject{currentpattern}{}%
\pgfsys@transformshift{1in}{0in}%
\pgfsys@useobject{currentpattern}{}%
\pgfsys@transformshift{1in}{0in}%
\pgfsys@transformshift{-2in}{0in}%
\pgfsys@transformshift{0in}{1in}%
\pgfsys@useobject{currentpattern}{}%
\pgfsys@transformshift{1in}{0in}%
\pgfsys@useobject{currentpattern}{}%
\pgfsys@transformshift{1in}{0in}%
\pgfsys@transformshift{-2in}{0in}%
\pgfsys@transformshift{0in}{1in}%
\pgfsys@useobject{currentpattern}{}%
\pgfsys@transformshift{1in}{0in}%
\pgfsys@useobject{currentpattern}{}%
\pgfsys@transformshift{1in}{0in}%
\pgfsys@transformshift{-2in}{0in}%
\pgfsys@transformshift{0in}{1in}%
\end{pgfscope}%
\begin{pgfscope}%
\pgfpathrectangle{\pgfqpoint{1.220385in}{0.906664in}}{\pgfqpoint{4.909615in}{3.312073in}}%
\pgfusepath{clip}%
\pgfsetbuttcap%
\pgfsetmiterjoin%
\definecolor{currentfill}{rgb}{0.819608,0.819608,0.819608}%
\pgfsetfillcolor{currentfill}%
\pgfsetlinewidth{1.003750pt}%
\definecolor{currentstroke}{rgb}{0.000000,0.000000,0.000000}%
\pgfsetstrokecolor{currentstroke}%
\pgfsetdash{}{0pt}%
\pgfpathmoveto{\pgfqpoint{4.631611in}{0.906664in}}%
\pgfpathlineto{\pgfqpoint{5.906836in}{0.906664in}}%
\pgfpathlineto{\pgfqpoint{5.906836in}{0.906664in}}%
\pgfpathlineto{\pgfqpoint{4.631611in}{0.906664in}}%
\pgfpathlineto{\pgfqpoint{4.631611in}{0.906664in}}%
\pgfpathclose%
\pgfusepath{stroke,fill}%
\end{pgfscope}%
\begin{pgfscope}%
\pgfsetbuttcap%
\pgfsetmiterjoin%
\definecolor{currentfill}{rgb}{0.819608,0.819608,0.819608}%
\pgfsetfillcolor{currentfill}%
\pgfsetlinewidth{1.003750pt}%
\definecolor{currentstroke}{rgb}{0.000000,0.000000,0.000000}%
\pgfsetstrokecolor{currentstroke}%
\pgfsetdash{}{0pt}%
\pgfpathrectangle{\pgfqpoint{1.220385in}{0.906664in}}{\pgfqpoint{4.909615in}{3.312073in}}%
\pgfusepath{clip}%
\pgfpathmoveto{\pgfqpoint{4.631611in}{0.906664in}}%
\pgfpathlineto{\pgfqpoint{5.906836in}{0.906664in}}%
\pgfpathlineto{\pgfqpoint{5.906836in}{0.906664in}}%
\pgfpathlineto{\pgfqpoint{4.631611in}{0.906664in}}%
\pgfpathlineto{\pgfqpoint{4.631611in}{0.906664in}}%
\pgfpathclose%
\pgfusepath{clip}%
\pgfsys@defobject{currentpattern}{\pgfqpoint{0in}{0in}}{\pgfqpoint{1in}{1in}}{%
\begin{pgfscope}%
\pgfpathrectangle{\pgfqpoint{0in}{0in}}{\pgfqpoint{1in}{1in}}%
\pgfusepath{clip}%
\pgfpathmoveto{\pgfqpoint{-0.500000in}{0.500000in}}%
\pgfpathlineto{\pgfqpoint{0.500000in}{1.500000in}}%
\pgfpathmoveto{\pgfqpoint{-0.333333in}{0.333333in}}%
\pgfpathlineto{\pgfqpoint{0.666667in}{1.333333in}}%
\pgfpathmoveto{\pgfqpoint{-0.166667in}{0.166667in}}%
\pgfpathlineto{\pgfqpoint{0.833333in}{1.166667in}}%
\pgfpathmoveto{\pgfqpoint{0.000000in}{0.000000in}}%
\pgfpathlineto{\pgfqpoint{1.000000in}{1.000000in}}%
\pgfpathmoveto{\pgfqpoint{0.166667in}{-0.166667in}}%
\pgfpathlineto{\pgfqpoint{1.166667in}{0.833333in}}%
\pgfpathmoveto{\pgfqpoint{0.333333in}{-0.333333in}}%
\pgfpathlineto{\pgfqpoint{1.333333in}{0.666667in}}%
\pgfpathmoveto{\pgfqpoint{0.500000in}{-0.500000in}}%
\pgfpathlineto{\pgfqpoint{1.500000in}{0.500000in}}%
\pgfusepath{stroke}%
\end{pgfscope}%
}%
\pgfsys@transformshift{4.631611in}{0.906664in}%
\end{pgfscope}%
\begin{pgfscope}%
\pgfsetbuttcap%
\pgfsetroundjoin%
\definecolor{currentfill}{rgb}{0.000000,0.000000,0.000000}%
\pgfsetfillcolor{currentfill}%
\pgfsetlinewidth{0.803000pt}%
\definecolor{currentstroke}{rgb}{0.000000,0.000000,0.000000}%
\pgfsetstrokecolor{currentstroke}%
\pgfsetdash{}{0pt}%
\pgfsys@defobject{currentmarker}{\pgfqpoint{0.000000in}{-0.048611in}}{\pgfqpoint{0.000000in}{0.000000in}}{%
\pgfpathmoveto{\pgfqpoint{0.000000in}{0.000000in}}%
\pgfpathlineto{\pgfqpoint{0.000000in}{-0.048611in}}%
\pgfusepath{stroke,fill}%
}%
\begin{pgfscope}%
\pgfsys@transformshift{2.081162in}{0.906664in}%
\pgfsys@useobject{currentmarker}{}%
\end{pgfscope}%
\end{pgfscope}%
\begin{pgfscope}%
\definecolor{textcolor}{rgb}{0.000000,0.000000,0.000000}%
\pgfsetstrokecolor{textcolor}%
\pgfsetfillcolor{textcolor}%
\pgftext[x=2.081162in,y=0.809442in,,top]{\color{textcolor}\sffamily\fontsize{18.000000}{21.600000}\selectfont 0}%
\end{pgfscope}%
\begin{pgfscope}%
\pgfsetbuttcap%
\pgfsetroundjoin%
\definecolor{currentfill}{rgb}{0.000000,0.000000,0.000000}%
\pgfsetfillcolor{currentfill}%
\pgfsetlinewidth{0.803000pt}%
\definecolor{currentstroke}{rgb}{0.000000,0.000000,0.000000}%
\pgfsetstrokecolor{currentstroke}%
\pgfsetdash{}{0pt}%
\pgfsys@defobject{currentmarker}{\pgfqpoint{0.000000in}{-0.048611in}}{\pgfqpoint{0.000000in}{0.000000in}}{%
\pgfpathmoveto{\pgfqpoint{0.000000in}{0.000000in}}%
\pgfpathlineto{\pgfqpoint{0.000000in}{-0.048611in}}%
\pgfusepath{stroke,fill}%
}%
\begin{pgfscope}%
\pgfsys@transformshift{3.675192in}{0.906664in}%
\pgfsys@useobject{currentmarker}{}%
\end{pgfscope}%
\end{pgfscope}%
\begin{pgfscope}%
\definecolor{textcolor}{rgb}{0.000000,0.000000,0.000000}%
\pgfsetstrokecolor{textcolor}%
\pgfsetfillcolor{textcolor}%
\pgftext[x=3.675192in,y=0.809442in,,top]{\color{textcolor}\sffamily\fontsize{18.000000}{21.600000}\selectfont 1}%
\end{pgfscope}%
\begin{pgfscope}%
\pgfsetbuttcap%
\pgfsetroundjoin%
\definecolor{currentfill}{rgb}{0.000000,0.000000,0.000000}%
\pgfsetfillcolor{currentfill}%
\pgfsetlinewidth{0.803000pt}%
\definecolor{currentstroke}{rgb}{0.000000,0.000000,0.000000}%
\pgfsetstrokecolor{currentstroke}%
\pgfsetdash{}{0pt}%
\pgfsys@defobject{currentmarker}{\pgfqpoint{0.000000in}{-0.048611in}}{\pgfqpoint{0.000000in}{0.000000in}}{%
\pgfpathmoveto{\pgfqpoint{0.000000in}{0.000000in}}%
\pgfpathlineto{\pgfqpoint{0.000000in}{-0.048611in}}%
\pgfusepath{stroke,fill}%
}%
\begin{pgfscope}%
\pgfsys@transformshift{5.269223in}{0.906664in}%
\pgfsys@useobject{currentmarker}{}%
\end{pgfscope}%
\end{pgfscope}%
\begin{pgfscope}%
\definecolor{textcolor}{rgb}{0.000000,0.000000,0.000000}%
\pgfsetstrokecolor{textcolor}%
\pgfsetfillcolor{textcolor}%
\pgftext[x=5.269223in,y=0.809442in,,top]{\color{textcolor}\sffamily\fontsize{18.000000}{21.600000}\selectfont other}%
\end{pgfscope}%
\begin{pgfscope}%
\definecolor{textcolor}{rgb}{0.000000,0.000000,0.000000}%
\pgfsetstrokecolor{textcolor}%
\pgfsetfillcolor{textcolor}%
\pgftext[x=3.675193in,y=0.511943in,,top]{\color{textcolor}\sffamily\fontsize{18.000000}{21.600000}\selectfont Events PMC5 counted}%
\end{pgfscope}%
\begin{pgfscope}%
\pgfsetbuttcap%
\pgfsetroundjoin%
\definecolor{currentfill}{rgb}{0.000000,0.000000,0.000000}%
\pgfsetfillcolor{currentfill}%
\pgfsetlinewidth{0.803000pt}%
\definecolor{currentstroke}{rgb}{0.000000,0.000000,0.000000}%
\pgfsetstrokecolor{currentstroke}%
\pgfsetdash{}{0pt}%
\pgfsys@defobject{currentmarker}{\pgfqpoint{-0.048611in}{0.000000in}}{\pgfqpoint{-0.000000in}{0.000000in}}{%
\pgfpathmoveto{\pgfqpoint{-0.000000in}{0.000000in}}%
\pgfpathlineto{\pgfqpoint{-0.048611in}{0.000000in}}%
\pgfusepath{stroke,fill}%
}%
\begin{pgfscope}%
\pgfsys@transformshift{1.220385in}{0.906664in}%
\pgfsys@useobject{currentmarker}{}%
\end{pgfscope}%
\end{pgfscope}%
\begin{pgfscope}%
\definecolor{textcolor}{rgb}{0.000000,0.000000,0.000000}%
\pgfsetstrokecolor{textcolor}%
\pgfsetfillcolor{textcolor}%
\pgftext[x=0.726556in, y=0.811693in, left, base]{\color{textcolor}\sffamily\fontsize{18.000000}{21.600000}\selectfont 0\%}%
\end{pgfscope}%
\begin{pgfscope}%
\pgfsetbuttcap%
\pgfsetroundjoin%
\definecolor{currentfill}{rgb}{0.000000,0.000000,0.000000}%
\pgfsetfillcolor{currentfill}%
\pgfsetlinewidth{0.803000pt}%
\definecolor{currentstroke}{rgb}{0.000000,0.000000,0.000000}%
\pgfsetstrokecolor{currentstroke}%
\pgfsetdash{}{0pt}%
\pgfsys@defobject{currentmarker}{\pgfqpoint{-0.048611in}{0.000000in}}{\pgfqpoint{-0.000000in}{0.000000in}}{%
\pgfpathmoveto{\pgfqpoint{-0.000000in}{0.000000in}}%
\pgfpathlineto{\pgfqpoint{-0.048611in}{0.000000in}}%
\pgfusepath{stroke,fill}%
}%
\begin{pgfscope}%
\pgfsys@transformshift{1.220385in}{1.522749in}%
\pgfsys@useobject{currentmarker}{}%
\end{pgfscope}%
\end{pgfscope}%
\begin{pgfscope}%
\definecolor{textcolor}{rgb}{0.000000,0.000000,0.000000}%
\pgfsetstrokecolor{textcolor}%
\pgfsetfillcolor{textcolor}%
\pgftext[x=0.567499in, y=1.427778in, left, base]{\color{textcolor}\sffamily\fontsize{18.000000}{21.600000}\selectfont 10\%}%
\end{pgfscope}%
\begin{pgfscope}%
\pgfsetbuttcap%
\pgfsetroundjoin%
\definecolor{currentfill}{rgb}{0.000000,0.000000,0.000000}%
\pgfsetfillcolor{currentfill}%
\pgfsetlinewidth{0.803000pt}%
\definecolor{currentstroke}{rgb}{0.000000,0.000000,0.000000}%
\pgfsetstrokecolor{currentstroke}%
\pgfsetdash{}{0pt}%
\pgfsys@defobject{currentmarker}{\pgfqpoint{-0.048611in}{0.000000in}}{\pgfqpoint{-0.000000in}{0.000000in}}{%
\pgfpathmoveto{\pgfqpoint{-0.000000in}{0.000000in}}%
\pgfpathlineto{\pgfqpoint{-0.048611in}{0.000000in}}%
\pgfusepath{stroke,fill}%
}%
\begin{pgfscope}%
\pgfsys@transformshift{1.220385in}{2.138834in}%
\pgfsys@useobject{currentmarker}{}%
\end{pgfscope}%
\end{pgfscope}%
\begin{pgfscope}%
\definecolor{textcolor}{rgb}{0.000000,0.000000,0.000000}%
\pgfsetstrokecolor{textcolor}%
\pgfsetfillcolor{textcolor}%
\pgftext[x=0.567499in, y=2.043863in, left, base]{\color{textcolor}\sffamily\fontsize{18.000000}{21.600000}\selectfont 20\%}%
\end{pgfscope}%
\begin{pgfscope}%
\pgfsetbuttcap%
\pgfsetroundjoin%
\definecolor{currentfill}{rgb}{0.000000,0.000000,0.000000}%
\pgfsetfillcolor{currentfill}%
\pgfsetlinewidth{0.803000pt}%
\definecolor{currentstroke}{rgb}{0.000000,0.000000,0.000000}%
\pgfsetstrokecolor{currentstroke}%
\pgfsetdash{}{0pt}%
\pgfsys@defobject{currentmarker}{\pgfqpoint{-0.048611in}{0.000000in}}{\pgfqpoint{-0.000000in}{0.000000in}}{%
\pgfpathmoveto{\pgfqpoint{-0.000000in}{0.000000in}}%
\pgfpathlineto{\pgfqpoint{-0.048611in}{0.000000in}}%
\pgfusepath{stroke,fill}%
}%
\begin{pgfscope}%
\pgfsys@transformshift{1.220385in}{2.754919in}%
\pgfsys@useobject{currentmarker}{}%
\end{pgfscope}%
\end{pgfscope}%
\begin{pgfscope}%
\definecolor{textcolor}{rgb}{0.000000,0.000000,0.000000}%
\pgfsetstrokecolor{textcolor}%
\pgfsetfillcolor{textcolor}%
\pgftext[x=0.567499in, y=2.659948in, left, base]{\color{textcolor}\sffamily\fontsize{18.000000}{21.600000}\selectfont 30\%}%
\end{pgfscope}%
\begin{pgfscope}%
\pgfsetbuttcap%
\pgfsetroundjoin%
\definecolor{currentfill}{rgb}{0.000000,0.000000,0.000000}%
\pgfsetfillcolor{currentfill}%
\pgfsetlinewidth{0.803000pt}%
\definecolor{currentstroke}{rgb}{0.000000,0.000000,0.000000}%
\pgfsetstrokecolor{currentstroke}%
\pgfsetdash{}{0pt}%
\pgfsys@defobject{currentmarker}{\pgfqpoint{-0.048611in}{0.000000in}}{\pgfqpoint{-0.000000in}{0.000000in}}{%
\pgfpathmoveto{\pgfqpoint{-0.000000in}{0.000000in}}%
\pgfpathlineto{\pgfqpoint{-0.048611in}{0.000000in}}%
\pgfusepath{stroke,fill}%
}%
\begin{pgfscope}%
\pgfsys@transformshift{1.220385in}{3.371004in}%
\pgfsys@useobject{currentmarker}{}%
\end{pgfscope}%
\end{pgfscope}%
\begin{pgfscope}%
\definecolor{textcolor}{rgb}{0.000000,0.000000,0.000000}%
\pgfsetstrokecolor{textcolor}%
\pgfsetfillcolor{textcolor}%
\pgftext[x=0.567499in, y=3.276033in, left, base]{\color{textcolor}\sffamily\fontsize{18.000000}{21.600000}\selectfont 40\%}%
\end{pgfscope}%
\begin{pgfscope}%
\pgfsetbuttcap%
\pgfsetroundjoin%
\definecolor{currentfill}{rgb}{0.000000,0.000000,0.000000}%
\pgfsetfillcolor{currentfill}%
\pgfsetlinewidth{0.803000pt}%
\definecolor{currentstroke}{rgb}{0.000000,0.000000,0.000000}%
\pgfsetstrokecolor{currentstroke}%
\pgfsetdash{}{0pt}%
\pgfsys@defobject{currentmarker}{\pgfqpoint{-0.048611in}{0.000000in}}{\pgfqpoint{-0.000000in}{0.000000in}}{%
\pgfpathmoveto{\pgfqpoint{-0.000000in}{0.000000in}}%
\pgfpathlineto{\pgfqpoint{-0.048611in}{0.000000in}}%
\pgfusepath{stroke,fill}%
}%
\begin{pgfscope}%
\pgfsys@transformshift{1.220385in}{3.987089in}%
\pgfsys@useobject{currentmarker}{}%
\end{pgfscope}%
\end{pgfscope}%
\begin{pgfscope}%
\definecolor{textcolor}{rgb}{0.000000,0.000000,0.000000}%
\pgfsetstrokecolor{textcolor}%
\pgfsetfillcolor{textcolor}%
\pgftext[x=0.567499in, y=3.892118in, left, base]{\color{textcolor}\sffamily\fontsize{18.000000}{21.600000}\selectfont 50\%}%
\end{pgfscope}%
\begin{pgfscope}%
\definecolor{textcolor}{rgb}{0.000000,0.000000,0.000000}%
\pgfsetstrokecolor{textcolor}%
\pgfsetfillcolor{textcolor}%
\pgftext[x=0.511943in,y=2.562700in,,bottom,rotate=90.000000]{\color{textcolor}\sffamily\fontsize{18.000000}{21.600000}\selectfont Frequency}%
\end{pgfscope}%
\begin{pgfscope}%
\pgfsetrectcap%
\pgfsetmiterjoin%
\pgfsetlinewidth{0.803000pt}%
\definecolor{currentstroke}{rgb}{0.000000,0.000000,0.000000}%
\pgfsetstrokecolor{currentstroke}%
\pgfsetdash{}{0pt}%
\pgfpathmoveto{\pgfqpoint{1.220385in}{0.906664in}}%
\pgfpathlineto{\pgfqpoint{1.220385in}{4.218737in}}%
\pgfusepath{stroke}%
\end{pgfscope}%
\begin{pgfscope}%
\pgfsetrectcap%
\pgfsetmiterjoin%
\pgfsetlinewidth{0.803000pt}%
\definecolor{currentstroke}{rgb}{0.000000,0.000000,0.000000}%
\pgfsetstrokecolor{currentstroke}%
\pgfsetdash{}{0pt}%
\pgfpathmoveto{\pgfqpoint{6.130000in}{0.906664in}}%
\pgfpathlineto{\pgfqpoint{6.130000in}{4.218737in}}%
\pgfusepath{stroke}%
\end{pgfscope}%
\begin{pgfscope}%
\pgfsetrectcap%
\pgfsetmiterjoin%
\pgfsetlinewidth{0.803000pt}%
\definecolor{currentstroke}{rgb}{0.000000,0.000000,0.000000}%
\pgfsetstrokecolor{currentstroke}%
\pgfsetdash{}{0pt}%
\pgfpathmoveto{\pgfqpoint{1.220385in}{0.906664in}}%
\pgfpathlineto{\pgfqpoint{6.130000in}{0.906664in}}%
\pgfusepath{stroke}%
\end{pgfscope}%
\begin{pgfscope}%
\pgfsetrectcap%
\pgfsetmiterjoin%
\pgfsetlinewidth{0.803000pt}%
\definecolor{currentstroke}{rgb}{0.000000,0.000000,0.000000}%
\pgfsetstrokecolor{currentstroke}%
\pgfsetdash{}{0pt}%
\pgfpathmoveto{\pgfqpoint{1.220385in}{4.218737in}}%
\pgfpathlineto{\pgfqpoint{6.130000in}{4.218737in}}%
\pgfusepath{stroke}%
\end{pgfscope}%
\begin{pgfscope}%
\pgfsetbuttcap%
\pgfsetmiterjoin%
\definecolor{currentfill}{rgb}{1.000000,1.000000,1.000000}%
\pgfsetfillcolor{currentfill}%
\pgfsetfillopacity{0.800000}%
\pgfsetlinewidth{1.003750pt}%
\definecolor{currentstroke}{rgb}{0.800000,0.800000,0.800000}%
\pgfsetstrokecolor{currentstroke}%
\pgfsetstrokeopacity{0.800000}%
\pgfsetdash{}{0pt}%
\pgfpathmoveto{\pgfqpoint{5.855000in}{3.893737in}}%
\pgfpathlineto{\pgfqpoint{5.955000in}{3.893737in}}%
\pgfpathquadraticcurveto{\pgfqpoint{6.005000in}{3.893737in}}{\pgfqpoint{6.005000in}{3.943737in}}%
\pgfpathlineto{\pgfqpoint{6.005000in}{4.043737in}}%
\pgfpathquadraticcurveto{\pgfqpoint{6.005000in}{4.093737in}}{\pgfqpoint{5.955000in}{4.093737in}}%
\pgfpathlineto{\pgfqpoint{5.855000in}{4.093737in}}%
\pgfpathquadraticcurveto{\pgfqpoint{5.805000in}{4.093737in}}{\pgfqpoint{5.805000in}{4.043737in}}%
\pgfpathlineto{\pgfqpoint{5.805000in}{3.943737in}}%
\pgfpathquadraticcurveto{\pgfqpoint{5.805000in}{3.893737in}}{\pgfqpoint{5.855000in}{3.893737in}}%
\pgfpathlineto{\pgfqpoint{5.855000in}{3.893737in}}%
\pgfpathclose%
\pgfusepath{stroke,fill}%
\end{pgfscope}%
\end{pgfpicture}%
\makeatother%
\endgroup%